\documentclass[useAMS,usenatbib]{mn2e}

\usepackage{graphicx}
\usepackage{natbib}
\usepackage{amsmath}
\usepackage{units}
\usepackage{fixltx2e}

\def\lesssim{\mathrel{\hbox{\rlap{\hbox{\lower5pt\hbox{$\sim$}}}\hbox{$<$}}}}
\def\gtrsim{\mathrel{\hbox{\rlap{\hbox{\lower5pt\hbox{$\sim$}}}\hbox{$>$}}}}
\def\crit{\textrm{crit}}
\def\out{\textrm{out}}
\def\esc{\textrm{esc}}

\newcommand{\gslaccuracy}{$10^{-14}$}

\newcommand{\bmax}{$b_{\textrm{max}}$}
\newcommand{\vcrit}{$v_{\textrm{crit}}$}
\newcommand{\ein}{$e_{\textrm{in}}$}
\newcommand{\eout}{e_{\textrm{out}}}
\newcommand{\Msun}{M$_{\odot}$}
\newcommand{\kms}{km s$^{-1}$}
\newcommand{\Rsun}{R$_{\odot}$}

\usepackage{color}

\title[Dynamical formation \& scattering of triples]{Dynamical formation \&
scattering of hierarchical triples: Cross sections, Kozai-Lidov
oscillations, and collisions}

\author[Antognini \& Thompson]{Joseph~M.~O.~Antognini$^{1, 2}$ and Todd
A.~Thompson$^{1, 2}$
\\
$^1$ Department of Astronomy, The Ohio State University, Columbus, Ohio
43210, USA\\
$^2$ Center for Cosmology and Astro-Particle Physics, The Ohio State
University, Columbus, Ohio 43210, USA\\
E-mail: antognini@astronomy.ohio-state.edu}

\begin{document}

\maketitle

\begin{abstract}

Dynamical scattering of binaries and triple systems of stars, planets, and
compact objects may produce highly inclined triple systems subject to
Kozai-Lidov (KL) oscillations, potentially leading to collisions, mergers,
Type Ia supernovae, and other phenomena. We present the results of more than
400 million gravitational scattering experiments of binary-binary,
triple-single, and triple-binary scattering. We compute the cross sections
for all possible outcomes and explore their dependencies on incoming
velocity, mass, semi-major axis, and eccentricity, including analytic fits
and discussion of the velocity dependence.  For the production of new triple
systems by scattering we find that compact triples are preferred, with
ratios of outer to inner semi-major axes of $\sim$few--100, flat or
quasi-thermal eccentricity distributions, and flat distributions in cosine
of the mutual inclination. Dynamically formed triples are thus subject to
strong KL oscillations, the ``eccentric Kozai mechanism,'' and non-secular
effects. For single and binary flyby encounters with triple systems, we
compute the cumulative cross section for changes to the mutual inclination,
eccentricity, and semi-major axis ratio. We apply these results to
scattering events in the field, open clusters, and globular clusters, and
explore the implications for Type Ia supernovae via collisions and mergers,
stellar collisions, and the lifetime and dynamical isolation of triple
systems undergoing KL oscillations. An Appendix provides an analysis of the
velocity dependence of the collision cross section in binary-single
scattering.

\end{abstract}

\section{Introduction}
\label{sec:intro}

Gravitational scattering events are common in globular clusters and are also
dynamically important for many systems in open clusters and the field
\citep{hills+day76}.  Although two-body scattering can be studied
analytically, three-body scattering is too complex to permit general,
practical analytic results \citep{poincare92, sundman07}.  The three-body
scattering problem was therefore not studied in detail until the development
of computers \citep[e.g.,][]{saslaw+74, heggie75, hut+bahcall83}.  Since
then binary-single and binary-binary scattering events have been studied
extensively \citep[e.g.,][and the references therein]{hut83, mikkola83,
hills91, valtonen+mikkola91, sigurdsson+phinney93, bacon+96, fregeau+04},
but little attention has been devoted to the scattering of triple systems or
to the orbital characteristics of triple systems formed from binary-binary
scattering \citep[though see][]{ivanova08, ivanova+08, leigh+11,
leigh+geller12, moeckel+bonnell13, leigh+geller13, leigh+geller15}.  Yet due
to the Kozai-Lidov (KL) mechanism \citep{lidov62, kozai62}, it is now
appreciated that the dynamics of triples may play an important role in a
wide variety of astrophysical phenomena \citep[e.g.,][]{holman+97, ford+00,
miller+hamilton02a, blaes+02, wu+murray03, wen03, fabrycky+tremaine07,
wu+07, ivanova+08, perets+fabrycky09, naoz+11, thompson11, naoz+12,
katz+dong12, shappee+thompson13, antonini+14, antognini+14}.  Furthermore,
it is now known that triple systems are not rare in our Galaxy; demographic
surveys of the field have revealed that triples constitute 10\% of all
stellar systems \citep{duquennoy+mayor91, raghavan+10, tokovinin14,
sana+14}.  

Given the prevalence of triple systems it is important to understand in
detail how triples interact with other stellar systems.  Observations of the
Tarantula Nebula have demonstrated that binary interactions strongly affect
the observed multiplicity fraction \citep{sana+13} so triple interactions
may contribute to the observed multiplicity fraction as well.  Moreover, the
large multiplicity fraction of high mass stars indicates that triple
dynamics may be even more important for these systems
\citep{leigh+geller13}; for example, scattering of triples with binaries may
be an important formation mechanism for quadruple systems.
\citet{raghavan+10} showed that quadruple systems are nearly as common as
triple systems and many even higher-order systems have been discovered
\citep[e.g.,][]{koo+14}, but it is unclear what fraction of such systems are
formed \emph{in situ} versus dynamically \citep{goodwin+kroupa05}.
Quadruple and higher-order systems may be even more dynamically important
than triples since numerical and semi-analytic experiments have demonstrated
that they exhibit stronger KL oscillations than triples over a wider region
of parameter space \citep{pejcha+13, hamers+15}. 

The perturbative influence of interloping stars on hierarchical triple
systems may have an important effect on the long-term evolution of triples.
The KL mechanism can drive the inner binary to very high eccentricities, but
the maximum eccentricity reached is sensitive to the orbital parameters, in
particular the mutual inclination and the outer eccentricity.  In the
standard quadrupole order KL mechanism the tertiary can drive the inner
binary to arbitrarily large eccentricities over a narrow range of
inclinations near 90$^{\circ}$ (the exact value depends on the particular
system).  However, at octupole order, in the so-called ``eccentric KL
mechanism,'' the range of inclinations over which the tertiary can drive the
inner binary to extreme eccentricities is significantly increased,
especially if the eccentricity of the orbit of the tertiary is large.  Thus,
perturbations to the inclination or eccentricity of the tertiary may produce
much stronger KL oscillations for star-star or star-planet systems.  If
these KL oscillations drive the inner binary to sufficiently high
eccentricities, the components of the inner binary will tidally interact
\citep[e.g.,][]{mazeh+shaham79, wu+murray03, fabrycky+tremaine07,
perets+fabrycky09, naoz+11, naoz+12, naoz+fabrycky14}, potentially
explaining the observation that nearly all close binaries
\citep{tokovinin+06} and certain subsets of warm and hot Jupiters
\citep{wu+murray03, wu+07, naoz+11, naoz+12, socrates+12b, dong+14} have
tertiary companions.

Tidal interactions may not always occur in highly inclined triple systems;
\citet{li+14a} showed that some coplanar systems can be driven to very large
eccentricities.  Moreover, certain systems undergo changes in eccentricity
so rapidly that the angular momentum of the inner orbit can change by an
order of magnitude in a single orbit \citep{antonini+perets12, katz+dong12,
seto13, antonini+14, antognini+14}.  These rapid eccentricity oscillations
present the possibility that the components of the inner orbit may be driven
to merger more rapidly than the tidal circularization process can
circularize the orbit and quench KL oscillations.  The components of the
inner binaries of such systems will then not merge gently, but will instead
collide head-on.  These head-on collisions may produce a variety of unusual
astrophysical phenomena depending on the objects comprising the inner
binary.  Most notably, if the inner binary components are two white dwarfs,
these collisions may produce a Type Ia supernova \citep[SN
Ia;][]{thompson11, katz+dong12, hamers+13, kushnir+13, prodan+13}. 

The possible connection between triple dynamics and SNe Ia should be
explored because, despite their crucial role in constraining cosmological
parameters \citep{riess+98, perlmutter+99}, it is unknown whether the
progenitor systems consist of one white dwarf \citep[the single degenerate
model;][]{whelan+iben73, nomoto82} or two \citep[the double degenerate
model;][]{iben+tutukov84, webbink84}.  Although observational evidence
currently favors the double degenerate model \citep{howell11,
maoz+mannucci12, maoz+14}, it is unclear how to drive white dwarf binaries
to merge at a rate large enough to be consistent with the observed SN Ia
rate \citep{ruiter+09, ruiter+11}.  One approach is by driving the binary to
high eccentricity through the perturbative influence of a tertiary via the
KL mechanism.  At high eccentricities the inner binary would then emit much
more gravitational radiation \citep{blaes+02, miller+hamilton02a, wen03},
thereby leading to more rapid coalescence of the WD-WD binary
\citep{thompson11}.  However, it is unclear how to prevent the inner binary
from merging or colliding while still on the main sequence.  KL oscillations
would bring the two stars into tidal contact and tidal circularization would
then shrink the orbit, greatly increasing the semi-major axis ratio and
``freezing'' the inner binary at an inclination close to the critical Kozai
angle \citep{fabrycky+tremaine07}.  These two effects would effectively shut
off further KL oscillations and prevent the inner binary from merging when
the stars evolve into white dwarfs.  The dynamical formation of
high-inclination triples from scattering may be one means to circumvent this
difficulty. 

Scattering of high-multiplicity systems may also be one channel to produce
free-floating planets.  Microlensing studies have revealed that there may be
as many as two free-floating planets for every bound planet in the Galaxy
\citep{zapatero+00, sumi+11}.  Moreover, there is some evidence that it is
difficult for planet-planet scattering to produce free floating planets in
the required numbers \citep{veras+raymond12}.  Dynamical scattering of field
stars off of binary stars may be an important component to the rate of
planet ejection.  

Scattering events involving high-order stellar systems could also be a
source of stellar collisions.  Stellar collisions are estimated to occur at
a rate of $\sim$0.5 yr$^{-1}$ in the Galaxy \citep{kochanek+14} and it is
unknown whether these collisions are due to KL oscillations, scattering
events, or a combination of the two.  Even when scattering does not lead to
a stellar collision the scattering event will change the orbital parameters
of the system.  This will lead to evolution in the distribution of orbital
parameters of the triples in the population.  This evolution would be
strongest in globular clusters due to their high stellar densities and large
ages.

This paper has three goals.  The first is to perform idealized
binary-binary, triple-single, and triple-binary scattering experiments in
order to derive general relationships between the cross sections of various
outcomes with the initial orbital parameters and the incoming velocity.  In
this way our paper is similar to \citet{hut+bahcall83} and \citet{mikkola83}
except that we study higher-order scattering.  The second goal of our paper
is to determine the orbital parameters of triple systems after scattering
events.  The final goal of our paper is to apply the cross sections we
derive in several contexts.  In particular we estimate the rate of formation
of WD-WD binaries in highly inclined triples, the ejection rate of planets
due to scattering events, the stellar collision rate, and the lifetime of
high inclination triples.  To meet these goals we have performed over 400
million scattering experiments.

Details of the numerical methods of this paper are presented in Section
\ref{sec:methods}.  We study triple scattering in detail in Section
\ref{sec:scattering} by performing numerical experiments and comparing them
with analytic approximations in Section \ref{sec:analytic}.  We then discuss
the distribution of orbital parameters of dynamically formed triples in
Section \ref{sec:dyntrip}.  We discuss the implications of these results for
SNe Ia, the longevity of triple systems undergoing KL oscillations, and
provide estimates for the rate of planetary ejection and stellar collisions
in Section \ref{sec:discussion}.  We conclude in Section
\ref{sec:conclusion}.

\section{Numerical methods}
\label{sec:methods}

We use the open source \textsc{Fewbody} suite to perform our scattering
experiments \citep{fregeau+04}.  \textsc{Fewbody} is optimized to
numerically compute the dynamics of systems with small numbers of components
($N \lesssim 10$).  \textsc{Fewbody} uses the ordinary differential
equations library of the GNU Scientific Library (GSL) for its underlying
integrator \citep{gough09}.  The GSL ordinary differential equations library
supports six integration algorithms with adaptive time steps, from which we
use eighth-order Runge-Kutta Prince-Dormand integration.  We find, however,
that the choice of integration algorithm makes little difference to the
results of the calculations because GSL's adaptive time steps are chosen to
target a specified relative and absolute accuracy (\gslaccuracy{} in our
experiments) regardless of the algorithm used.  \textsc{Fewbody} also
supports the use of Kustaanheimo-Stiefel (KS) regularization
\citep{kustaanheimo+stiefel65}, a coordinate transformation which removes
the singularities in the gravitational force present in ordinary $N$-body
integration.  Throughout this paper we use KS regularization as it has the
particular advantage of making eccentric orbits much easier to compute.  

\subsection{Notation}
\label{subsec:notation}

Throughout this paper we use the same notation as \citet{fregeau+04} and
\textsc{Fewbody} in which each star in a system with $n$ stars is labelled
with a unique index running from 0 to $n-1$.  Bound pairs are denoted by
square brackets surrounding the pair and collisions between two stars are
denoted by colons between the pair.  For example, a hierarchical triple with
an unbound interloping star is notated \texttt{[[0~1]~2]~3} (stars 0 and 1
form the inner binary, star 2 is the tertiary, and star 3 is the interloping
star), and a hierarchical triple in which the two stars of the inner binary
have collided is notated \texttt{[0:1~2]}.

Orbital parameters (e.g., semi-major axis or eccentricity) of the component
binaries of the system are notated by subscripts in a top-down fashion.
Leftmost indices in subscripts represent the outermost binaries of the
separate hierarchies, and rightward indices represent inner binaries in
those hierarchies.  For example, in a triple-binary scattering event, the
semi-major axis of the outer binary of the triple is $a_1$, the semi-major
axis of the inner binary of the triple is $a_{11}$, and the semi-major axis
of the interloping binary is $a_2$.  Similarly, the mass of the second star
in the innermost binary of the triple is $m_{112}$, the mass of the tertiary
is $m_{12}$, and the mass of the first star of the interloping binary is
$m_{21}$.  We also combine masses in the subscripts.  Thus the total mass in
the inner binary of the triple is $m_{11} = m_{111} + m_{112}$ and the total
mass of the triple is $m_1 = m_{11} + m_{12}$. 

For clarity we occasionally refer to the orbital parameters of hierarchical
triples using the subscripts ``in'' and ``out.''  Thus the eccentricity of
the inner orbit of a triple may be referred to equivalently by $e_{11}$ or
$e_{\textrm{in}}$ and the eccentricity of the outer orbit by $e_1$ or
$e_{\textrm{out}}$. 

\subsection{Cross sections}
\label{subsec:crosssections}

Throughout this paper we provide cross sections for the outcomes of
scattering events involving triple or higher-order hierarchical systems.
Our treatment of these cross sections and their uncertainties follows that
of \citet{hut+bahcall83}.  We use the usual definition of the cross section for a
particular outcome, $X$:
\begin{equation}
\label{eq:crosssection}
\sigma_X = \pi b_{\max}^2 \frac{n_X}{n_{\textrm{tot}}},
\end{equation}
where $b_{\max}$ is the maximum impact parameter in a set of experiments,
$n_X$ is the number of experiments with outcome $X$, and $n_{\textrm{tot}}$
is the total number of experiments performed.  Throughout this paper we
present our results using normalized cross sections, $\hat{\sigma}$.  The
cross sections from triple-single scattering experiments are normalized to
the area of the outer binary of the triple:
\begin{equation}
\label{eq:normtripsing}
\hat{\sigma} \equiv \frac{\sigma}{\pi a_1^2} \quad \textrm{(triple-single)}.
\end{equation}
The cross sections for binary-binary experiments are normalized to the sum
of the areas of the two binaries and similarly in triple-binary experiments
the cross sections are normalized to the sum of the areas of the outer
binary of the triple and the interloping binary:
\begin{equation}
\label{eq:normtripbin}
\hat{\sigma} \equiv \frac{\sigma}{\pi(a_1^2 + a_2^2)} \quad
\textrm{(binary-binary, triple-binary)}.
\end{equation}

There are two uncertainties in the calculation of the cross sections.  The
first is the statistical uncertainty due to the finite number of experiments
performed:\footnote{In cases where the fraction of outcomes with the outcome
$X$ is close to unity or zero it is better to use a confidence interval like
the Wilson score interval rather than the normal approximation in order to
capture the asymmetry of the statistical uncertainty \citep{wilson27}.
However, we perform enough experiments that the difference between the
Wilson score interval and the normal approximation is almost always
negligible.  Where it is not, we use the Wilson score interval.}
\begin{equation}
\label{eq:statisticaluncertainty}
\Delta_{\textrm{stat}} \sigma_X = \frac{\sigma_X}{\sqrt{n_X}}.
\end{equation}
The second source of uncertainty is due to the fact that certain systems
require prohibitively long computation times to resolve.  For example, while
it is impossible for a single star scattering off of a triple system to
produce a stable quadruple system \citep{chazy29, littlewood52, heggie75,
heggie+hut03}, certain systems can enter into a ``metastable'' state where a
quadruple system is produced that takes an exceedingly long time to
dissociate.  Such unresolved systems produce a separate systematic
uncertainty in the calculation of the cross sections given by 
\begin{equation}
\label{eq:systematicuncertainty}
\Delta_{\textrm{sys}} \sigma_X = \pi b_{\textrm{max}}^2
\frac{n_{\textrm{unres}}}{n_{\textrm{tot}}},
\end{equation}
where $n_{\textrm{unres}}$ is the number of experiments with unresolved
outcomes. 

It is important to note that the systematic uncertainty is completely
asymmetric and only serves to increase the cross section of any outcome.
That is to say, our estimates of the cross sections are lower bounds on the
true cross sections.

\subsection{Initial conditions and halting criteria}
\label{subsec:initialconditions}

Each scattering experiment begins with the interloping system at a large,
but finite, distance from the target system.  We choose the initial
separation between the two systems to be the distance at which the tidal
force on the outer binary of the target system is some small fraction,
$\delta$, of the relative force between the two components of the outer
binary when at apocenter.  Hence, the initial separation, $r$, is given by
\begin{equation}
\label{eq:delta}
\frac{F_{\textrm{tid}}}{F_{\textrm{rel}}} = \delta,
\end{equation}
where
\begin{equation}
\label{eq:tide}
F_{\textrm{tid}} = \frac{2G(m_{11} + m_{12})m_2}{r^3} a(1+e),
\end{equation}
and
\begin{equation}
\label{eq:relativeforce}
F_{\textrm{rel}} = \frac{Gm_{11} m_{12}}{\left[ a (1+e) \right]^2},
\end{equation}
where $a$ and $e$ refer to the semi-major axis and eccentricity,
respectively, of the outer binary of the target system.  Throughout this
paper we use $\delta = 10^{-5}$, the same choice as \citet{fregeau+04}.
Smaller choices of $\delta$ do not change the results but increase the
running time.  The interloping system is then analytically brought along a
hyperbolic orbit from infinity with the $b_{\textrm{max}}$ and velocity at
infinity, $v_{\infty}$, that has been fixed for that particular experiment.  

It is also essential to choose the appropriate \bmax{} when computing cross
sections.  Too small a choice of \bmax{} will result in in an underestimate of
the cross section, whereas too large a choice of \bmax{} will result in few
experiments producing outcomes of interest, thereby leading to large
statistical uncertainties.  We adopt the choice of \bmax{} similar to that
of \citet{hut+bahcall83} of:
\begin{equation}
\label{eq:bmax}
b_{\textrm{max}} = \left( \frac{4 v_{\textrm{crit}}}{v_{\inf}} + 3 \right) 
a_1,
\end{equation}
where $v_{\inf}$ is the incoming velocity at infinity and \vcrit{} is the
critical velocity at which the total energy of the system is zero.  For
binary-binary scattering, \vcrit{} is
\begin{equation}
\label{eq:vcrit_binbin}
v_{\textrm{crit, bin-bin}}^2 = \frac{G (m_1 + m_2)}{m_1 m_2} \left(
\frac{m_{11} m_{12}}{a_1} + \frac{m_{21} m_{22}}{a_2} \right),
\end{equation}

for triple-single scattering, \vcrit{} is 
\begin{equation}
\label{eq:vcrit_tripsing}
v_{\textrm{crit, trip-sing}}^2 = 
\frac{G(a_1 m_{111} m_{112} + a_{11} m_{11} m_{12}) (m_1 +
m_2)}{a_1 a_{11} m_1 m_2},
\end{equation}

and for triple-binary scattering \vcrit{} is
\begin{multline}
\label{eq:vcrit_tripbin}
v_{\textrm{crit, trip-bin}}^2 = 
\frac{G (m_1 + m_2)}{a_1 a_{11} a_2 m_1 m_2} \\
\times \left[a_{11} a_2 m_{11} m_{12} + a_1 (a_2 m_{111} m_{112} + 
a_{11} m_{21} m_{22})\right].
\end{multline}

\begin{figure*}
\centering
\includegraphics[width=17cm]{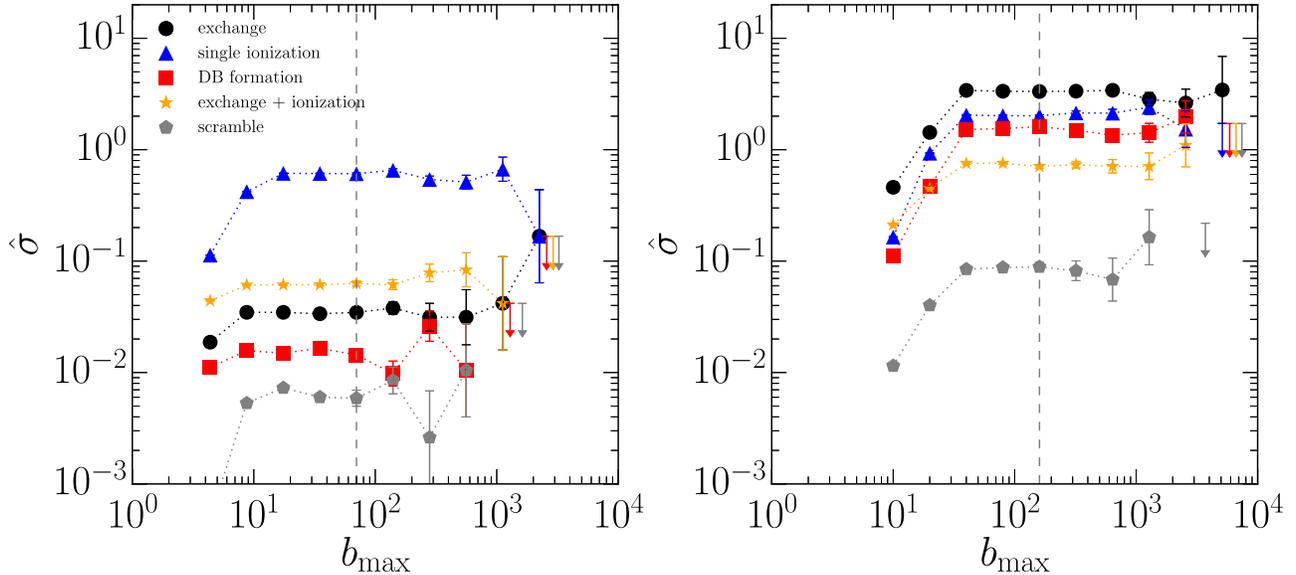}

\caption{Convergence of the scattering cross sections with \bmax{} for
triple-single scattering.  We performed $3 \times 10^5$ scattering
experiments at each choice of \bmax.  \emph{Left panel:} All masses are
equal point masses and the orbits of the initial triple were circular with a
semi-major axis ratio of 10.  The cross sections for four categories of
outcomes are presented: exchanges (black circles), single ionization (blue
triangles), double binary formation (red squares), and exchange-ionization
(orange stars).  (See Section \ref{subsec:parameters} for a description of
the categories.)  \emph{Right panel:} Same as the left panel, but the mass
of one star in the inner binary is set to be a test particle.  The choice
for \bmax{} we adopt in this paper (dashed line) is given by
equation~(\ref{eq:bmax}) and is similar to that of \citet{hut+bahcall83}.
Smaller \bmax's do not probe the full area over which non-flyby outcomes
occur, whereas the statistical power of larger \bmax's is reduced because of
the small fraction of non-flyby outcomes.  The choice we adopt is converged
but is small enough to have strong statistics.} 

\label{fig:bmax}
\end{figure*}

We will frequently refer to velocities in terms of the normalized
velocity, $\hat{v}$, which is the velocity, $v$, scaled to \vcrit{},
\begin{equation}
\label{eq:vhatdef}
\hat{v} \equiv \frac{v}{v_{\textrm{crit}}}.
\end{equation}
We test our choice of $b_{\max}$ by computing the cross sections for four
categories of outcome in triple-single scattering over a broad range of
\bmax.  We demonstrate that the cross sections are well converged in
Fig.~\ref{fig:bmax}.

The calculation halts when one of four conditions is met: (1) if the system
consists of some number of stable hierarchical systems as determined by the
Mardling stability criterion \citep{mardling+aarseth01}:
\begin{equation}
\label{eq:mardling}
\frac{a_1}{a_{11}}(1 - e_1) > 2.8 \left[ \left( 1 + \frac{m_{11}}{m_{12}}
\right) \frac{1 + e_1}{\sqrt{1 - e_1}} \right]^{2/5} \left(1 -
\frac{0.3}{\pi} i \right)
\end{equation}
(where the mutual inclination, $i$, is in radians), and all the systems are
far enough apart that the ratios between their tidal forces to their
relative forces is less than $\delta$; (2) the system integrates for $10^6$
times the initial orbital period of the outer binary of the target system;
(3) the system integrates for one hour; or (4) the total energy or angular
momentum of the system changes from its initial value by more than one part
in $10^3$.  If conditions (2) or (3) are met the outcome is classified as
unresolved and included in our systematic
uncertainty.\footnote{\citet{portegieszwart+boekholt14} have argued that
typical energy conservation standards in $N$-body studies are extremely
conservative and that maintaining energy conservation to better than one
part in 10 is sufficient to preserve the statistical properties of dynamical
experiments.  These results have been confirmed by
\citet{boekholt+portegieszwart15}.  Since very few of our experiments
violate our energy conservation standard, this finding does not change our
results, but does imply that we may have slightly overestimated our
systematic uncertainties.}

\section{Scattering experiments}
\label{sec:scattering}

Hierarchical triples can be formed through scattering in one of three ways:
(1) binary-binary scattering, (2) scattering of triples, or (3) scattering
of higher-order systems.  Binary-binary scattering has been studied
extensively, but only rarely have any studies addressed the formation of
triple systems and, moreover, scattering of higher-order systems in general
has received little attention at all.  We study triple and binary scattering
in detail in this section.  We do not study the scattering of higher-order
systems (e.g., triple-triple, quadruple-single) because such scattering
events are rare in most environments \citep{leigh+geller13}.

\begin{figure*}
\centering
\includegraphics[width=17.75cm]{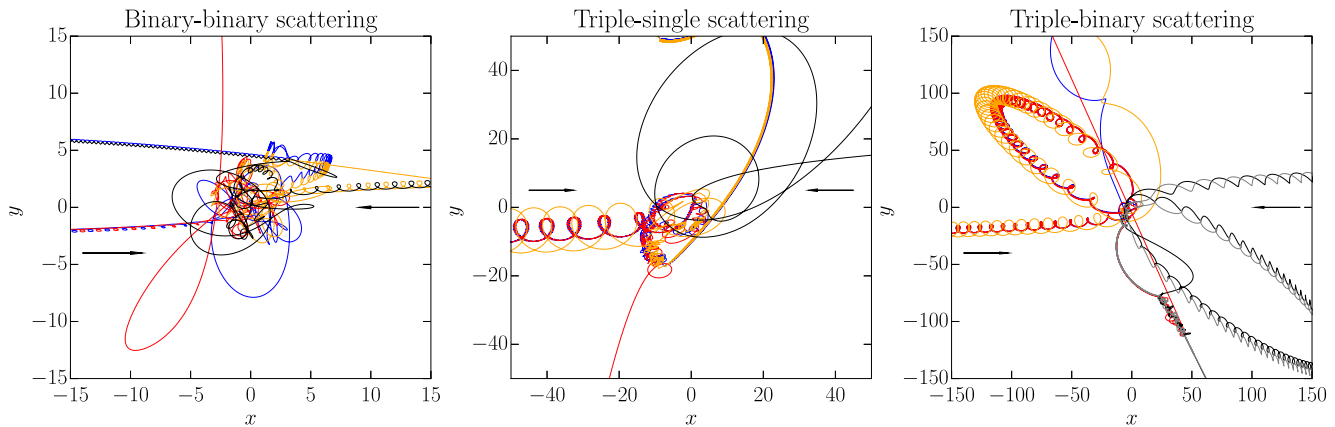}

\caption{A sample binary-binary, triple-single, and triple-binary scattering
experiment projected onto the $xy$-plane.  The incoming velocity is
0.1\vcrit{} in the case of triple scattering and 0.2\vcrit{} in the case of
binary-binary scattering.  The semi-major axis ratio of the triple in the
case of triple scattering is 10 and the semi-major axis of the incoming
binary is equal to the semi-major axis of the outer binary in the case of
triple-binary scattering.  In the case of binary-binary scattering the
semi-major axes are equal.  In the case of triple scattering the triple
approaches from the left and the interloping systems approaches from the
right. \emph{Left panel:} In this binary-binary scattering event one star
from each binary combine to form a new binary, and the two remaining stars
are ejected.  \emph{Middle panel:} In this triple-single scattering event
one star from the inner binary of the triple is ejected and the interloping
star is captured as a tertiary.  \emph{Right panel:} In this triple-binary
scattering event one star from the inner binary is ejected, leaving two
binaries and a single star.}

\label{fig:sample}
\end{figure*}

We compute the cross sections for the outcomes of both single and binary
stars scattering off of hierarchical triples along with binaries scattering
off of binaries in the point mass limit using Newtonian gravity.  The masses
of all stars in the system are equal unless we explicitly vary the mass of
one star of the inner binary of the triple.  In an individual scattering
event, the semi-major axis and eccentricity of all component binaries (both
binaries of the triple and, in the case of triple-binary scattering, that of
the interloping binary) in the system are fixed, but the argument of
pericenter and the mean anomaly are chosen from a uniform distribution
between 0 and $2\pi$.  The component binaries are then oriented randomly by
pointing each binary's angular momentum vector toward a randomly chosen
point on a sphere.  We present a sample binary-binary, triple-single, and
triple-binary scattering experiment projected in the $xy$-plane in
Fig.~\ref{fig:sample}.

We first perform scattering experiments on a model system in Section
\ref{subsec:model} and then vary several of the initial orbital and physical
parameters of this system one at a time to determine the dependence of the
cross sections on these parameters  in Section \ref{subsec:parameters}.  

\subsection{Cross sections of model systems}
\label{subsec:model}

Our model system for triple-single scattering is a moderately hierarchical
triple system with semi-major axis ratio, $\alpha \equiv a_1/a_{11} = 10$.
This choice of $\alpha$ is large enough that the triple is stable
\citep{mardling+aarseth01}, but is still small enough that there will be
dynamical interactions between the inner and outer binaries.  In this model
system, the interloping star has an incoming velocity of
$v_{\infty}/v_{\textrm{crit}} = 1$, which is approximately the velocity
dispersion of stars in the thin disk of the Milky Way ($\sim$40 \kms{})
relative to the critical velocity of a triple system consisting of three 1
\Msun{} stars with $a_1 = 10$ AU and $a_{11} = 1$ AU
\citep[p.~656]{binney+merrifield98}.  We take both orbits to be circular.  

Our model system for triple-binary scattering is identical, but with the
interloping star replaced by an interloping binary with semi-major axis
equal to the outer semi-major axis of the triple, $a_2 = a_1 = 10 a_{11}$.
The incoming velocity is again taken to be \vcrit, but with \vcrit{} now
calculated using equation~(\ref{eq:vcrit_tripbin}) instead of
equation~(\ref{eq:vcrit_tripsing}).  Again, all stars are of equal mass, so
now the mass ratio of the interloping system to the triple is 2/5 instead of
1/3. 

Lastly, our model system for binary-binary scattering is identical to the
case of triple-binary scattering but with the inner binary of the triple
replaced by a point mass of 1 \Msun{}.  The incoming velocity is taken to be
\vcrit, but with \vcrit{} calculated using equation~(\ref{eq:vcrit_binbin}).
In this case the mass ratio of the interloping system to the target system
is 1. 

We perform $10^6$ scattering experiments for each model system.  We present
the cross sections for all possible outcomes with their statistical and
total uncertainties in Table~\ref{tbl:tripsing} for the triple-single case,
Table~\ref{tbl:tripbin} for the triple-binary case, and
Table~\ref{tbl:binbin} for the binary-binary case.

\begin{table}
\centering

\caption{Normalized cross sections for the outcomes of triple-single
scattering (see equation~\ref{eq:normtripsing} for the normalization).  The
initial conditions are described in Section \ref{sec:scattering} and the
initial hierarchy is \texttt{[[0 1] 2] 3}.  We present both the statistical
uncertainty and the systematic uncertainty.  Note that the systematic
uncertainty only represents an uncertainty toward larger cross sections.
That is, the cross sections presented are lower limits and may be larger by
the systematic uncertainty---see Section~\ref{subsec:crosssections}.  We
only present cross sections for which the statistical uncertainty is less
than 50 per cent.  The cross section for the outcome equal to the initial
hierarchy (i.e., a flyby) is not well defined and so is not included.  Such
interactions can change the orbital parameters of the triple system,
however, and are discussed in Section~\ref{subsec:flybydist}.  The outcome
classes are defined in Section~\ref{subsec:parameters}.  We also note if the
scattering event produces a new triple with a hierarchy distinct from the
original hierarchy.}

\begin{tabular}{llll}

\textsc{Triple-single} & & & \\

\hline

Outcome & 
$\hat{\sigma}$ & 
$\Delta_{\textrm{stat}} \hat{\sigma}$ &
Outcome class \\

\hline

\texttt{[0 1] 3 2} & 1.309 & 0.013 & Single ionization \\ \relax
\texttt{[0 2] 1 3} & 0.092 & 0.003 & Single ionization \\ \relax
\texttt{0 [1 2] 3} & 0.080 & 0.003 & Single ionization \\ \relax
\texttt{0 [1 3] 2} & 0.076 & 0.003 & Exchange ionization \\ \relax
\texttt{[0 3] 1 2} & 0.074 & 0.003 & Exchange ionization \\ \relax
\texttt{[[0 1] 3] 2} & 0.033 & 0.002 & Exchange, new triple \\ \relax
\texttt{[[0 3] 2] 1} & 0.027 & 0.002 & Exchange, new triple \\ \relax
\texttt{0 [[1 3] 2]} & 0.021 & 0.002 & Exchange, new triple \\ \relax
\texttt{[0 1] [2 3]} & 0.017 & 0.001 & Double binary \\ \relax
\texttt{[0 2] [1 3]} & 0.011 & 0.001 & Double binary \\ \relax
\texttt{3 [[1 2] 0]} & 0.010 & 0.001 & Scramble, new triple \\ \relax
\texttt{[0 3] [1 2]} & 0.010 & 0.001 & Double binary \\ \relax
\texttt{[[0 2] 1] 3} & 0.008 & 1e-03 & Scramble, new triple \\ \relax
\texttt{0 1 [2 3]} & 0.005 & 8e-04 & Exchange ionization \\ \relax
 & & \\
$\Delta_{\textrm{sys}} \hat{\sigma}$ & 0.032 & 0.002 \\

\hline

\label{tbl:tripsing}
\end{tabular}
\end{table}

\begin{table}
\centering

\caption{Normalized cross sections for the outcomes of triple-binary
scattering (see equation~\ref{eq:normtripbin} for the normalization).  The
initial conditions are described in Section \ref{sec:scattering} and the
initial hierarchy is \texttt{[[0 1] 2] [3 4]}.  We present both the
statistical uncertainty and the systematic uncertainty.  Note that the
systematic uncertainty only represents an uncertainty toward larger cross
sections.  That is, the cross sections presented are lower limits and may be
larger by the systematic uncertainty---see Section
\ref{subsec:crosssections}.  We only present cross sections for which the
statistical uncertainty is less than 50 per cent.  Flyby outcomes are not
included.  Such interactions can change the orbital parameters of the triple
system, however, and are discussed in Section~\ref{subsec:flybydist}. The
outcome classes are defined in Section~\ref{subsec:parameters}.  We also
note if the scattering event produces a new triple with a hierarchy distinct
from the original hierarchy.}

\begin{tabular}{llll}

\textsc{Triple-binary} & & & \\

\hline

Outcome & 
$\hat{\sigma}$ & 
$\Delta_{\textrm{stat}} \hat{\sigma}$ &
Outcome class \\

\hline

\texttt{[0 1] 4 2 3} & 1.110 & 0.016 & Double ionization \\ \relax
\texttt{[0 1] [3 4] 2} & 0.990 & 0.015 & Double binary \\ \relax
\texttt{[[0 1] 2] 4 3} & 0.911 & 0.015 & Binary disruption \\ \relax
\texttt{[0 1] 3 [2 4]} & 0.072 & 0.004 & Double binary \\ \relax
\texttt{[0 1] 4 [2 3]} & 0.070 & 0.004 & Double binary \\ \relax
\texttt{[0 2] 1 4 3} & 0.048 & 0.003 & Double ionization \\ \relax
\texttt{0 [1 2] [3 4]} & 0.048 & 0.003 & Double binary \\ \relax
\texttt{0 [1 2] 4 3} & 0.048 & 0.003 & Double ionization \\ \relax
\texttt{[0 2] 1 [3 4]} & 0.045 & 0.003 & Double binary \\ \relax
\texttt{[[0 1] 4] 3 2} & 0.025 & 0.002 & Bin. disruption, new triple \\ \relax
\texttt{0 [1 3] 2 4} & 0.025 & 0.002 & Double ionization \\ \relax
\texttt{[0 4] 1 2 3} & 0.022 & 0.002 & Double ionization \\ \relax
\texttt{[[0 1] 3] 4 2} & 0.021 & 0.002 & Bin. disruption, new triple \\ \relax
\texttt{[0 3] 1 2 4} & 0.021 & 0.002 & Double ionization \\ \relax
\texttt{0 [1 4] 2 3} & 0.018 & 0.002 & Double ionization \\ \relax
\texttt{[[0 1] 3] [2 4]} & 0.011 & 0.002 & Exchange, new triple \\ \relax
\texttt{[[0 1] 4] [2 3]} & 0.011 & 0.002 & Exchange, new triple \\ \relax
\texttt{[[0 2] 1] [3 4]} & 0.006 & 0.001 & Scramble, new triple \\ \relax
\texttt{[[0 3] 2] 1 4} & 0.005 & 0.001 & Bin. disruption, new triple \\ \relax
\texttt{[0 4] [1 3] 2} & 0.005 & 0.001 & Double binary \\ \relax
\texttt{[3 4] [[1 2] 0]} & 0.005 & 0.001 & Exchange, new triple \\ \relax
\texttt{0 1 [2 3] 4} & 0.004 & 1e-03 & Double ionization \\ \relax
\texttt{0 1 2 [3 4]} & 0.004 & 1e-03 & Double ionization \\ \relax
\texttt{[0 3] [1 2] 4} & 0.004 & 1e-03 & Double binary \\ \relax
\texttt{[[0 2] 1] 3 4} & 0.004 & 9e-04 & Bin. disruption, new triple \\ \relax
\texttt{[[0 4] 2] 1 3} & 0.004 & 9e-04 & Bin. disruption, new triple \\ \relax
\texttt{3 [[1 2] 0] 4} & 0.004 & 9e-04 & Bin. disruption, new triple \\ \relax
\texttt{0 [[1 3] 2] 4} & 0.003 & 9e-04 & Bin. disruption, new triple \\ \relax
\texttt{[0 2] [1 3] 4} & 0.003 & 9e-04 & Double binary \\ \relax
\texttt{[0 4] [1 2] 3} & 0.003 & 9e-04 & Double binary \\ \relax
\texttt{[0 3] [1 4] 2} & 0.003 & 9e-04 & Double binary \\ \relax
\texttt{[[0 4] 3] 1 2} & 0.003 & 8e-04 & Bin. disruption, new triple \\ \relax
\texttt{0 1 [2 4] 3} & 0.003 & 8e-04 & Double ionization \\ \relax
\texttt{0 [[1 4] 2] 3} & 0.003 & 8e-04 & Bin. disruption, new triple \\ \relax
\texttt{[0 2] [1 4] 3} & 0.002 & 7e-04 & Double binary \\ \relax
\texttt{0 [[1 4] 3] 2} & 0.002 & 7e-04 & Bin. disruption, new triple \\ \relax
\texttt{[0 3] 1 [2 4]} & 0.002 & 7e-04 & Double binary \\ \relax
\texttt{[0 2] [[1 4] 3]} & 0.002 & 7e-04 & Exchange, new triple \\ \relax
\texttt{[0 4] 1 [2 3]} & 0.002 & 7e-04 & Double binary \\ \relax
\texttt{[0 4] [[1 3] 2]} & 0.001 & 6e-04 & Exchange, new triple \\ \relax
\texttt{[[0 3] 2] [1 4]} & 0.001 & 6e-04 & Exchange, new triple \\ \relax
\texttt{0 [1 3] [2 4]} & 0.001 & 6e-04 & Double binary \\ \relax
\texttt{[[0 4] 2] [1 3]} & 0.001 & 5e-04 & Exchange, new triple \\ \relax
\texttt{[0 3] [[1 4] 2]} & 0.001 & 5e-04 & Exchange, new triple \\ \relax
\texttt{[[0 2] 3] 1 4} & 0.001 & 5e-04 & Bin. disruption, new triple \\ \relax
\texttt{0 [[1 3] 4] 2} & 0.001 & 5e-04 & Bin. disruption, new triple \\ \relax
\texttt{3 [[1 4] 0] 2} & 0.001 & 5e-04 & Bin. disruption, new triple \\ \relax
\texttt{[[0 3] 4] 1 2} & 0.001 & 5e-04 & Bin. disruption, new triple \\ \relax
\texttt{[[0 4] 1] 3 2} & 0.001 & 5e-04 & Bin. disruption, new triple \\ \relax
 & & \\
$\Delta_{\textrm{sys}} \hat{\sigma}$ & 0.045 & 0.003 \\

\hline

\label{tbl:tripbin}
\end{tabular}
\end{table}

\begin{table}
\centering

\caption{Normalized cross sections for the outcomes of binary-binary
scattering.  The initial conditions are described in Section
\ref{sec:scattering} and the initial hierarchy is \texttt{[0 1] [2 3]}.  The
cross sections have been normalized to the sum of the areas of the two
binaries.  We present both the statistical uncertainty and the systematic
uncertainty.  Note that the systematic uncertainty only represents an
uncertainty toward larger cross sections.  That is, the cross sections
presented are lower limits and may be larger by the systematic
uncertainty---see Section \ref{subsec:crosssections}.  We only present cross
sections for which the statistical uncertainty is less than 50\%.  The cross
section for the outcome equal to the initial hierarchy (i.e., a flyby) is
not well defined and so is not included.}

\begin{tabular}{llll}

\textsc{Binary-binary} & & & \\

\hline

Outcome & 
$\hat{\sigma}$ & 
$\Delta_{\textrm{stat}} \hat{\sigma}$ &
Outcome class \\

\hline

\texttt{0 1 [2 3]} & 1.335 & 0.016 & Single ionization \\ \relax
\texttt{[0 1] 3 2} & 1.310 & 0.016 & Single ionization \\ \relax
\texttt{[0 2] 1 3} & 0.829 & 0.013 & Exchange ionization \\ \relax
\texttt{[0 3] 1 2} & 0.823 & 0.013 & Exchange ionization \\ \relax
\texttt{0 [1 3] 2} & 0.807 & 0.013 & Exchange ionization \\ \relax
\texttt{0 [1 2] 3} & 0.800 & 0.013 & Exchange ionization \\ \relax
\texttt{[0 2] [1 3]} & 0.152 & 0.005 & Exchange \\ \relax
\texttt{[0 3] [1 2]} & 0.150 & 0.005 & Exchange \\ \relax
\texttt{[[0 3] 2] 1} & 0.003 & 8e-04 & Triple formation \\ \relax
\texttt{[[0 1] 2] 3} & 0.002 & 6e-04 & Triple formation \\ \relax
\texttt{[[0 3] 1] 2} & 0.002 & 6e-04 & Triple formation \\ \relax
\texttt{0 [[1 2] 3]} & 0.002 & 6e-04 & Triple formation \\ \relax
\texttt{0 [[1 3] 2]} & 0.002 & 6e-04 & Triple formation \\ \relax
\texttt{[[2 3] 0] 1} & 0.002 & 6e-04 & Triple formation \\ \relax
\texttt{[[0 2] 1] 3} & 0.002 & 6e-04 & Triple formation \\ \relax
\texttt{2 [[1 3] 0]} & 0.002 & 6e-04 & Triple formation \\ \relax
\texttt{[[0 2] 3] 1} & 0.001 & 5e-04 & Triple formation \\ \relax
\texttt{[[0 1] 3] 2} & 0.001 & 5e-04 & Triple formation \\ \relax
\texttt{3 [[1 2] 0]} & 0.001 & 5e-04 & Triple formation \\ \relax
\texttt{0 [[2 3] 1]} & 0.001 & 5e-04 & Triple formation \\ \relax
 & & \\
$\Delta_{\textrm{sys}} \hat{\sigma}$ & 0.002 & 0.001 \\

\hline

\label{tbl:binbin}
\end{tabular}
\end{table}

Because there are a large number of possible outcomes in binary-binary,
triple-single, and triple-binary scattering (22, 22, and 161, respectively),
many of which are qualitatively similar, we group these outcomes into broad
classes.  In the case of triple-single scattering we define six classes: (1)
flybys, in which the hierarchical structure of the system remains the same,
although the orbital parameters may have changed; (2) exchanges, in which
the interloping star replaces one of the stars in the triple; (3) double
binary formation, in which one star of the triple is ionized and binds to
the interloping star; (4) single ionization, in which one star from the
triple is ionized, leaving a binary and two unbound stars; (5) full
ionization, in which all stars become unbound from each other; and (6) a
scramble, in which the tertiary exchanges with one of the stars of the inner
binary.  

In the case of binary-binary scattering we define six classes: (1) flybys,
defined as in the triple-single case; (2) exchanges, in which one star from
each binary exchanges places with the other; (3) triple formation, in which
a stable triple is formed, leaving a single unbound star; (4) single
ionization, in which one star is ionized, leaving a binary and two unbound
stars; (5) full ionization, defined as in the triple-single case; and (6)
exchange + ionization, in which one star from each binary binds to form a
new binary, and the two other stars remain unbound.  

In the case of triple-binary scattering we define seven classes: (1) flybys,
defined as in the triple-single case; (2) exchanges, in which one star from
the interloping binary exchanges with one star from the interloping binary;
(3) double binary formation, in which one star from the triple is ionized,
resulting in two binaries and an unbound star; (4) binary disruption, in
which the two stars of the interloping binary become unbound; (5) triple
disruption, in which the three stars of the triple become unbound from each
other; (6) quadruple formation, in which one star from the interloping
binary becomes bound to the triple, forming a quadruple; (7) and a scramble,
defined as in the triple-single case.  

Note that in triple-single and triple-binary scattering several of these
outcome classes produce a `new' triple (i.e., the final hierarchy differs
from the original hierarchy).  (In binary-binary scattering triple formation
is its own class and is mutually exclusive with the other classes.)  The
cross sections for these classes for triple-single, triple-binary, and
binary-binary scattering are displayed in Table \ref{tbl:cat}.  We include
there the cross section for new triple formation, though for triple-single
and triple-binary scattering it is not independent of the cross sections for
other outcome classes.

In the case of triple scattering the systematic uncertainty is dominated by
marginally unstable triples (according to Equation~\ref{eq:mardling}) for
which the integration time exceeded the one-hour CPU time limit; 0.0263 per
cent of the triple-single scattering experiments failed to complete, and of
these 81 per cent failed to complete due to the CPU time limit (the rest
failed because they violated the energy conservation limit).  Because the
Mardling stability criterion of equation~(\ref{eq:mardling}) is not a hard
boundary \citep{petrovich15b} it is possible that these triples are, in
fact, stable and should be classified as flybys.  If, however, they are
unstable on longer timescales than our integrations permit they should be
classified as single ionization events since no other outcome is possible.
Thus the systematic uncertainty of triple scattering should be considered as
an uncertain contribution to the cross section for single ionization.  In
the case of binary-binary scattering the systematic uncertainty is dominated
by systems that violated the energy conservation criterion during an
extremely close passage.  This is a much rarer occurrence than the formation
of a marginally unstable triple during a triple scattering event, so the
systematic uncertainty of binary-binary scattering is much lower than that
of triple scattering (0.0012 per cent of systems failed to resolve, of which
all were due to violations of the energy conservation limit).

\begin{table}
\centering

\caption{Cross sections for outcome classes in triple-single, triple-binary,
and binary-binary scattering at $\hat{v} = 1$.  Cross sections for
``Exchange + ionization'' are not included in either ``Exchange'' or
``Single ionization.''  However, the cross sections for new triples are not
independent of the other cross sections.  (I.e., they are a sum of subsets
from the other classes.)  See Section \ref{subsec:parameters} for
definitions of the outcome classes.}

\begin{tabular}{lll}

\textsc{Triple-single} & & \\

\hline

Outcome class & $\hat{\sigma}$ & $\Delta_{\textrm{stat}} \hat{\sigma}$ \\

\hline

Exchange & 0.082 & 0.003 \\ \relax
Single ionization & 1.481 & 0.013 \\ \relax
Full ionization & 0.000 & 0.000 \\ \relax
Double binary formation & 0.038 & 0.002 \\ \relax
Exchange + ionization & 0.155 & 0.004 \\ \relax
Scramble & 0.018 & 0.001 \\ \relax
New triple & 0.100 & 0.003 \\ \relax
 & & \\ \relax
$\Delta_{\textrm{sys}} \hat{\sigma}$ & 0.032 & 0.002 \\

\hline

& & \\

\textsc{Triple-binary} & & \\
\hline
Exchange & 0.036 & 0.003 \\ \relax
Quadruple formation & 0.000 & 2e-04 \\ \relax
Scramble & 0.006 & 0.001 \\ \relax
Binary disruption & 0.996 & 0.016 \\ \relax
Double binary formation & 1.250 & 0.017 \\ \relax
Triple disruption & 1.303 & 0.018 \\ \relax
Full ionization & 0.000 & 2e-04 \\ \relax
New triple & 0.127 & 0.006 \\ \relax
 & & \\ \relax
$\Delta_{\textrm{sys}} \hat{\sigma}$ & 0.045 & 0.003 \\

\hline

& & \\

\textsc{Binary-binary} & & \\
\hline
Exchange & 0.302 & 0.008 \\ \relax
Triple formation & 0.020 & 0.002 \\ \relax
Single ionization & 2.646 & 0.023 \\ \relax
Full ionization & 0.000 & 0.000 \\ \relax
Exchange + ionization & 3.259 & 0.025 \\ \relax
 & & \\ \relax
$\Delta_{\textrm{sys}} \hat{\sigma}$ & 0.002 & 0.001 \\

\hline

\label{tbl:cat}
\end{tabular}
\end{table}

\subsection{Dependence on initial parameters}
\label{subsec:parameters}

We next explore the dependence of the cross sections on the initial
parameters of the system.  In particular we separately vary the semi-major
axis ratio, the incoming velocity, the eccentricities, and the masses.

\subsubsection{Semi-major axis ratio}
\label{subsubsec:alpha}

\begin{figure*}
\centering
\includegraphics[width=18cm]{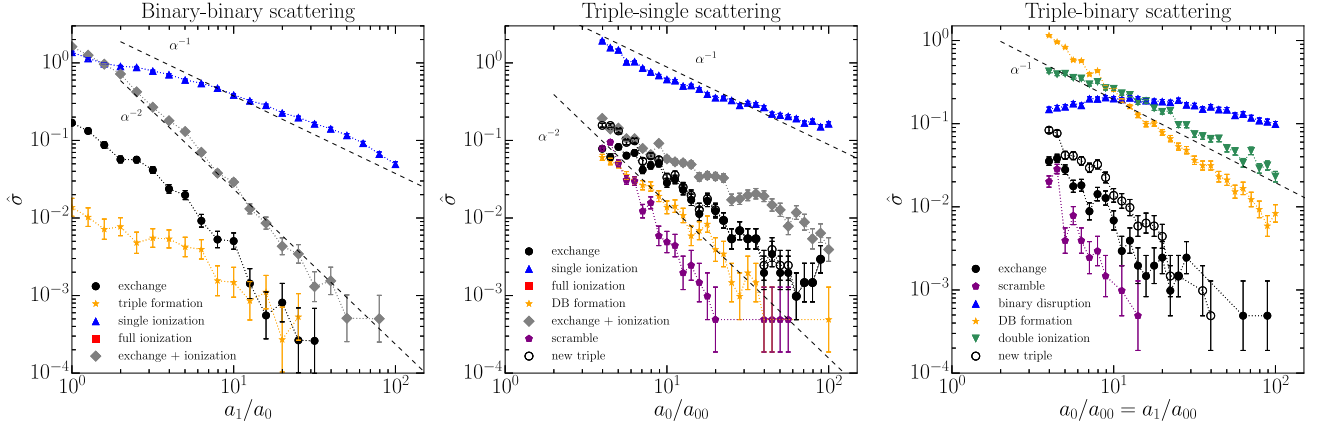}

\caption{Cross sections of various classes of outcomes as a function of
semi-major axis ratio, $\alpha$, for binary-binary (left panel),
triple-single, (middle panel), and triple-binary (right panel) scattering.
In the limit of large semi-major axis ratios for triple-single scattering,
interactions involving the inner binary should scale like the area of the
inner orbit (i.e., an inverse square dependence on the semi-major axis).
Indeed, the cross sections for scrambles and exchange + ionization (both of
which require the interloper interact with the inner binary) are consistent
with the inverse square dependence shown (black dashed line).  Single
ionization is the dominant outcome of triple-single scattering at all
semi-major axes.  At small $\alpha$ the triple becomes unstable to
ionization from weak perturbations due to the interloper and at large
$\alpha$ the tertiary is weakly bound relative to the inner binary so weak
perturbations from the interloper lead to ionization.  See Section
\ref{subsubsec:alpha} for a discussion of the origin of the $\alpha$
dependence of different cross sections.}

\label{fig:alpha}
\end{figure*}

We first vary the semi-major axis ratio from $\alpha \equiv a_2 / a_1 =
10^{0.6} - 10^2$.  At choices of $\alpha$ much below our lower limit the
triples become unstable and at choices of $\alpha$ much larger than our
upper limit the computational time becomes prohibitively long because the
time step is set by the orbital period of the inner binary, but the crossing
time is set by the size of the outer orbit.  We hold all other parameters
(e.g., eccentricities, masses) fixed to their values in the model system
(Section \ref{subsec:model}).  The normalized cross sections for
binary-binary, triple-single, and triple-binary scattering are shown in
Fig.~\ref{fig:alpha}.  Single ionization is the dominant outcome of
triple-single scattering for all $\alpha$.  At the low-$\alpha$ end this is
because the triples are only marginally stable so minor perturbations from
the interloper tend to ionize one member of the triple.  At the
high-$\alpha$ end this is because the tertiary becomes weakly bound to the
inner binary.  Weak perturbations from the interloper are therefore likely
to ionize the tertiary.  Other classes of outcomes generally require some
interaction with the stars of the inner binary.  The probability of such an
interaction scales with the orbital area of the inner binary, so we observe
that $\hat{\sigma} \propto \alpha^{-2}$.  

For the case of triple-binary scattering we fix the initial value of $a_1$
to be equal to $a_2$ as we vary the semi-major axis, $\alpha$.  At large
$\alpha$, binary disruption is the dominant outcome whereas at small
$\alpha$ double binary formation dominates.  In the low-$\alpha$ limit the
triple comes closer to instability, so single ionization from the triple is
also likely to occur in addition to the disruption of the interloping
binary, leading to double binary formation.  In the large $\alpha$ case the
scattering problem approaches that of binary-binary scattering in which one
binary is more massive than the other.  Earlier experiments have shown
\citep[e.g.,][]{fregeau+04} that the less massive binary is more likely to
be disrupted, in accordance with our results here.  In this limit we might
na\"ively expect no dependence on $\alpha$ since we fix $a_1 = a_2$.
Instead we observe a dependence of $\sigma \propto \alpha^{-1}$.  This is
because as $\alpha$ increases the incoming velocity (which, in the
large-$\alpha$ limit is comparable to the orbital velocity of the inner
binary of the triple) becomes larger relative to the orbital velocities of
the outer binary and interloping binary.  In the limit of large incoming
velocities $\hat{\sigma} \propto \hat{v}^{-2}$ \citep{hut+bahcall83}, and
since $\hat{v} \propto a^{-1/2}$, we have that $\hat{\sigma} \propto a_{11}
\propto \alpha^{-1}$.  For the same reason the scaling of the single
ionization cross section in triple-single scattering exhibits the same
dependence. 

\subsubsection{Eccentricity}
\label{subsubsec:eccentricity}

\begin{figure*}
\centering
\includegraphics[width=18cm]{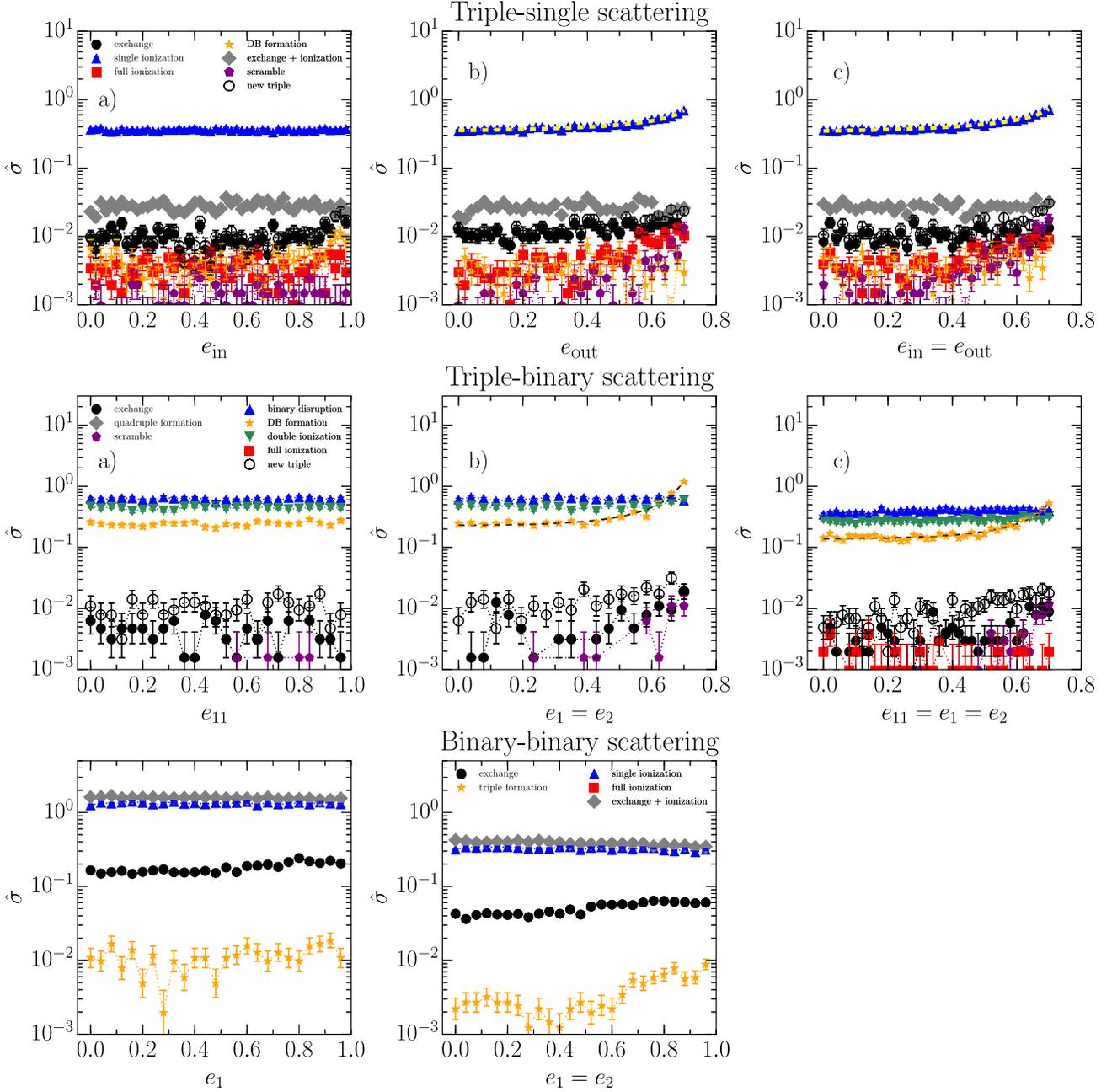}

\caption{Cross sections of outcome classes for triple-single (top row),
triple-binary (middle row), and binary-binary (bottom row) scattering as a
function of eccentricity.  For triple scattering we vary the inner
eccentricity alone (panels a), the outer eccentricity alone (panels b), and
the inner and outer eccentricities simultaneously (panels c).  In the case
of triple-binary scattering we set the eccentricity of the incoming binary
to be equal to that of the tertiary.  In the case of binary-binary
scattering we vary the eccentricity of one binary only (left panel) and both
binaries simultaneously (right panel).  In binary-binary scattering the
cross sections are independent of the eccentricities.  In triple scattering
the cross sections are independent of the inner eccentricity and are
independent of the outer eccentricity for small outer eccentricities.  At
large outer eccentricities there is a small increase in the cross section
for single ionization in triple-single scattering and double binary
formation in triple-binary scattering.  This is because these triples start
closer to the Mardling instability boundary, so small perturbations to the
outer eccentricity can destabilize them.  We fit these cross sections with
the functional form of equation~(\ref{eq:ecc_approx}) (dashed lines) and
find excellent agreement.  The best-fitting parameters are shown in
Table~\ref{tbl:ecc_fit}.}

\label{fig:ecc}
\end{figure*}

We next vary the eccentricity of binaries in the systems, both individually
and together while holding constant all other parameters (e.g., semi-major
axis ratio and masses).  To increase the range over which we can vary the
outer eccentricity in triple scattering we set the outer semi-major axes to
20 AU (i.e., we use $\alpha$ of 20 instead of 10 as before).  We vary the
outer eccentricity of the triples ($e_1$) from 0 to 0.7 and the eccentricity
of the inner binary ($e_{11}$) from 0 to 0.98.  Even with the larger value
of $\alpha$ at outer eccentricities larger than $\sim$0.7 the triple
violates the Mardling stability criterion (equation~\ref{eq:mardling}).  In
the case of triple-binary scattering we fix $e_1 = e_2$.  In the case of
binary-binary scattering we vary the eccentricities of either one binary or
both binaries from 0 to 0.98.  The cross sections for triple-single,
triple-binary, and binary-binary scattering are shown in Fig.~\ref{fig:ecc}.
In the case of binary-binary scattering the cross sections are completely
independent of the eccentricity.  In the case of triple scattering the cross
sections are independent of the inner eccentricity and are generally
independent of the outer eccentricity, although there is a modest increase
in the cross section for single ionization in the case of triple-scattering
and in the cross section for double binary formation in the case of
triple-binary scattering.  

The physical mechanism behind these dependences is twofold.  First, there is
some baseline cross section for the outcome which is independent of any
eccentricity.  \citet{hut+bahcall83} argue that this is because the cross
sections for a particular orbit are only dependent on the average velocity
of the orbit, which is independent of the eccentricity as a result of the
virial theorem.  This baseline cross section applies to all outcomes.
Second, in the case of triple scattering there is an additional contribution
to the cross section of ionization-like outcomes (i.e., single ionization in
the case of triple-single scattering and double binary formation in the case
of triple-binary scattering) due to perturbations to the eccentricity of the
outer orbit of the triple, $e_1$.  If $e_1$ exceeds some critical
eccentricity, $e_{\textrm{crit}}$, the triple will violate the Mardling
stability criterion and will become unstable.  As we show in Section
\ref{subsec:flybydist}, and in particular in panels b) of
Fig.~\ref{fig:cumu}, the logarithm of the cross section for cumulative
changes in the eccentricity follows the inverse of the Gompertz function:
\begin{equation}
\label{eq:inv_gompertz_app}
\Delta e_1 = \exp \left(-c_1 \exp \left(c_2 \ln \hat{\sigma}\right)\right) =
\exp \left(-c_1 \hat{\sigma}^{c_2}\right),
\end{equation}
where $c_1$ and $c_2$ are positive constants.  This implies that the cross
section to undergo an ionization-like reaction takes the form
\begin{equation}
\label{eq:ecc_approx}
\hat{\sigma} = \hat{\sigma}_0 + (-c_1^{-1} \ln(e_{\textrm{crit}} -
e_1))^{1/c_2},
\end{equation}
where $\hat{\sigma}_0$ is the baseline cross section described above.
Taking $e_{\crit}$ from the Mardling stability criterion ($e_{\crit} \approx
0.733$), we find an excellent match between this functional form and the
cross sections for ionization-like outcomes in Fig.~\ref{fig:ecc}.
Best-fitting parameters are shown in Table~\ref{tbl:ecc_fit}.

\begin{table}
\centering

\caption{Best-fitting parameters of the eccentricity cross sections in
Fig.~\ref{fig:ecc} to the inverse Gompertz function.  See equation
(\ref{eq:ecc_approx}) for the definition of the parameters.}

\begin{tabular}{lllll}

\hline

Scattering type & panel & $\hat{\sigma}_0$ & $c_1$ & $c_2$ \\

\hline

Triple-single & b & 0.35 & 6.21 & 0.57 \\ \relax
Triple-single & c & 0.34 & 6.05 & 0.60 \\ \relax
Triple-binary & b & 0.23 & 3.45 & 0.35 \\ \relax
Triple-binary & c & 0.14 & 4.88 & 0.39 \\

\hline

\end{tabular}
\label{tbl:ecc_fit}
\end{table}

Because \ein{} does not influence the stability of the triple, we do not see
a similar effect when \ein{} is varied.  This result that the cross sections
are independent of the eccentricities, except when related to the stability
of the triple, is consistent with the results of binary-single scattering
experiments, which have also found that cross sections are independent of
eccentricity \citep[e.g.,][]{hut+bahcall83}.  We note, however, that is only
the cross sections of strong interactions that are independent of
eccentricity.  \citet{heggie+rasio96} found that the magnitude of small
secular perturbations to the eccentricity of a binary from a distant flyby
is proportional to $e \sqrt{1 - e^2}$. 

\subsubsection{Mass ratio}
\label{subsubsec:massratio}

\begin{figure*}
\centering
\includegraphics[width=17.75cm]{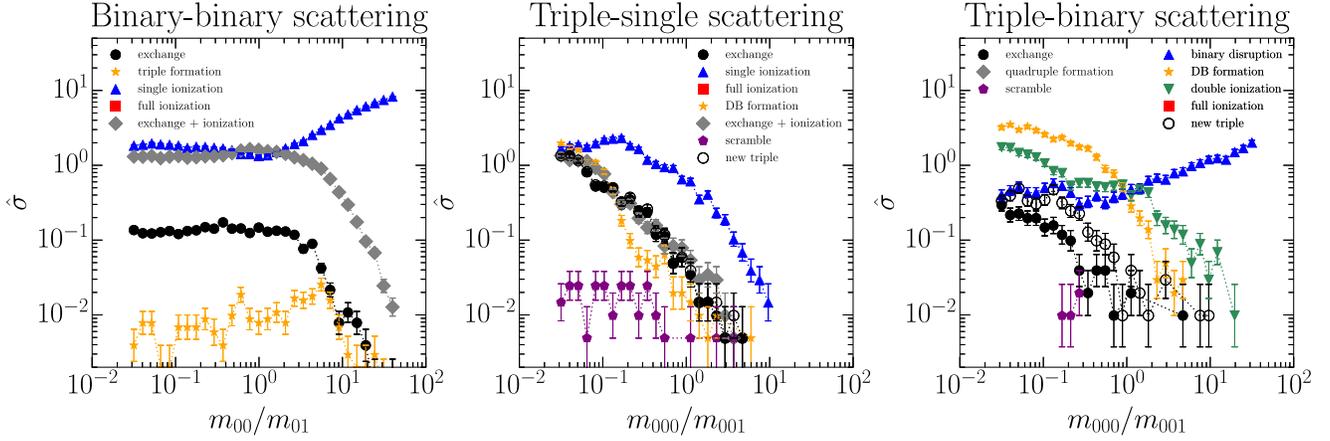}

\caption{Cross sections of outcome classes for binary-binary (left panel),
triple-single (middle panel) and triple-binary (right panel) scattering as a
function of the mass ratio of one component of the binary (in the case of
binary-binary scattering) or of one component of the inner binary (in the
case of triple scattering) to all other components of the system.  The
masses of all other components of the system remain fixed and equal to each
other.  In the high mass limit the more massive system tends to disrupt the
incoming system.  This leads to an increase in the cross section for single
ionization in binary-binary and triple-binary scattering.  In the case of
triple-single scattering all cross sections decrease because the incoming
system cannot disrupt.}

\label{fig:massratio}
\end{figure*}

We show in Fig.~\ref{fig:massratio} the effect of changing the mass of one
component of the system in binary-binary, triple-single, and triple-binary
scattering.  In the case of triple scattering the mass of one component of
the inner binary of the triple is varied.  All other parameters (e.g.,
$\alpha$, eccentricities, and $\hat{v}$) are held fixed.  The mass of the
one star is varied from 0.03 $M_{\odot}$ to 40 $M_{\odot}$.  For small mass
ratios (i.e., as the test particle limit is approached) the cross section
for all outcomes in triple scattering increases by roughly an order of
magnitude.  In the high mass limit single ionization becomes the dominant
outcome of binary-binary scattering, rising roughly with the square root of
mass, and binary disruption becomes the dominant outcome of triple-binary
scattering with roughly the same mass dependence.  Conversely, although
single ionization is the dominant outcome at high mass for triple-single
scattering, the cross section decreases, nearly with the square of the mass.  

\citet{hut83} derived the mass dependence of the ionization cross section
for binary-single scattering in the large $\hat{v}$ limit and found it to be
proportional to $m^{-2}$ when the most massive body resides in the binary
and proportional to $m^2$ when the most massive body is the interloping star
(see equation~\ref{eq:hut_ion}).  In the large mass limit of binary-binary and
triple-binary scattering, the mass dependence should be similar to the case
of binary-single scattering with a massive interloping star.  Likewise, the
case of triple-single scattering should exhibit a similar mass dependence to
binary-single scattering with a massive binary.  However, we are not in the
high velocity limit ($\hat{v} = 1$) so we expect the mass dependencies to
soften, which is what we observe.  Nevertheless, the trend of increased
ionization cross sections for low-mass interloping binaries and decreased
ionization cross sections for low-mass interloping single stars holds even
in the moderate velocity regime. 

\subsubsection{Incoming velocity}
\label{subsubsec:vinf}

We finally examine the effect of the velocity at infinity of the interloping
system.  We measure these velocities relative to the critical velocity of
the system (see equations~\ref{eq:vcrit_tripsing} and
\ref{eq:vcrit_tripbin}) and explore a range of two dex centered around
\vcrit{}.  It is difficult to explore a much larger range because velocities
much larger than 10\vcrit{} yield very small cross sections and therefore
have large statistical uncertainties.  Velocities much smaller than
0.1\vcrit{} require an excessive amount of computing time because the low
total energy of the system means that it takes a long time for the system to
random walk to a region of phase space where one object has a sufficient
amount of energy to escape and leave behind a stable system.  

\begin{figure*}
\centering
\includegraphics[width=17.75cm]{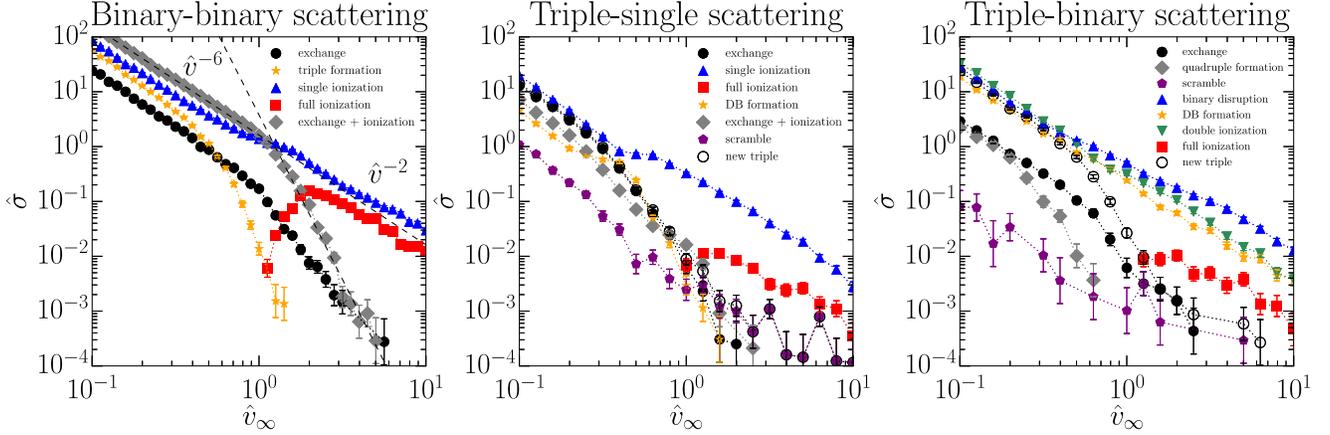}

\caption{Cross sections of outcome classes for binary-binary (left panel),
triple-single (middle panel) and triple-binary (right panel) scattering as a
function of the incoming velocity.  The velocities are written relative to
the critical velocity of the system (see equations~\ref{eq:vcrit_tripsing}
and \ref{eq:vcrit_tripbin}).  Outcomes which require an ionization from one
system follow a $v^{-2}$ dependence at high velocities, whereas those
requiring an exchange follow a steeper dependence.  The kink in the cross
section for single ionization is explained in
Section~\ref{subsec:fourbody}.}

\label{fig:vinf}
\end{figure*}

The cross sections are presented in Fig.~\ref{fig:vinf}.  \citet{hut83}
found two high-velocity limits in binary-single scattering: the cross
section for ionization, which has a dependence of $\hat{\sigma} \propto
\hat{v}^{-2}$ and the cross section for exchange, which has a dependence of
$\hat{\sigma} \propto \hat{v}^{-6}$.  Although there are many more possible
outcomes in binary-binary, triple-single, and triple-binary scattering, the
behavior of the cross sections is qualitatively similar to binary-single
scattering, in that at high velocity the outcomes may be grouped into those
with a $\hat{v}^{-2}$ dependence (``ionization-like'' outcomes), and those
with a steeper dependence (``exchange-like'' outcomes).  In some cases
exchange-like outcomes follow the same $\hat{v}^{-6}$ dependence as in
binary-single scattering (e.g., triple formation in binary-binary scattering
and quadruple formation in triple-binary scattering), but in others the
velocity dependence is shallower (e.g., exchanges in binary-binary
scattering, which display a $\hat{v}^{-4}$ dependence).  We discuss these
results in more depth in Section~\ref{subsec:fourbody}.

\section{Analytic approximations}
\label{sec:analytic}

We here develop analytic approximations to the cross sections computed in
Section \ref{sec:scattering}.  We first review binary-single scattering, and
then generalize the theory of binary-single scattering to four-body
scattering.

\subsection{Three-body scattering}
\label{subsec:binsingapprox}

The analytic theory of binary-single scattering of point masses was
developed by \citet{heggie75} and \citet{hut83} and has been found to agree
with numerical scattering experiments quite well \citep{hut+bahcall83}.
Although the derivation is complicated, the results are straightforward to
summarize.  In binary-single scattering of point masses only three classes
of outcomes are possible: (1) a flyby, (2) an exchange, or (3) an
ionization.  It is not possible to form a stable triple from binary-single
scattering \citep{chazy29, littlewood52, heggie75, heggie+hut03}.  The cross
section of a flyby is not well defined, but \citet{hut83} explicitly
calculated the cross section for the other two classes in the high velocity
limit ($\hat{v} \gg 1$) and found them to be
\begin{equation}
\label{eq:hut_ion}
\hat{\sigma}_{\textrm{ion}} = \frac{40}{3} \frac{m_2^3}{m_{11} m_{12}
(m_{11} + m_{12} + m_2)} \frac{1}{\hat{v}^2}, \quad \hat{v} \gg 1
\end{equation}
and
\begin{equation}
\hat{\sigma}_{\textrm{ex}} = \frac{20}{3} \frac{m^2 (m + m_2)^4}{m_2^3 (2m +
m_2)^3} \frac{1}{\hat{v}^6}, \quad \hat{v} \gg 1
\end{equation}
where the latter is only valid for equal mass binaries, $m \equiv m_{11} =
m_{12}$.  \citet{hut83} found that both cross sections are independent of
eccentricity.  

In the low velocity limit ($\hat{v} \ll 1$) the cross sections cannot be
computed with precise numerical coefficients.  Nevertheless, the velocity
scaling in the low velocity case is simple because ionization is not
possible and the only mechanism to change the cross section is gravitational
focusing.  This is because at low velocities the speed of the interloping
star when it is close enough to interact strongly with the binary will be
very close to the escape speed.  Thus for any low velocity system, the
incoming velocities when the system begins strong interactions will be
nearly the same, and the cross section will scale as
\begin{equation}
\hat{\sigma} \propto \frac{1}{\hat{v}^2}, \quad \hat{v} \ll 1.
\end{equation}

\begin{figure}
\centering
\includegraphics[width=8cm]{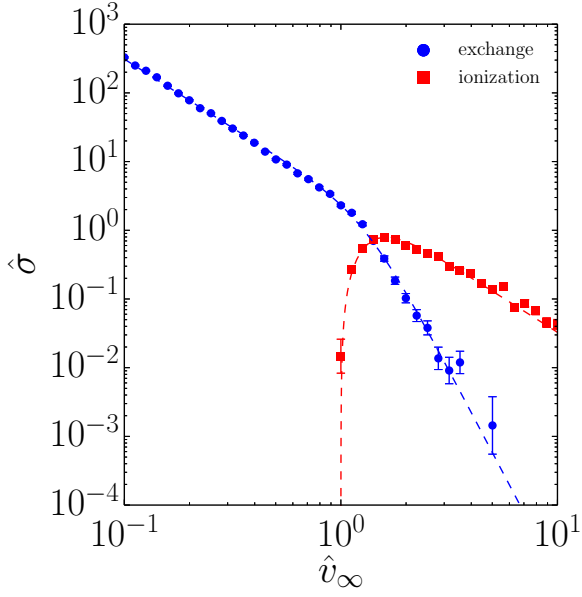}

\caption{Normalized cross sections for exchange (blue circles) and
ionization (red squares) in equal-mass binary-single scattering as a
function of incoming velocity \citep[cf.~Fig.~5 of][]{hut+bahcall83}.  The
dashed lines are an empirical fit to equations~(\ref{eq:exanalytic}) and
(\ref{eq:ionanalytic}).  The analytic approximation is based on a
combination of the low velocity limits ($\hat{v} \ll 1$ for exchange and
$\hat{v} - 1 \ll 1$ for ionization) and the high velocity limits ($\hat{v}
\gg 1$).  Values for the normalization parameters $\hat{\sigma}_a$,
$\hat{\sigma}_b$, etc.~are estimated from the numerical data with
best-fitting values of $\sigma_a \approx 9.76$, $\sigma_b \approx 28.9$,
$\sigma_c \approx 12.8$, and $\sigma_d \approx 10.3$.  The good agreement
between the data and the approximation for all $\hat{v}$ implies that the
two limits capture all the relevant physics of binary-single scattering.}

\label{fig:binsing}
\end{figure}

The cross sections may be combined by taking the reciprocal of the sum of
the reciprocals of the cross sections in the two extreme limits.  This is
because different physical effects limit the rate of production of
particular outcomes and the combined effect of rate limiting processes is
generally the reciprocal of the sum of the reciprocals.  Because the mass
terms are in general not known, we introduce normalization factors
($\hat{\sigma}_a$, $\hat{\sigma}_b$, etc.), which can be determined
empirically.  The cross section for exchange is then
\begin{equation}
\label{eq:exanalytic}
\frac{1}{\hat{\sigma}_{\textrm{ex}}} = \frac{\hat{v}^2}{\hat{\sigma}_a} +
\frac{\hat{v}^6}{\hat{\sigma}_b}.
\end{equation}
The validity of this functional form is demonstrated in
Fig.~\ref{fig:binsing} where we fit equation~(\ref{eq:exanalytic}) to
numerical exchange cross sections (blue dashed line) for binary-single
scattering of equal mass stars.

The cross section for full ionization is necessarily zero for $\hat{v} <$
\vcrit{}.  For velocities slightly in excess of \vcrit{} ($v =
v_{\textrm{crit}} + \sqrt{2 \varepsilon}$, where $\varepsilon \ll 1$), the
cross section for full ionization in equal mass systems scales as
\citep{heggie+sweatman91}
\begin{equation} 
\label{eq:sigmaex_smallv}
\hat{\sigma}_{\varepsilon \ll 1} \propto \varepsilon^{\sqrt{13} -
1} \propto (\hat{v} - 1)^{(\sqrt{13} - 1)/2}.
\end{equation}
Therefore the general cross section for ionization is
\begin{equation}
\label{eq:ionanalytic}
\frac{1}{\hat{\sigma}_{\textrm{ion}}} = \frac{(\hat{v} - 1)^{(1 -
\sqrt{13})/2}}{\hat{\sigma}_c} + \frac{\hat{v}^2}{\hat{\sigma}_d}.
\end{equation}
Note, however, that the exponent on the $\hat{\sigma}_c$ term is only valid
in the case of approximately equal masses.  If one object contains more than
$\sim$$\nicefrac{1}{2}$ of the mass, the dominant orbital configuration
leading to full ionization will change, leading to a different exponent
\citep{sweatman07}.  \citet{sweatman07} lists exponents for other
configurations, but this exponent could also be determined empirically.  The
validity of the combined cross section for ionization is also illustrated in
Fig.~\ref{fig:binsing} (red dashed line).

\subsection{Four-body scattering}
\label{subsec:fourbody}

Four-body scattering can occur through binary-binary or triple-single
scattering.\footnote{In principle a simultaneous interaction between a
binary and two independent single stars or a simultaneous interaction
between four single stars are also possible, but such events are rare and so
we do not study them here.}  Because the number of objects interacting is
the same in both cases, they generally may be studied in the same framework.
Although there are 22 distinct possible outcomes of four-body scattering, we
here examine the velocity dependences of three broad classes of outcomes:
(1) ionization, including single and full ionization; (2) exchange; and (3)
triple formation.

It is instructive to study the velocity dependence of ionization in the
extreme semi-major axis ratio limit.  If the semi-major axes between the two
binaries are very different or the semi-major axis ratio of the triple is
very large, then the problem can be considered to be a three-body scattering
problem in two limits: (1) at low velocities the smaller binary can be
considered a point mass and (2) at high velocities one component of the
larger binary can be ignored.  There are therefore two critical velocities
in the four-body scattering problem: the critical velocity of the wide
binary scattering off of interloping system, $\hat{v}_{\textrm{crit}, 1}$,
and the critical velocity of the interloping system scattering off of one
component of the wide binary, $\hat{v}_{\textrm{crit}, 2}$. 

\begin{figure}
\centering
\includegraphics[width=8cm]{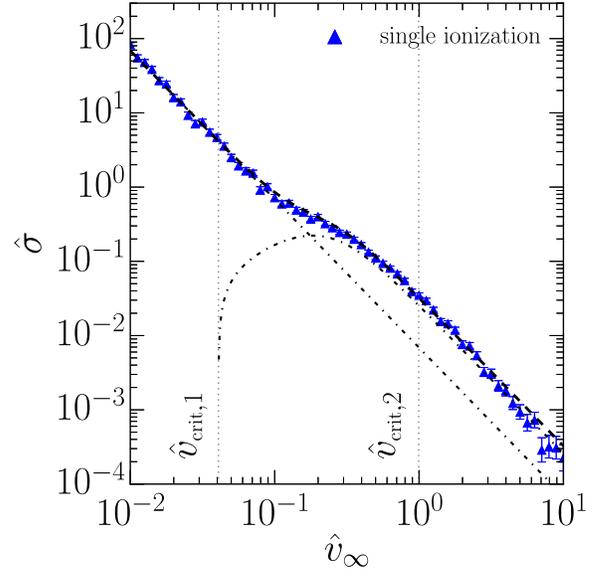}

\caption{Cross sections for outcomes of equal mass binary-binary scattering
with a semi-major axis ratio of 100 and circular orbits as a function of
incoming velocity.  We fit the cross sections for single ionization to
equation~(\ref{eq:bbion}) (dashed line).  The best-fitting parameters are
$\hat{\sigma}_a \approx 2.16 \times 10^{-2}$, $\hat{\sigma}_b \approx 3.47$,
and $\hat{\sigma}_c \approx 8.08 \times 10^{-2}$.  The ionization cross
section consists of two components: a component similar to the full
ionization cross section in binary-single scattering (curved dot-dashed
line), and a component due to hardening of the inner binary (straight
dot-dashed line).  When summed, these two components produce a kink at
velocities intermediate between $\hat{v}_{\crit, 1}$ and $\hat{v}_{\crit,
2}$ (dotted lines).}

\label{fig:binbin}
\end{figure}

In the limit that the semi-major axis of the smaller binary approaches zero,
the cross section for single ionization will appear like the cross section
for ionization in three-body scattering given by
equation~(\ref{eq:ionanalytic}) with the only difference being that one
object in the three-body scattering event has a mass twice as great as the
others because it is a binary.  The effect of this is to change the exponent
on the $(\hat{v} - 1)$ term from $\sqrt{13} - 1$ to $(2 + \sqrt{10})/8$
\citep[section 4.3 of][]{heggie+sweatman91}, and to change the base from
$(\hat{v} - 1)$ to $(\hat{v} - \hat{v}_{\crit, 1})$.  In the limit that the
semi-major axis of one binary approaches zero the cross section for single
ionization approaches zero for $\hat{v} < v_{\textrm{crit}, 1}$.  But the
cross section will always be non-zero because it is possible for single
ionization to occur by hardening the smaller binary.  The cross section for
a wide encounter with a hard binary to harden the binary enough to ionize
the interloping object takes the form \citep[][equation~5.44]{heggie75}
\begin{equation}
\label{eq:lowvbbion}
\hat{\sigma}_{\hat{v} < v_{\crit, 1}} = \frac{\hat{\sigma}_a}{\hat{v}^2},
\end{equation}
where $\hat{\sigma}_a$ is a normalization factor and sub-linear terms (i.e.,
terms with only a logarithmic dependence on the velocity) have been
neglected.  The total cross section for single ionization in either
triple-single or binary-binary scattering is then given by
\begin{equation}
\label{eq:bbion}
\hat{\sigma}_{\textrm{ion}} = \frac{\hat{\sigma}_a}{\hat{v}^2} + \left(
\frac{(\hat{v} - \hat{v}_{\crit, 1})^{(2 + \sqrt{10})/8}}{\hat{\sigma}_b} +
\frac{\hat{v}^2}{\hat{\sigma}_c} \right)^{-1}.
\end{equation}
In Fig.~\ref{fig:binbin} we present the cross section for single ionization
for binary-binary scattering with a semi-major axis ratio of 100, equal
masses, and circular orbits.  We fit the numerical data to the functional
form of equation~(\ref{eq:bbion}) and find an excellent match.  (See the
caption of Fig.~\ref{fig:binbin} for the best-fitting parameters.)

To study exchange and triple formation we return to more moderate semi-major
axis ratios: unity for binary-binary scattering, and 10 for triple-single
scattering.  As discussed in Section~\ref{subsec:binsingapprox},
\citet{hut83} derived the high-velocity exchange cross section to be
proportional to $\hat{v}^{-6}$ for binary-single scattering.  We find from
our numerical experiments in Section~\ref{subsubsec:vinf} that the
high-velocity exchange cross section for exchange is shallower, being
instead proportional to $\hat{v}^{-4}$ for binary-binary and triple-single
scattering (Fig.~\ref{fig:vinf}, black points).  

\begin{figure}
\centering
\includegraphics[width=8cm]{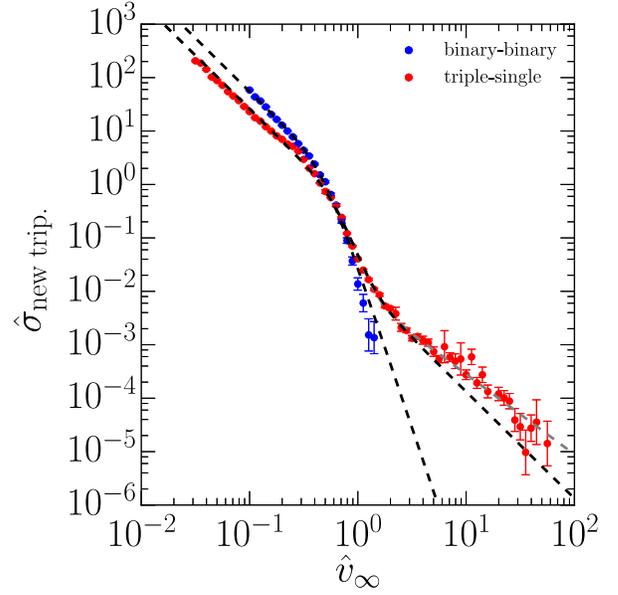}

\caption{The velocity dependence of the cross section for new triple
formation for binary-binary and triple-single scattering of our model system
(see Section~\ref{subsec:model}).  We fit the velocity dependence to
equations~(\ref{eq:vnewtrip_bb}) and (\ref{eq:vnewtrip_ts}) (black dashed
lines).  At high velocities the triple formation cross section from
binary-binary scattering is proportional to $\hat{v}^{-6}$.  At velocities
slightly above the critical velocity, the new triple formation cross section
from triple-single scattering carries the same $\hat{v}^{-6}$ dependence,
but at higher velocities, new triple formation is dominated by scrambles,
which are ionization-like and would therefore carry a $\hat{v}^{-2}$
dependence.  We note, however, that the triple-single data are better fit by
a $\hat{v}^{-3/2}$ dependence (gray dashed line).  For binary-binary
scattering our best fitting parameters are $\hat{\sigma}_a \approx 0.567$
and $\hat{\sigma}_b \approx 0.0279$.  For triple-single scattering our best
fitting parameters assuming a $\hat{v}^{-2}$ dependence for scrambles is
$\hat{\sigma}_a \approx 0.247$, $\hat{\sigma}_b \approx 0.0402$, and
$\hat{\sigma}_c \approx 0.0133$.  Assuming a $\hat{v}^{-3/2}$ dependence for
scrambles the best fitting parameters are $\hat{\sigma}_a \approx 0.257$,
$\hat{\sigma}_b \approx 0.044$, $\hat{\sigma}_c \approx 8.89 \times
10^{-3}$.}

\label{fig:vnewtrip}
\end{figure}

In binary-binary scattering new triple formation occurs via an exchange.  We
find empirically that the velocity dependence for the new triple cross
section is proportional to $\hat{v}^{-6}$ in the high velocity limit for
binary-binary scattering (blue points, Fig.~\ref{fig:vnewtrip}).  In
triple-single scattering, exchanges are the dominant channel for new triple
formation at low velocities and velocities slightly larger than the critical
velocity.  In the high velocity limit the cross section for new triple
formation via exchange is likewise proportional to $\hat{v}^{-6}$ for
triple-single scattering. 

In the case of triple-single scattering new triple formation may also occur
via a scramble.  In the high velocity limit the cross section for scrambles
is proportional to $\hat{v}^{-2}$ (red points, Fig.~\ref{fig:vnewtrip}).
This is because a scramble is, in a sense, an incomplete ionization.  The
interloping star imparts enough energy to one star of the inner binary such
that its semi-major axis becomes much larger than that of the tertiary of
the original system, but not enough energy to ionize it.  At low velocities
a scramble is a rare outcome, but due to the shallower velocity dependence,
at sufficiently high velocities it becomes the dominant mechanism by which
to form new triples in triple-single scattering. 

We may combine these velocity dependences to obtain the general velocity
dependence of the cross section for triple formation.  For binary-binary
scattering, the cross section has an identical form to the exchange cross
section in binary-single scattering, except that the
break occurs near $\hat{v}_{\crit, 1}$ rather than $\hat{v} = 1$ (this is
reflected in the smaller best-fitting value for $\hat{\sigma_b}$):
\begin{equation}
\label{eq:vnewtrip_bb}
\frac{1}{\hat{\sigma}_{\textrm{new trip.}}} =
\frac{\hat{v}^2}{\hat{\sigma}_a} + \frac{\hat{v}^6}{\hat{\sigma}_b}.
\end{equation}
For triple-single scattering the cross section is similar, except that an
additional high velocity $\hat{v}^{-2}$ term must be added to account for
scrambles:
\begin{equation}
\label{eq:vnewtrip_ts}
\hat{\sigma}_{\textrm{new trip.}} = \frac{\hat{\sigma}_a}{\hat{v}^2} +
\left(\frac{\hat{v}^2}{\hat{\sigma}_b} + \frac{\hat{v}^6}{\hat{\sigma}_c}.
\right)^{-1}
\end{equation}
We present a fit of the new triple cross section to
equations~(\ref{eq:vnewtrip_bb}) and (\ref{eq:vnewtrip_ts}) in
Figure~\ref{fig:vnewtrip}.  The match between
equation~(\ref{eq:vnewtrip_bb}) and the binary-binary scattering cross
sections is excellent.  The match between equation~(\ref{eq:vnewtrip_ts})
and the triple-single scattering cross sections is good, but we note that at
high velocities the scramble cross section appears to be better fit with a
$\hat{v}^{-3/2}$ dependence.  We present this alternative velocity
dependence in Figure~\ref{fig:vnewtrip} in gray.  At present we do not have
an analytic understanding of this shallower slope.

While the normalization parameters $\hat{\sigma}_a$, $\hat{\sigma}_b$, etc.,
should be determined by fitting equations~(\ref{eq:bbion}),
(\ref{eq:vnewtrip_bb}), and (\ref{eq:vnewtrip_ts}), for systems not too
dissimilar from those presented in Figures~\ref{fig:binbin} and
\ref{fig:vnewtrip} the ratios between the normalization parameters should be
approximately constant.  The overall scale can then be estimated from the
dependences of the cross sections in
Figures~\ref{fig:alpha}--\ref{fig:massratio}.  For systems far from the
systems presented in Figures~\ref{fig:binbin} and \ref{fig:vnewtrip}, the
normalization parameter may be quite different from these estimates because
they may depend on each other in a nonlinear way.

\section{Orbital parameter distributions after scattering}
\label{sec:dyntrip}

Scattering events can affect the orbital parameter distribution of triples
in two ways: by perturbing the orbital elements of an existing triple in a
flyby, and by creating a new triple system.  We are particularly interested
in determining the distribution of orbital parameters that govern the
strength and timescale of KL oscillations, namely $\cos i$, $\alpha$, and
$e_{\textrm{out}}$.  We first examine the case of flybys in Section
\ref{subsec:flybydist} and then examine newly formed triples in Section
\ref{subsec:dyntrip}.

\subsection{Changes to the orbital parameters from flybys}
\label{subsec:flybydist}

\begin{figure*}
\centering
\includegraphics[width=18cm]{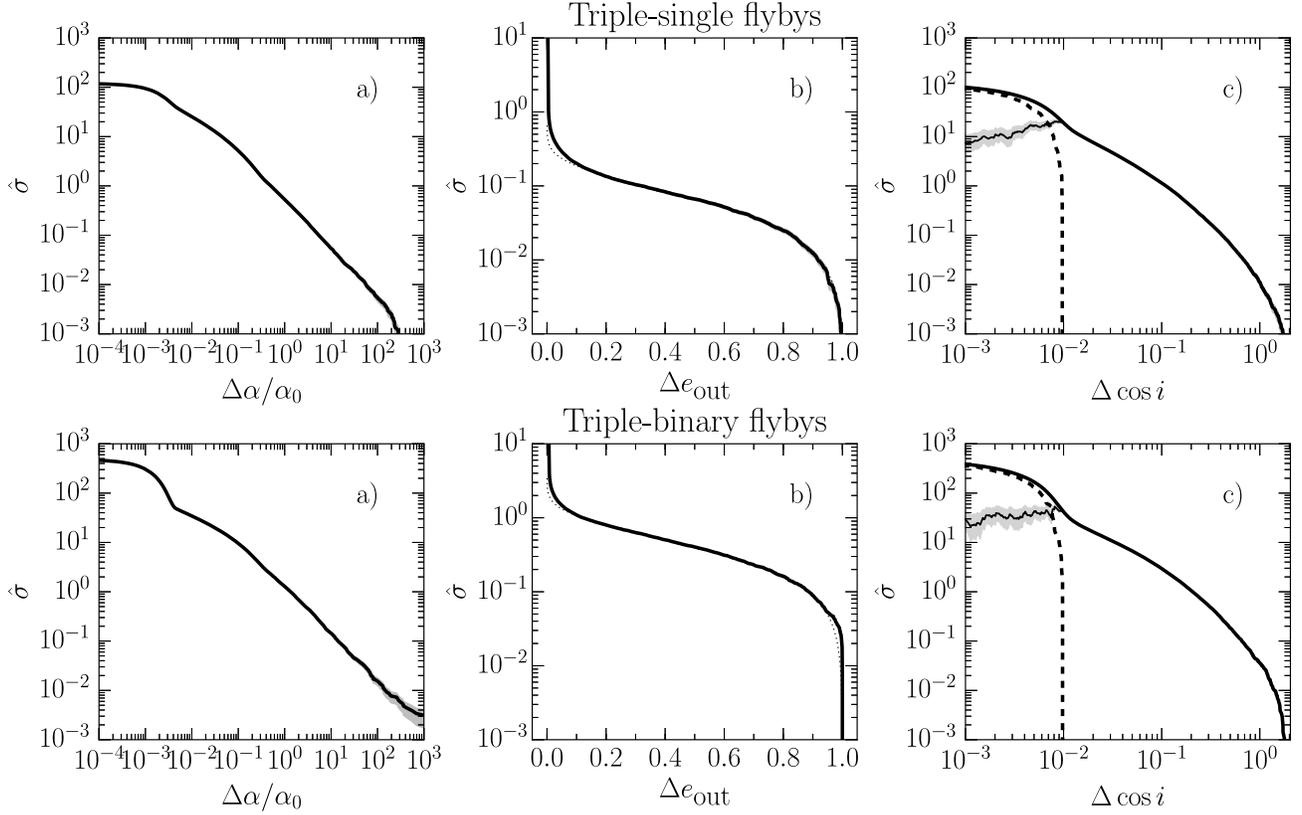}

\caption{The cumulative cross section for changes to the orbital parameters
from triple-single flybys (top row) and triple-binary flybys (bottom row)
with incoming velocities of $\hat{v} = 1$.  We show here changes to the
semi-major axis ratio (panels a), the outer eccentricity (panels b), and the
mutual inclination (panels c).  The cross sections presented are the cross
sections for a change in the orbital parameter of that magnitude \emph{or
greater}.  The 1-$\sigma$ confidence interval is shown by the gray shaded
region, although typically these uncertainties are smaller than the width of
the lines.  We fit the $\Delta e_1$ cumulative cross sections to a Gompertz
function (panels b, dotted line) and find an extremely close match.  Except
for very small or large $\Delta e_1$ the difference between the numerical
data and the fit is smaller than the width of the line.  The best-fitting
parameters are $c_1 \approx 0.0626$ and $c_2 \approx 0.866$ for
triple-single flybys and $c_1 \approx 0.469$ and $c_2 \approx 0.810$ for
triple-binary flybys.  Since isolated triples can undergo changes to their
inclination due to KL oscillations (an effect which is necessarily present
when we perform our scattering experiments), we compare the magnitude of
this effect to the cross sections due to scattering by calculating the
evolution of an identical set of triples for an identical length of time
(panels c, dashed line).  The difference between the scattering cross
section and the cross section from the isolated triples is shown by the thin
black line with 1-$\sigma$ uncertainties.  KL oscillations are unable to
produce large changes in the inclination because the flyby timescale is much
shorter than the KL timescale.}

\label{fig:cumu}
\end{figure*}

Cross sections for flybys are not well defined because they diverge for
infinitesimally small perturbations.  As such, we instead calculate the
cumulative cross section, $\sigma(X > x)$, defined to be the cross section
for a change in parameter $X$ by at least $x$.  From our experiments
involving the model system (see Section \ref{subsec:model}) we select those
which result in a flyby.  We then calculate the cumulative cross section for
generating changes in the orbital parameters of a given magnitude or larger.
These cross sections are shown in Fig~\ref{fig:cumu}.  The normalized cross
section for inducing a change in the semi-major axis of order unity is
$\hat{\sigma} \approx$ 0.54\unskip, which is very
close to the cross section for single ionization of
1.48\unskip.  That the two cross sections should be
comparable is to be expected since there is only a small difference in
energy in changing $\alpha$ by order unity and in ionizing the object.  The
logarithm of the cumulative cross section for changes to $e_1$ is well
described by the inverse of a Gompertz function\footnote{A Gompertz function
is defined to be any function of the form
\begin{equation*}
f(t) = a e^{-b e^{-c t}},
\end{equation*}
where $a$ is the asymptote and $b$ and $c$ are positive constants.}
(cf.~equation~\ref{eq:inv_gompertz_app}):
\begin{equation}
\label{eq:gompertz}
\Delta e_1 = \exp \left(-c_1 \exp \left(c_2 \ln \hat{\sigma}\right)\right)
= \exp \left(-c_1 \hat{\sigma}^{c_2}\right).
\end{equation}
We present a fit to the cumulative cross sections in Fig.~\ref{fig:cumu},
but the difference between the two is typically smaller than the width of
the lines (the best-fitting parameters are presented in the caption to
Fig.~\ref{fig:cumu}).  The cross section for large changes to the
eccentricity is small ($\hat{\sigma} \ll 1$ for $1 - \Delta e_1 \ll 1$)
because to remain stable such systems need to gain a large amount of energy
and are therefore more likely to be ionized than remain bound.  The cross
section for moderate changes to the eccentricity ($\Delta e_1 \sim 0.5$) is
of order unity ($\hat{\sigma} \sim 1$), similar to the case for a change in
$\alpha$.  The cumulative cross section for changes to $\eout$ drops more
steeply for $\Delta \eout \gtrsim 0.7$.  This is due to the fact that the
initial triple is unstable for $\eout \sim 0.7$ (see
equation~\ref{eq:mardling}).  To produce $\Delta e \gtrsim 0.7$ two things
are required: (1) the semi-major axis ratio must increase by enough that at
the final $\eout$ the triple is no longer unstable, and (2) $\eout$ must
change by $\Delta \eout$.  The addition of requirement (1) induces the
sharper drop in the cross section for $\Delta \eout \gtrsim 0.7$.  To check
if the form of equation~(\ref{eq:gompertz}) is independent on the assumption
of equal masses we repeated the calculations used to produce panel b) of the
upper row of Figure~\ref{fig:ecc} and panel b) of the upper row of
Figure~\ref{fig:cumu} using the following sets of masses: $(m_{111},
m_{112}, m_{12}, m_2) \in$ \{(4, 3, 2, 1), (1, 2, 3, 4), and (1, $10^{-6}$,
1, 1)\}.  We found equation~(\ref{eq:gompertz}) to be an excellent fit in
all cases.  While this does not prove that equation~(\ref{eq:gompertz}) is
necessarily valid for all possible mass ratios, it demonstrates that the
form of equation~(\ref{eq:gompertz}) is not strongly dependent on the
assumption of equal masses.

The change in the inclination is particularly interesting due to its
relevance for inducing KL oscillations in a triple system.  However there is
a confounding effect in the cumulative cross section for this orbital
parameter.  Since some triples will begin at high inclinations, their
inclination can change in isolation due to KL oscillations.  This is a
difficult effect to disentangle from changes that are solely due to
scattering or perturbations from interloping stars.  To estimate the
magnitude of this effect, we evolve a sample of triple systems identical to
those used in our model systems but without any interloping stars for the
same length of time as our flyby calculations.  We then calculate the
cumulative distribution of the change in inclination to compare its
magnitude to the distribution when an interloping star is present.  Small
changes in inclination can be accounted for almost entirely by KL
oscillations.  Because these scattering calculations complete in $\sim$10\%
of the KL timescale, $t_{\textrm{KL}}$, KL oscillations cannot produce large
changes in inclination.  Thus, the cumulative cross sections for large
changes in inclination are entirely due to the interloping star.  In more
compact systems, however, the KL timescale may be comparable to the
timescale of the scattering event, in which case either scattering or KL
oscillations would be able to produce large changes in inclination.  The
distributions of positive and negative changes in $\cos i$ are equal to
within the statistical uncertainty of our results.

\begin{figure}
\centering
\includegraphics[width=8cm]{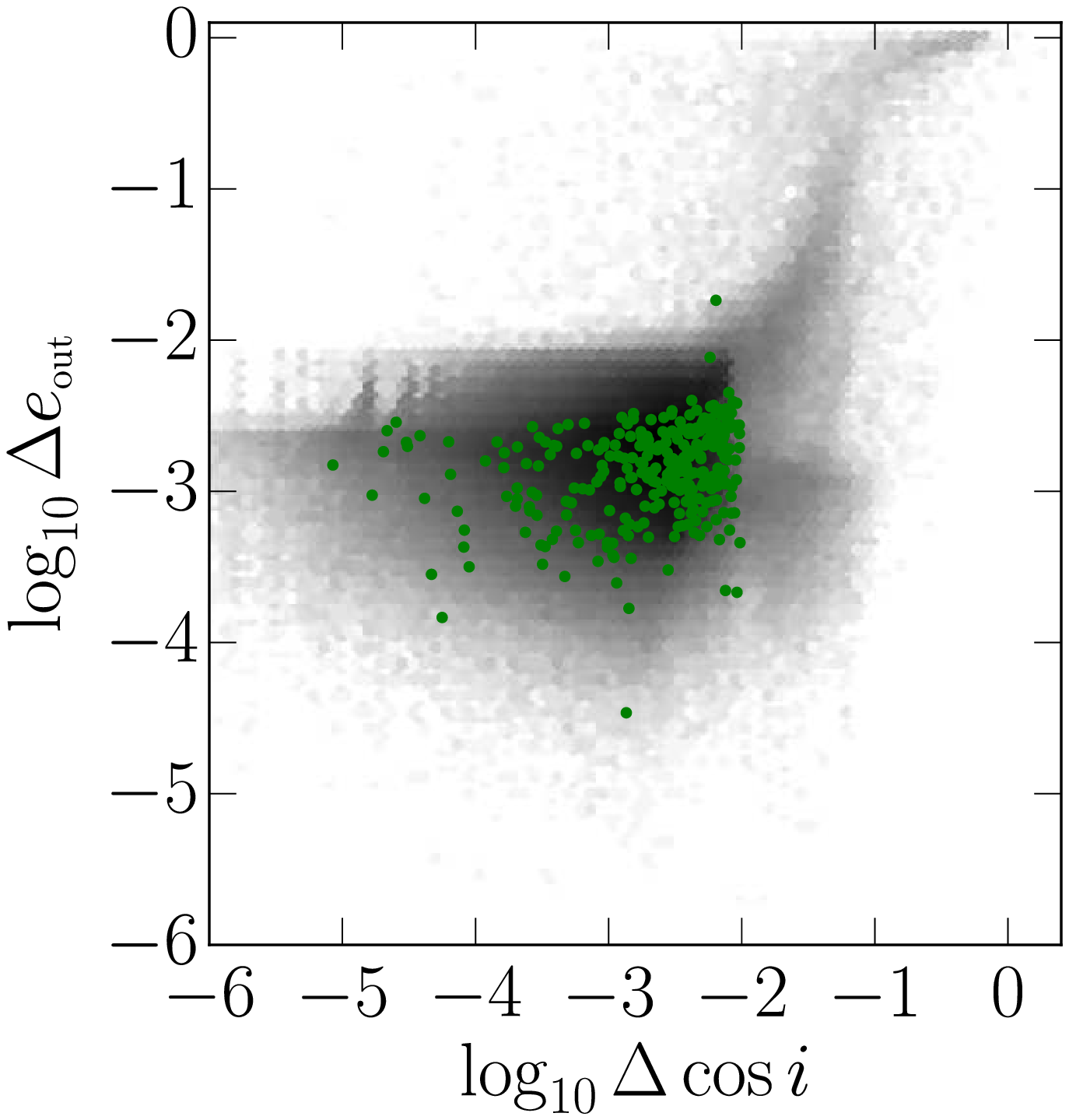}

\caption{The relationship between changes to the mutual inclination and
changes to $\eout$ from flyby events.  The darkness of the hexagonal bins
corresponds to the number of events, binned logarithmically.  To demonstrate
the magnitude of KL oscillations we include this relationship for several
isolated triples (green points).  For events in which there is a change to
$\cos i$ less than $\sim$10$^{-2}$ there is not a strong correlation
between $\Delta \cos i$ and $\eout$.  Changes to the orbital parameters are
due primarily to KL oscillations rather than the dynamical effect of the
interloping star.  For events in which there is a change to $\cos i$ greater
than $\sim$10$^{-2}$ there is a strong correlation between $\Delta \cos i$
and $\Delta \eout$, though with substantial scatter.}

\label{fig:cosi_ecc}
\end{figure}

We additionally show the relationship between changes to the mutual
inclination and changes to $\eout$ in Figure~\ref{fig:cosi_ecc}.  To
demonstrate the magnitude of the effect of KL oscillations we also include
the relationship between changes to the inclination and $\eout$ from the
isolated triples.  Flyby events for which $\Delta \cos i \lesssim 10^{-2}$
do not exhibit a strong correlation between the change to the inclination
and $\eout$.  In these systems the dynamics of the isolated triple therefore
overwhelm the contributions of the interloping star.  Flyby events for which
$\Delta \cos i \gtrsim 10^{-2}$ do exhibit a strong correlation between the
change in the inclination and $\eout$, albeit with substantial scatter.
This result is consistent with other dynamical studies
\citep[e.g.,][]{li+adams15}.

\subsubsection{A note on convergence}
\label{subsubsec:flyby_convergence}

\begin{figure}
\centering
\includegraphics[width=8cm]{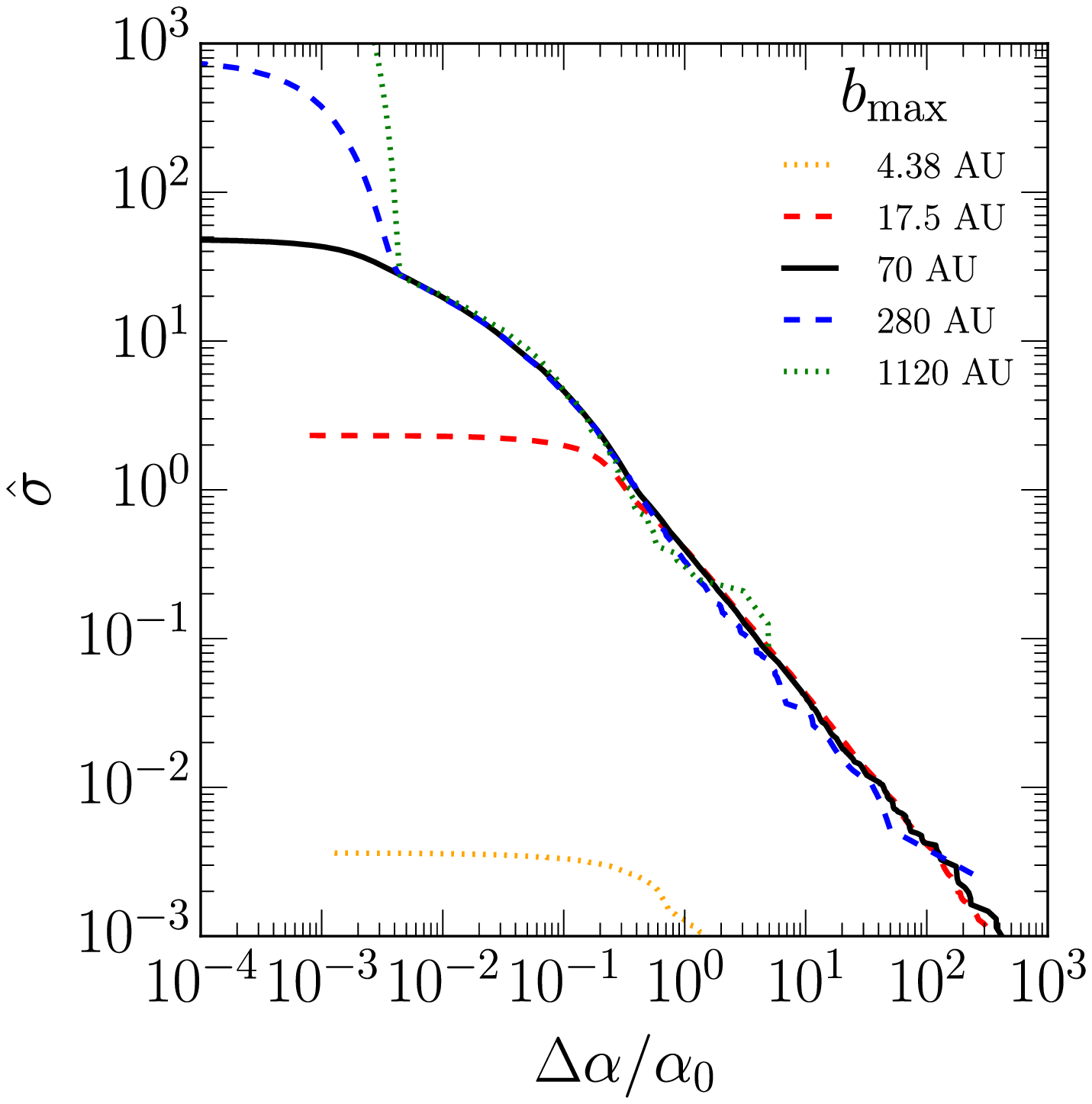}

\caption{The cumulative cross section for changes to the semi-major axis
ratio, $\alpha$, for a variety of choices of \bmax{}.  For changes larger
than $\Delta \alpha / \alpha_0 \sim 3 \times 10^{-3}$ our calculations using
\bmax{} = 70 AU are converged.  For changes to $\alpha$ smaller than $\sim$3
$ \times 10^{-3}$ the cross sections diverge with larger \bmax{} because
$\alpha$ in an isolated triple will vary on this scale over time (see
Figure~\ref{fig:alpha_variability}).  Choices of \bmax{} less than 70 AU are
not converged down to the limit of $\Delta \alpha / \alpha_0 \sim 3 \times
10^{-3}$, resulting in a smaller asymptotic cross section.}

\label{fig:flybyconv}
\end{figure}

The issue of whether our calculations are converged in the maximum impact
parameter, \bmax{}, is subtler in flyby events than the convergence analysis
presented in Section~\ref{subsec:initialconditions}, so we revisit it here.
In Section~\ref{subsec:initialconditions} we considered the convergence of
the cross sections of outcomes from ``strong'' interactions (i.e.,
interactions for which the hierarchy changed).  In flyby events, however,
arbitrarily distant encounters can have arbitrarily weak dynamical effects,
so our calculations can never be truly converged in \bmax{}.  We can,
however, determine the magnitude of the change of orbital parameters down to
which our calculations are converged by calculating the cumulative cross
sections using a variety of \bmax{}'s.  We present the results of this
calculation for changes to the semi-major axis ratio, $\alpha$, in
Figure~\ref{fig:flybyconv}.

\begin{figure}
\centering
\includegraphics[width=8cm]{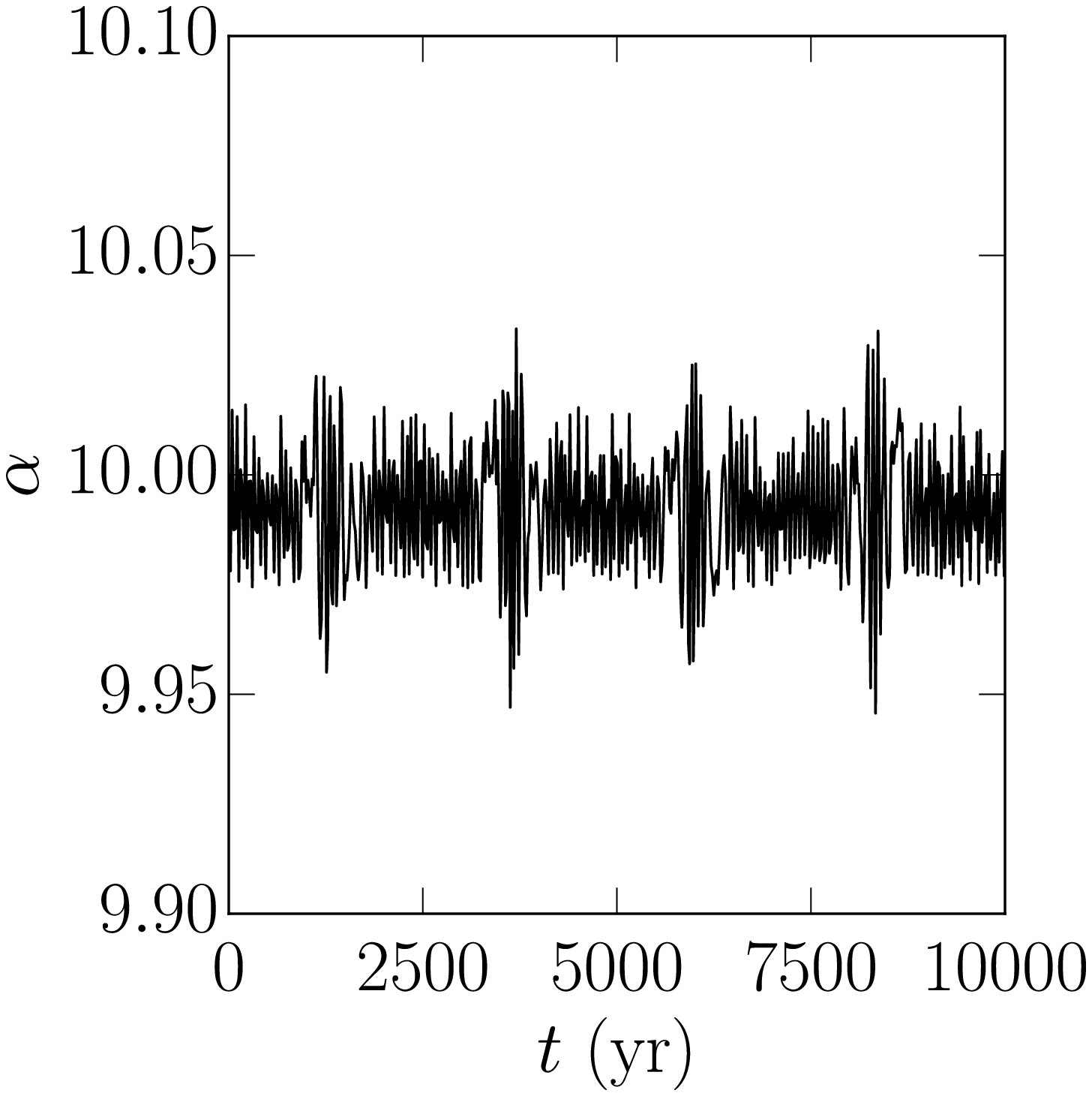}

\caption{The semi-major axis ratio, $\alpha$ over time for an isolated
triple with initial orbital parameters of: $a_{11} = 1$ AU, $a_1 = 10$ AU,
$e_{11} = e_1 = 0$, $i = 70^{\circ}$.  The relative variation of $\alpha$ is
$\sim$3 $\times 10^{-3}$, in agreement with the point of divergence seen
in Figure~\ref{fig:flybyconv}.}

\label{fig:alpha_variability}
\end{figure}

Our results are converged down to relative changes in $\alpha$ of $\sim$3
$\times 10^{-3}$.  The cross section for smaller changes to $\alpha$
diverges, however.  This is due to the fact that the semi-major axes of the
inner and outer orbits of an isolated triple will exhibit small variations
over time.  We demonstrate the scale of this variability by integrating an
isolated triple in Figure~\ref{fig:alpha_variability}.  For the triples we
are studying here the scale of this variability is $\Delta \alpha / \alpha_0
\sim 3 \times 10^{-3}$.  This scale corresponds with the point at which the
cross sections diverge in Figure~\ref{fig:flybyconv}.  Note, however, that
the scale of this variability is dependent on the on the orbital parameters
of the triples being studied.  In particular, as $\alpha \to \infty$, the
relative size of this variability will approach zero.  Calculation of the
cross sections of changes to the orbital parameters from flybys of triples
with very large $\alpha$ may therefore require a larger \bmax{} than given
by equation~(\ref{eq:bmax}).

\subsection{Distribution of orbital parameters in dynamically formed
triples}
\label{subsec:dyntrip}

The scattering experiments presented in Section~\ref{sec:scattering}
demonstrated that some fraction of binary-binary, triple-single, and
triple-binary scattering events produce new triple systems.  Because this
fraction is relatively small, we here run three larger suites of scattering
experiments to obtain better statistics on the distribution of orbital
parameters of dynamically formed triples.  One suite consists of
binary-binary scattering experiments, another of triple-single experiments,
and the last of triple-binary experiments.  In the case of triple-binary
scattering the interloping binary has a semi-major axis equal to the outer
binary of the triple.  We set the semi-major axis ratio of the triples to be
10 and set $a_1 = a_2$ in the case of binary-binary scattering.  We
furthermore set all masses to be equal and set an incoming velocity of $v =
v_{\textrm{crit}}$.  We do not include the orbital parameters of triples
resulting from flybys in these distributions as these were studied in
Section~\ref{subsec:flybydist}.  To ensure that the resulting triples are
stable we run them in isolation for 100 outer orbital periods and reject
those which do not maintain the same hierarchy.  (E.g.~we reject systems
which begin in the configuration \texttt{[[0 1] 2]} but end in the
configuration \texttt{[[0 2] 1]}).  This is similar to the stability test
used by \citet{mardling+aarseth99}.

We must be careful when calculating the inclination distribution.  Because
the calculations can often run for times comparable to $t_{\textrm{KL}}$, KL
oscillations can bias the inclination distribution towards the Kozai angles
of 39$^{\circ}$ and 141$^{\circ}$.  To correct for this we run the triples
in isolation for half the length of time that the scattering event was
computed; this is approximately the length of time that the triple has
existed for.  We then use the inclination closest to 90$^{\circ}$ in the
inclination distribution because the triple spends most of its time at this
inclination.  (For example, an equal mass triple with an initial inclination
of 85$^{\circ}$ spends over 80 per cent of its time within 10$^{\circ}$ of
its initial inclination.)

\begin{figure*}
\centering
\includegraphics[width=18cm]{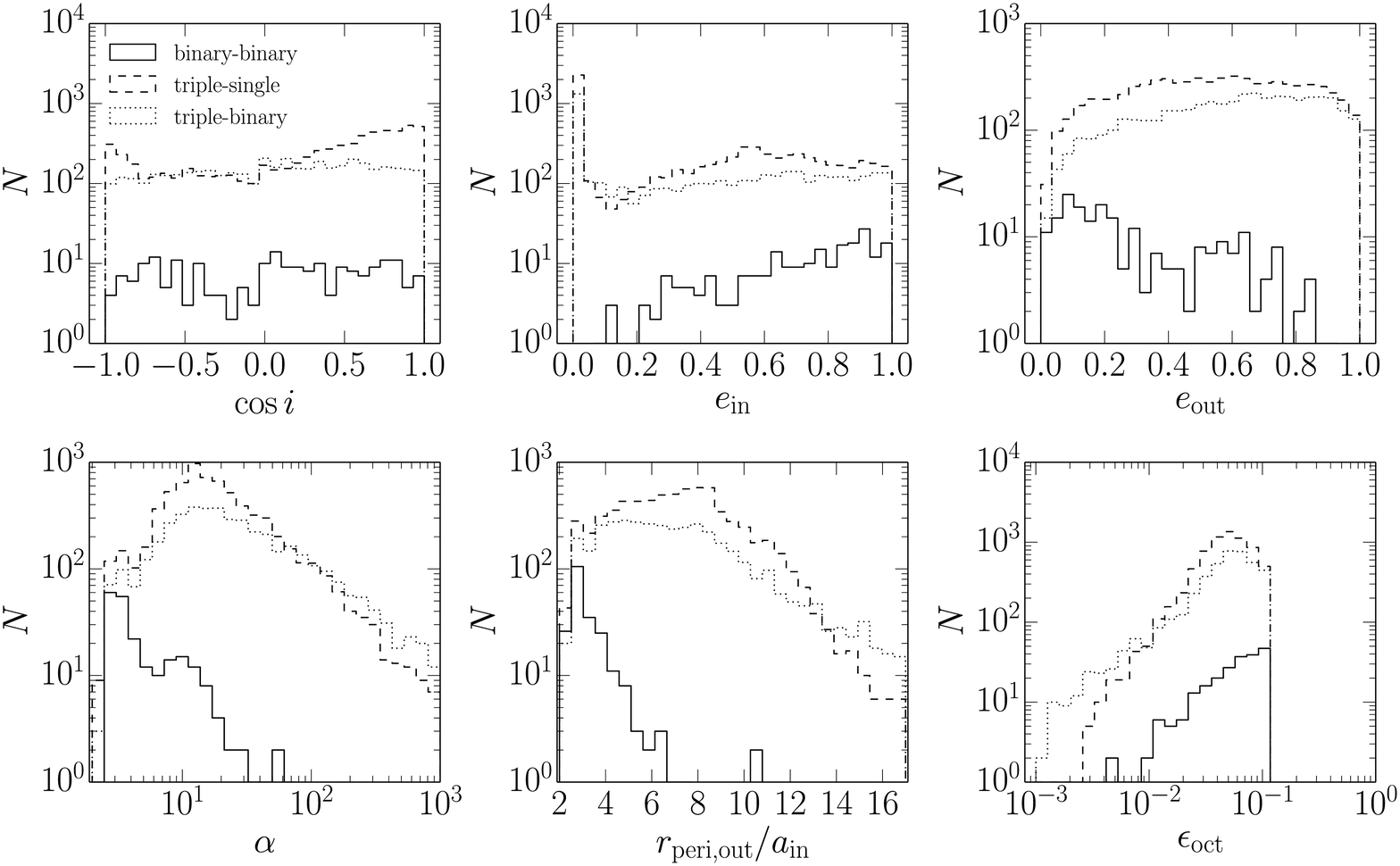}

\caption{Orbital parameters of dynamically formed triples from binary-binary
(solid lines), triple-single (dashed lines), and triple-binary (dotted
lines) scattering.  The systems consist of 1 \Msun{} stars in circular
orbits with a semi-major axis of 1 AU in the case of binary-binary
scattering, and $a_{\textrm{in}} = 1$ AU and $a_{\textrm{out}} = 10$ AU in
the case of triple scattering.  We have excluded triples whose final
hierarchy is identical to their initial hierarchy.  \emph{Upper left:} the
inclination distribution for dynamically formed triples.  We correct the
inclinations for KL oscillations; this panel shows the minimum $\cos i$ that
is achieved by KL oscillations.  The distribution is approximately uniform
in $\cos i$.  \emph{Upper middle:} the distribution of inner eccentricities.
Both triple-single and triple-binary scattering produce a peak at low
eccentricities due to our choice of initial conditions of circular inner
orbits.  Both also produce a flat distribution of $e_{\textrm{in}}$ whereas
binary-binary scattering produces a distribution of $e_{\textrm{in}}$ which
is closer to thermal.  \emph{Upper right}: The distribution of outer
eccentricities.  Triples produced from binary-binary scattering are biased
towards small $e_{\textrm{out}}$, but triple scattering produces triples
that are slightly biased towards large $e_{\textrm{out}}$.  \emph{Lower
left:} the distribution of semi-major axis ratios.  \emph{Lower middle:} the
distribution of distances at periapsis relative to the inner semi-major
axis.  Dynamically formed triples, particularly those formed from
binary-binary scattering, are extremely compact.  Most are close to the
stability boundary.  \emph{Lower right:} the distribution of
$\epsilon_{\textrm{oct}}$, a measure of the strength of the octupole order
term in the Hamiltonian.  Generally, if $\epsilon_{\textrm{oct}} > 10^{-2}$
the EKM will operate and produce flips for a substantial range of initial
inclinations.  Although these triples will not exhibit the EKM because all
objects are equal mass, dynamically formed triples with unequal masses
produce a similar $\epsilon_{\textrm{oct}}$ distribution. (See
Fig.~\ref{fig:poptrip}.)}

\label{fig:dyntrip}
\end{figure*}

The distribution of orbital parameters is shown in Fig.~\ref{fig:dyntrip}.
The most important feature of these distributions is that scattering
produces triples which are extremely compact, in that the ratio of the
distance at periapsis of the outer orbit, $r_{\textrm{peri, out}}$, to
$a_{\textrm{in}}$ is quite small, typically $\sim$10.  Indeed, the vast
majority of these triples are on the verge of instability.  

Because these triples are so compact, they have extremely large
$\epsilon_{\textrm{oct}}$.  This parameter is a measure of the strength of
octupole-order contributions to the KL oscillations and is defined to be
\citep{lithwick+naoz11, naoz+13a}
\begin{equation}
\label{eq:epsilonoct}
\epsilon_{\textrm{oct}} \equiv \frac{1}{\alpha} \left(
\frac{e_{\textrm{out}}}{1 - e_{\textrm{out}}^2} \right).
\end{equation}
Triples with larger $\epsilon_{\textrm{oct}}$ have stronger eccentric KL
oscillations.  In particular, \citet{lithwick+naoz11} note that for
$\epsilon_{\textrm{oct}} > 10^{-2}$ the parameter space for orbital flips
(and hence arbitrarily large eccentricities in the inner binary) widens
dramatically.  Our results indicate that nearly all dynamically formed
triples have $\epsilon_{\textrm{oct}} > 10^{-2}$.  We note that the
particular triples formed from these experiments do not undergo typical
eccentric KL oscillations because they are of equal mass, so the octupole
order terms of the Hamiltonian vanish \citep[e.g.,][]{ford+00}.
Nevertheless, in extremely compact triples non-secular effects can drive the
inner binary to arbitrarily large eccentricities
\citep[e.g.,][]{antonini+14, bode+wegg14, antognini+14}.  Moreover, as we
show in Section~\ref{subsec:pop}, these results for the dynamical formation
of triples are generic and apply to unequal mass triples.  

The inclination distributions appear to be approximately uniform, but are
clearly biased towards inclinations less than 90$^{\circ}$, i.e., prograde
orbits.  This is curious because we began with a distribution of
inclinations uniform in $\cos i$ (i.e., as many retrograde orbits as
prograde), but many studies have found that retrograde orbits are generally
more stable than prograde orbits \citep[e.g.,][]{harrington72,
mardling+aarseth01, morais+giuppone12}.  It is possible that scattering is
biased towards producing triples in prograde orbits and there remains an
excess of prograde orbits after the unstable triples have dissociated. 

The inner eccentricity distributions have two components, one of which is
approximately thermal for high $e_{\textrm{in}}$.  In the case of triple
scattering there is also a strong component peaked near $e_{\textrm{in}}
\approx 0$.  This is an artifact of the fact that we initialized triples
with $e_{\textrm{in}} = 0$ and many new triples were formed from an exchange
between the tertiary and the interloping star.  Since these exchange
reactions do not strongly affect the inner binary there remains a peak at
$e_{\textrm{in}} \approx 0$.  

\begin{figure}
\centering
\includegraphics[width=8cm]{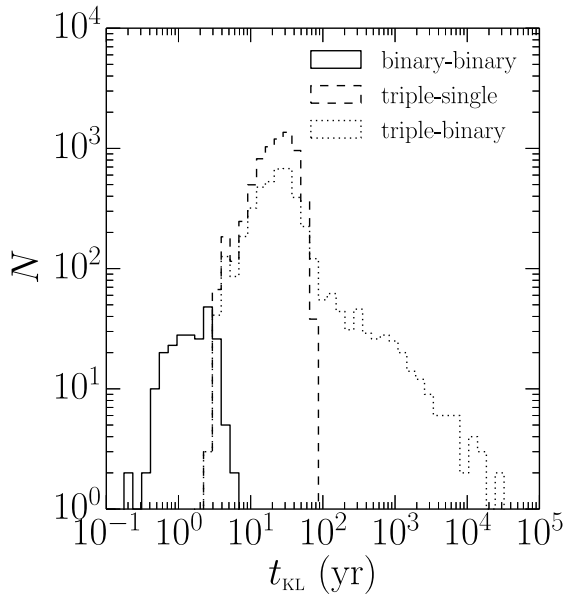}

\caption{The distribution of the timescale of KL oscillations of dynamically
formed triples in a set of model experiments.  In the case of binary-binary
scattering the initial semi-major axes were set to 1 AU, and in the case of
triple scattering we set $a_{\textrm{in}} = 1$ AU and $a_{\textrm{out}} =
10$ AU.  The masses are all 1 \Msun{}.  Triple scattering generally
produces larger triples and hence longer KL timescales because there is a
larger scale in the problem, namely $a_{\textrm{out}}$.  The distributions
of $t_{\textrm{KL}}$ do not extend much beyond a decade, except in the case
of triple-binary scattering which has a longer tail towards large
$t_{\textrm{KL}}$.}

\label{fig:tkdist}
\end{figure}

It is clear from this orbital parameter distribution that a significant
fraction of dynamically formed triples undergo strong KL oscillations.  The
period of these oscillations, $t_{\textrm{KL}}$, for equal mass systems is
given approximately by \citep{holman+97, innanen+97, antognini15}
\begin{equation}
\label{eq:tkl}
t_{\textrm{KL}} = \frac{8}{15 \pi}
\frac{P_{\textrm{out}}^2}{P_{\textrm{in}}} \left( 1 - e_{\textrm{out}}^2
\right)^{3/2}.
\end{equation}
Although our calculations are unitless, if we fix the initial semi-major
axes of the binaries to 1 AU in the case of binary-binary scattering and in
the case of triple scattering we fix the initial semi-major axis of the
inner binary to 1 AU and set all masses to 1 \Msun{} we can calculate the
resulting distribution of $t_{\textrm{KL}}$ in years.  This distribution is
presented in Fig.~\ref{fig:tkdist}.  There are two interesting features: the
first is that they are fairly narrow; the vast majority of dynamically
formed triples span only one decade in $t_{\textrm{KL}}$.  The second is
that $t_{\textrm{KL}}$ is comparable to the initial dynamical timescale of
the systems.  In the case of binary-binary scattering, the initial periods
of the binaries are 0.7 yr and the KL timescales are peaked at approximately
3 yr.  Likewise in the case of triple scattering the outer period is 18 yr
and the KL timescales have a peak at around 20--30 yr.  In the case of
triple-single scattering there is an additional peak at even shorter
timescales.  Such short KL timescales are a consequence of two facts: (1)
dynamically formed triples are extremely compact, and (2) at the critical
velocity triples are more easily formed by hardening binaries than by
softening them, thereby leading to smaller semi-major axes and shorter
periods.

\subsection{Population study}
\label{subsec:pop}

The above analysis reveals that triples produced from our model scattering
experiments are extremely compact.  It is possible that this result is an
artifact from the fact that in binary-binary scattering we scattered
binaries with equal semi-major axes and in triple scattering we scattered
relatively compact triples.  Binary and triple systems in the Galaxy span
many decades of semi-major axis, so the components of typical scattering
events will not be of comparable scales.  

To examine the distribution of orbital parameters of triples produced from a
more realistic population of binary and triple systems in the Galaxy we run
$10^6$ apiece of binary-binary, triple-single, and triple-binary scattering
experiments.  In each experiment we draw a primary mass from the IMF
provided by \citet{maschberger13} which is an analytic approximation to the
IMFs found by \citet{kroupa01} and \citet{chabrier03} We then draw secondary
and, if applicable, tertiary masses from a uniform distribution bounded by
0.08 \Msun{} and the primary mass.  We next draw semi-major axes from a
distribution uniform in $\ln a$ from 0.01 to $10^5$ AU.  In the case of a
triple, we draw two semi-major axes and take the smaller to be the
semi-major axis of the inner binary.  We then draw eccentricities from a
uniform distribution \citep[e.g.,][]{raghavan+10, duchene+kraus13}.  Lastly
we determine the stability of the triple by applying the Mardling stability
criterion (equation~\ref{eq:mardling}).  If the triple is unstable we throw
out the entire experiment and resample all parameters.  We set the incoming
velocity to 40 km s$^{-1}$, which is approximately the velocity dispersion
of older stars in the thin disk \citep{edvardsson+93, dehnen+binney98,
binney+merrifield98}.  Because we are only interested in the orbital
parameters here, and not the cross sections for collisions, we assume all
stars to be point masses.

\begin{figure*}
\centering
\includegraphics[width=18cm]{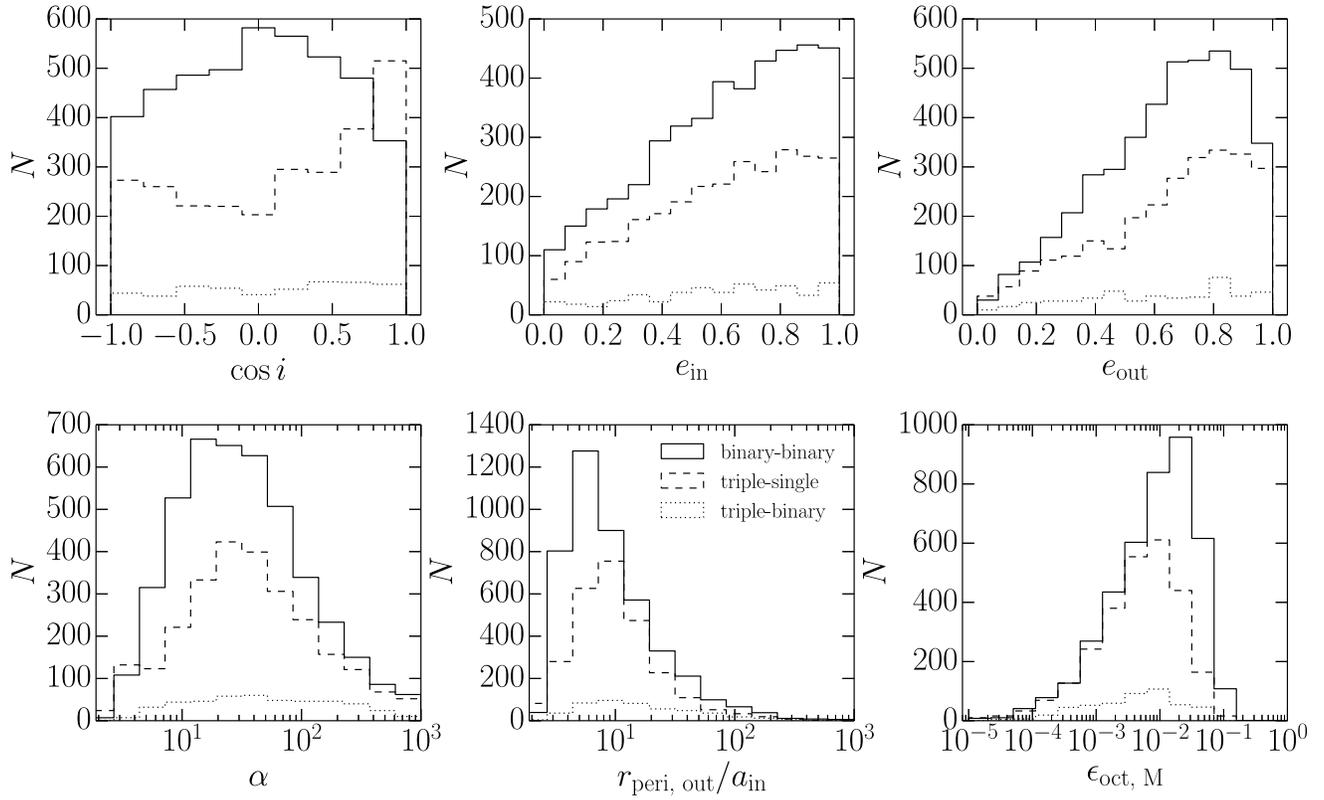}

\caption{Orbital parameters of new triples resulting from scattering
experiments in our population study (Section~\ref{subsec:pop}).  The panels
are as in Fig.~\ref{fig:dyntrip}, except for the bottom right panel, where
we have used $\epsilon_{\textrm{oct}, M}$ instead of
$\epsilon_{\textrm{oct}}$ (see equation~\ref{eq:eoctM}).  As in the model
case, triples formed from scattering are extremely compact and have large
$\epsilon_{\textrm{oct}}$.}

\label{fig:poptrip}
\end{figure*}

The distribution of orbital parameters of new triples is shown in
Fig.~\ref{fig:poptrip}.  Instead of $\epsilon_{\textrm{oct}}$
(equation~\ref{eq:epsilonoct}) we calculate $\epsilon_{\textrm{oct}, M}$,
defined to be \citep{naoz+13a}
\begin{equation}
\label{eq:eoctM}
\epsilon_{\textrm{oct}, M} = \left( \frac{m_{111} - m_{112}}{m_{111} +
m_{112}} \right) \epsilon_{\textrm{oct}},
\end{equation}
where $m_{111}$ and $m_{112}$ are the masses of the two components of the
inner binary.  This term accounts for the fact that the octupole term in the
Hamiltonian has a mass term which vanishes in the case of an equal mass
inner binary.  Thus, $\epsilon_{\textrm{oct}, M}$ indicates the strength of
eccentric KL oscillations.  We estimate the timescale for eccentric KL
oscillations (i.e., the time between orbital flips) to be
\citep{antognini15}
\begin{equation}
\label{eq:tekm}
t_{\textrm{EKM}} = \frac{128 \sqrt{10}}{15 \pi \sqrt{\epsilon_{\textrm{oct},
M}}} t_{\textrm{KL}}.
\end{equation}

\begin{figure*}
\centering
\includegraphics[width=16cm]{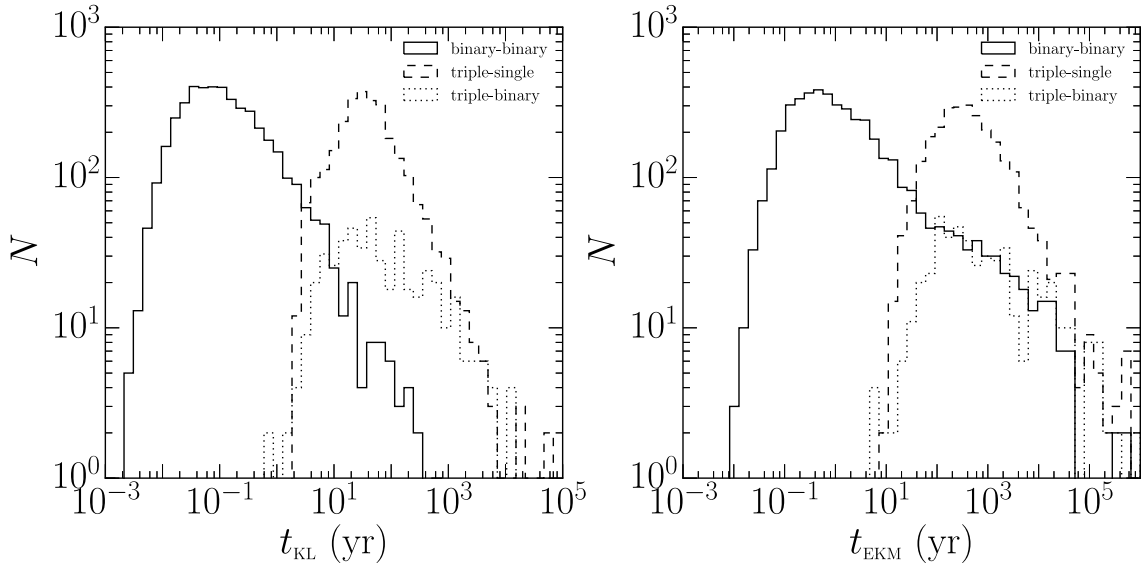}

\caption{Distribution of the timescales of KL oscillations (left panel) and
the eccentric KL mechanism (right panel) in dynamically formed triples in
our population study (Section~\ref{subsec:pop}).  Dynamically formed triples
have very short KL timescales, with nearly all KL timescales being less than
1000 yr.  This is a consequence of the fact that the triples formed are
compact and very wide systems are soft, and therefore have low interaction
cross sections.  Because the triples are so compact the timescale for the
eccentric KL mechanism is also very short, nearly always less than $10^5$
yr.}

\label{fig:poptkl}
\end{figure*}

These results indicate that dynamically formed triples in the field are, as
in our model system, extremely compact.  The peak of the
$\epsilon_{\textrm{oct}, M}$ distribution is broader than the
$\epsilon_{\textrm{oct}}$ distribution, but centered on $\sim$10$^{-2}$,
indicating that the vast majority of dynamically formed triples at high
inclination undergo strong eccentric KL oscillations.  We additionally show
the KL timescales and eccentric KL mechanism (EKM) timescales for these
triples in Fig.~\ref{fig:poptkl}.  Despite the wide range of orbital periods
in the initial systems (with outer periods up to $\sim$10$^7$ yr), the KL
timescales are all quite short.  Nearly all of them are less than 1000 yr.
This is due to two facts: (1) the dynamically formed triples are all
extremely compact, and so don't have KL timescales much longer than the
outer orbital period; and (2) above periods of $\sim$1000 yr the binaries
become soft and so the cross section for triple formation drops dramatically
(see Fig.~\ref{fig:binbin}).  Furthermore, because $\epsilon_{\textrm{oct},
M}$ is so large, the EKM timescales are generally not substantially larger
than the KL time.

\section{Discussion}
\label{sec:discussion}

Here we apply our results on binary-binary, triple-single and triple-binary
scattering in four contexts.  In Section \ref{subsec:typeia} we compare the
production rate of WD-WD binaries with highly inclined tertiaries via
scattering to the SN Ia rate.  In Section \ref{subsec:planet} we estimate
the number of planets in multiple systems ejected due to scattering and
compare to the estimated number of free-floating planets.  We then refine
our previous calculations by modeling stars with non-zero radii to estimate
the rate of stellar collisions in Section \ref{subsec:collisions}.  Finally,
in Section \ref{subsec:kl_lifetime} we address the question of how long KL
oscillations persist for triple systems in globular clusters and in the
field until they are disrupted by scattering events.

\subsection{Application to Type Ia Supernovae}
\label{subsec:typeia}

The progenitor systems of SNe Ia are currently unknown.  The two most viable
models are the single degenerate model, in which a WD accretes mass from a
main sequence or giant star \citep{whelan+iben73, nomoto82}, and the double
degenerate model, in which two WDs merge \citep{iben+tutukov84, webbink84}.
In the past several years the double degenerate model has received much
observational support \citep{bloom+12, schaefer+pagnotta12, edwards+12,
gonzalezhernandez+12, maoz+14, shappee+13}, but population synthesis models
demonstrate that while it may be possible to match the rate of
sub-Chandrasekhar mass WD-WD mergers to the SN Ia rate \citep{ruiter+09,
ruiter+11, maoz+12}, hydrodynamical simulations suggest that a large
fraction of these mergers may not produce SNe Ia \citep{nomoto+97,
maoz+mannucci12, maoz+14}.

\citet{thompson11} provided the first analysis of WD-WD mergers in triple
systems and showed that the perturbative influence of the tertiary can
decrease the merger time by several orders of magnitude
\citep{miller+hamilton02a, blaes+02}.  \citet{thompson11} argued that most
SNe Ia may occur in triple systems due to the relatively large triple
fraction and the larger fraction of triple systems that merge rapidly
relative to isolated WD-WD binaries.  Moreover, \citet{thompson11}
speculated that some of these mergers may result in a collision.
\citet{katz+dong12} first showed that clean, head-on collisions of two WDs
are possible in triple systems using direct N-body calculations and claimed
that $\sim$5\% of WD-WD binaries with tertiaries produce a direct collision
as a result of non-secular dynamics.  The triple scenario for SNe Ia is
attractive because collisions potentially soften the requirement that the
merger mass exceed the Chandrasekhar mass \citep{rosswog+09, raskin+10,
kushnir+13}.  Tertiaries might also hide WD-WD progenitor systems
\citep{thompson11}.  Since the rate of sub-Chandrasekhar mass mergers is
comparable to the SN Ia rate \citep{vankerkwijk+10, maoz+mannucci12,
maoz+12, maoz+14}, this model may solve several problems with the double
degenerate model.

The triple scenario has an important obstacle, however.  The same KL
mechanism that drives the WD-WD binaries to rapidly merge or collide would
drive the stars to tidal contact or merger on the main sequence.  The main
sequence stars would then have either tidally circularized at a few stellar
radii or have undergone common envelope evolution.  In both cases, the
semi-major axis ratio between the inner binary and tertiary would likely
increase significantly, rendering the KL mechanism potentially ineffective
by the time the stars evolved into WDs.  Furthermore, mass loss from one of
the stars generally exacerbates this problem by decreasing the mass ratio of
the inner binary and triggering the eccentric KL mechanism
\citep{shappee+thompson13}.  The eccentric KL mechanism can drive the inner
binary to much higher eccentricities \citep{lithwick+naoz11, katz+11}; thus,
even stars that had escaped tidal circularization on the main sequence may
be brought into tidal contact after the primary evolves into a WD.  While
this may be an important source of single degenerate systems, the resulting
double degenerate system when the smaller mass star evolves into a WD may be
too wide to merge within a Hubble time.  It is, however, possible that some
fraction of these systems might evolve into tight WD-WD binaries during
common envelope evolution, though it is currently very difficult to make
estimates of this fraction \citep{ivanova+13}.

Na\"ively, for the triple scenario to work, it must be possible for a binary
to begin its life in a system that does not undergo KL oscillations and then
efficiently change into a system that undergoes KL oscillations after
evolving into a WD-WD binary.  Scattering is a candidate by which this can
occur.  Although a detailed, comprehensive calculation of the rate is left
for future work, we can present an estimate to determine if scattering is a
viable channel for the production of high inclination triples.  In this
estimate we consider three environments: the field, open clusters and
globular clusters.  We study the field and open clusters separately because
the SN Ia delay time distribution is often decomposed into two separate
parts: a prompt component which explodes within a few hundred Myr of star
formation, and a delayed component which explodes a Gyr or more after star
formation \citep[e.g.,][]{scannapieco+bildsten05, mannucci+06,
maoz+badenes10}.  Since the KL timescale is much shorter than the scattering
timescale (see Section~\ref{subsec:kl_lifetime}), any KL-induced collisions
may occur rapidly after the scattering event \citep[e.g., figure 1
of][]{katz+dong12}.  The maximum lifetime of open clusters is generally only
a few hundred Myr \citep{delafuentemarcos97, lamers+05}, so it is plausible
that scattering events that contribute to the prompt component occur in open
clusters and scattering events that contribute to the delayed component
occur in the field.  However, WD-WD mergers due to KL oscillations do not
necessarily occur instantaneously.  Particularly in the case of
gravitational wave driven mergers, KL oscillations can take more than a Gyr
to drive the WD-WD binary to merger \citep{thompson11}.  It is therefore
possible that scattering events in open clusters could produce SN Ia
progenitors which eventually contribute to the delayed component of the SN
Ia delay time distribution rather than the prompt component. 

We additionally consider scattering in globular clusters.  The SN Ia rate in
globular clusters is poorly constrained.  Based on a lack of detections of
SNe Ia in globular clusters, \citet{voss+nelemans12} placed an upper limit
on the globular cluster SN Ia rate, but this limit was larger than the
Galactic SN Ia rate by a factor of 50.  \citet{graham+15} found one hostless
SN Ia which they claimed was likely to be in a globular cluster.  Based on
this detection they estimated that the SN Ia rate might be enhanced in
globular clusters by a factor of $\sim$25. 

The Galactic SN Ia rate is $\Gamma_{\textrm{Ia}} \sim 5 \times 10^{-3}$
yr$^{-1}$ \citep{maoz+mannucci12}.  If scattering in the field is the primary
channel to form high-inclination triple SN Ia progenitors, then the rate at
which scattering events produce high-inclination triples must be at least as
large as the SN Ia rate.  The rate of dynamical formation of new triples may
be written
\begin{equation}
\Gamma_{\textrm{new triple}} = N n \sigma v,
\end{equation}
where $N$ is the total number of binaries in the case of binary-binary
scattering or the total number of triples in the case of triple scattering,
$n$ is the number density of stars, and $\sigma$ is the cross section to
produce a new triple.  If we consider triples or binaries in a small range
of semi-major axes, $da$, then the differential rate is
\begin{equation}
\label{eq:differential_rate}
d\Gamma(a) = \frac{dN}{da} n v \sigma(a) \, da.
\end{equation}
For our calculation we assume \"Opik's law \citep{opik24},
\begin{equation}
\label{eq:opik}
\frac{dN}{da} = \frac{N_0}{a},
\end{equation}
where $N_0$ is a normalization constant set by the total number of binaries
or triples, $N$:
\begin{equation}
\label{eq:N0}
N_0 = N \left[ \ln \left( \frac{a_{\max}}{a_{\min}} \right) \right]^{-1}.
\end{equation}
Although a wide log-normal distribution is more accurate
\citep[e.g.,][]{duquennoy+mayor91}, the flat distribution is simpler to work
with analytically and provides an upper limit on the number of systems at
the widest semi-major axes. 

The cross section for collision can be written in terms of $\hat{\sigma}$ by
noting that
\begin{equation}
\label{eq:cross_section_written_out}
\sigma(a) = \pi a^2 \hat{\sigma}(\hat{v}).
\end{equation}
We found in Section~\ref{subsec:fourbody} that the normalized cross section
for new triple formation from exchange is proportional to $\hat{v}^{-2}$ for
$\hat{v} \ll 1$ and is proportional to $\hat{v}^{-6}$ for $\hat{v} \gg 1$.
We may therefore write the cross section as
\begin{equation}
\label{eq:sigmatoalowv}
\sigma = \pi a^2 \hat{\sigma}_0 \frac{v^2_{\crit}}{v^2} = \frac{\pi a
\hat{\sigma}_0 \eta^2}{v^2}, \quad \hat{v} < 1
\end{equation}
and
\begin{equation}
\label{eq:sigmatoabigv}
\sigma = \pi a^2 \hat{\sigma}_0 \frac{v_{\crit}^6}{v^6} = \frac{\pi
\hat{\sigma}_0 \eta^6}{a v^6}, \quad \hat{v} > 1
\end{equation}
where $\hat{\sigma}_0$ is the normalized cross section for high-inclination
triple formation at the critical velocity and where we have defined $\eta^2$
to be
\begin{equation}
\label{eq:etadef}
\eta^2 \equiv a v_{\crit}^2
\end{equation}
so as to separate out the dependence of $v_{\crit}$ on $a$.  $\eta$ is the
same for binaries or triples of any size as long as the masses and
semi-major axis ratio remain constant (cf.~equations~\ref{eq:vcrit_binbin}
and~\ref{eq:vcrit_tripsing}).  For the triples we are considering (0.6
\Msun{} and $\alpha = 10$) we have $\eta^2 \sim 10^3$ AU km$^2$ s$^{-2}$.  

We additionally found in Section~\ref{subsec:fourbody} that at very high
velocities the cross section for new triple formation from triple-single
scattering is dominated by scrambles, which have a $\hat{v}^{-2}$
dependence.  This contribution to the overall cross section for new triple
formation may be written
\begin{equation}
\label{eq:sigma_scramble}
\sigma = \pi a^2 \hat{\sigma}_{0, \textrm{sc.}} \frac{v_{\crit}^2}{v^2} =
\frac{\pi a \hat{\sigma}_{0, \textrm{sc.}} \eta^2}{v^2},
\end{equation}
where $\hat{\sigma}_{0, \textrm{sc.}}$ is the normalized cross section for
scrambles at the critical velocity. 

Substituting equations (\ref{eq:opik}),
(\ref{eq:cross_section_written_out}), (\ref{eq:sigmatoalowv}), and
(\ref{eq:sigmatoabigv}) into equation (\ref{eq:differential_rate}) and
integrating with respect to $a$, we find that for binary-binary scattering
\begin{equation}
\Gamma_{a < a_{\crit}} = \frac{\pi N_0 n \hat{\sigma}_0 \eta^2}{v}
\left(a_{\crit} - a_{\min} \right),
\end{equation}
and
\begin{equation}
\Gamma_{a > a_{\crit}} = \frac{\pi N_0 n \hat{\sigma}_0 \eta^6}{v^5}
\left(\frac{1}{a_{\crit}} - \frac{1}{a_{\max}} \right),
\end{equation}
where we have defined $a_{\crit}$ such that the incoming velocity is equal
to $v_{\crit}$,
\begin{equation}
\label{eq:acrit}
a_{\crit} = \frac{G m_{11} m_{12}}{\mu v^2}.
\end{equation}
For triple-single scattering $\Gamma_{a < a_{\crit}}$ remains unchanged, but
we instead have
\begin{equation}
\Gamma_{a > a_{\crit}} = \frac{\pi N_0 n \eta^2}{v} \left[ \hat{\sigma}_0
\left( \frac{\eta}{v} \right)^4 \left( \frac{1}{a_{\crit}} -
\frac{1}{a_{\max}} \right) + \hat{\sigma}_{0, \textrm{sc.}} a_{\max}
\right],
\end{equation}
where we have assumed that $a_{\max} \gg a_{\min}$.  We may then combine
these to obtain the overall rate of new triple formation, given by
\begin{multline}
\Gamma_{\textrm{new triple}} = \frac{\pi N_0 n \hat{\sigma}_0 \eta^2}{v}
\\
\times \left[ \left(a_{\crit} - a_{\min} \right) + \left( \frac{\eta^4}{v^4}
\right) \left(\frac{1}{a_{\crit}} - \frac{1}{a_{\max}} \right) \right]
\end{multline}
for binary-binary scattering, and given by
\begin{multline}
\Gamma_{\textrm{new triple}} = \frac{\pi N_0 n \hat{\sigma}_0 \eta^2}{v}
\\
\times \left[ \left(a_{\crit} - a_{\min} \right) + \left( \frac{\eta^4}{v^4}
\right) \left(\frac{1}{a_{\crit}} - \frac{1}{a_{\max}} \right) +
\left(\frac{\hat{\sigma}_{0, \textrm{sc.}}}{\hat{\sigma}_0} \right) a_{\max}
\right]
\end{multline}
for triple-single scattering.  In most cases we have $a_{\min} \ll a_{\crit}
\ll a_{\max}$, so this may be further simplified to
\begin{equation}
\label{eq:Gamma_newtrip}
\Gamma_{\textrm{new triple}} = \frac{\pi N_0 n \hat{\sigma}_0 \eta^2}{v}
\left[ a_{\crit} + \left( \frac{\eta}{v} \right)^4 \frac{1}{a_{\crit}}
\right]
\end{equation}
for binary-binary scattering, and
\begin{multline}
\Gamma_{\textrm{new triple}} = \frac{\pi N_0 n \hat{\sigma}_0 \eta^2}{v}
\\
\times \left[ a_{\crit} + \left( \frac{\eta}{v} \right)^4
\frac{1}{a_{\crit}} + \left( \frac{\hat{\sigma}_{0,
\textrm{sc.}}}{\hat{\sigma}_0} \right) a_{\max} \right]
\end{multline}
for triple-single scattering.  This implies that nearly all binary-binary
scattering events which produce new triples occur for systems in which the
outer semi-major axis is such that the critical velocity of the outer binary
is close to the velocity dispersion of the surrounding stars.  By contrast,
new triple formation from triple-single scattering events is dominated by
scrambles in systems near the maximum semi-major axis.  

Using our scattering experiments from Section~\ref{subsubsec:vinf} we
calculate $\hat{\sigma}_0$ for binary-binary scattering by calculating the
cross section for outcomes in which a triple is formed with a mutual
inclination of $80^{\circ} < i < 110^{\circ}$.  Although the range of
inclinations that will produce WD-WD collisions is dependent on the
semi-major axis (larger systems will produce collisions for a narrower range
of inclinations), the dependence of the required inclination on the
semi-major axis is not known.  We simply assume the inclination range of
$80^{\circ} < i < 110^{\circ}$ found by \citet{katz+dong12} to produce WD-WD
collisions to hold for all scales.  The cross sections we use should
therefore be considered to be accurate only at the order of magnitude level.
We find that for binary-binary scattering $\hat{\sigma}_0 = 2 \times
10^{-5}$.  For triple scattering we calculate $\hat{\sigma}_0$ in a similar
way, but we only select triples which were initialized with inclinations
less than the Kozai angle ($i < 39.2^{\circ}$ or $i > 141.8^{\circ}$).  We
find that for triple scattering (both triple-single and triple-binary)
$\hat{\sigma}_0 = 2 \times 10^{-3}$.  We likewise calculate the cross
section for high-inclination new triple formation from scrambles at the
critical velocity in a similar way and find $\hat{\sigma}_{0, \textrm{sc.}}
= 7 \times 10^{-5}$.  

\subsubsection{Estimate for the field}

To estimate $\Gamma_{\textrm{new triple}}$ in the field we make the
following assumptions: (1) the total number of white dwarfs in the Galaxy is
$10^{10}$ \citep{napiwotzki09}; (2) the fraction of white dwarfs in WD-WD
binaries is 10 per cent \citep{holberg09}; (3) the fraction of WD-WD
binaries with tertiary companions is the same as the fraction of stellar
binaries with tertiary companions, i.e., 20 per cent \citep{raghavan+10}; (4)
the volume density of stars in the field is 0.09 \Msun{} pc$^{-3}$
\citep{flynn+06}; (5) the mean star mass is 0.36 \Msun{}
\citep{maschberger13}; and (6) the velocity dispersion of stars in the thin
disk is 40 km s$^{-1}$ \citep[p.~656]{edvardsson+93, binney+merrifield98}.

Under these assumptions the total number of WD-WD binaries is
$N_{\textrm{bin}} = 10^9$, the total number of WD-WD binaries with tertiary
companions is $N_{\textrm{trip}} = 2 \times 10^8$, and the number density in
the field is $n = 0.25$ pc$^{-3}$.  We additionally take $a_{\max} = 10^4$
AU and $a_{\min} = 10^{-2}$ AU.  From equation (\ref{eq:N0}) we then have
$N_{0, \textrm{bin}} = 7 \times 10^7$, and $N_{0, \textrm{trip}} = 10^7$,
where $N_{0, \textrm{bin}}$ is the normalization for WD-WD binaries and
$N_{0, \textrm{trip}}$ is the normalization for WD-WD binaries with tertiary
companions.  Using our data from Section~\ref{subsubsec:vinf} we calculate
the normalized cross section for new triple formation to be $\hat{\sigma}_0
= 2 \times 10^{-3}$ at the critical velocity of the tertiary for triple
scattering and $\hat{\sigma}_0 = 2 \times 10^{-5}$ for binary-binary
scattering.  For systems consisting of solar mass stars with an incoming
velocity of 40 km s$^{-1}$, $a_{\crit} = 0.8$ AU.  With these numbers,
equation (\ref{eq:Gamma_newtrip}) gives $\Gamma_{\textrm{new triple}} \sim 5
\times 10^{-9}$~yr$^{-1}$ for triple scattering and $\Gamma_{\textrm{new
triple}} \sim 2 \times 10^{-13}$~yr$^{-1}$ for binary-binary scattering in
the Galaxy.  The rates from triple scattering are several orders of
magnitude larger than the rates from binary-binary scattering because the
rate from triple scattering is dominated by scrambles at large semi-major
axes, which are not a possible outcome of binary-binary scattering.
However, rates from both binary-binary and triple scattering are far below
$\Gamma_{\textrm{Ia}}$.  In elliptical galaxies the rates are further
depressed relative to the estimate in the Galactic field due to the lower
number densities and higher velocity dispersions (and hence lower
$a_{\crit}$).

\subsubsection{Estimate for open clusters}

To estimate $\Gamma_{\textrm{new triple}}$ for open clusters we make the
following assumptions: (1) the typical number density of an open cluster is
10 pc$^{-3}$ \citep{piskunov+07}; (2) the typical velocity dispersion of an
open cluster is 0.3 km s$^{-1}$; (3) the binary fraction is 50 per cent and
the triple fraction is 10 per cent \citep{raghavan+10}; (4) the total number
of open clusters in the Galaxy is $10^5$ \citep{piskunov+06}; and (5) the
typical system is in an open cluster with 300 members \citep{porras+03}.  

Under these assumptions the total number of triples in open clusters is
$\sim$3 $\times 10^6$ and the total number of binaries in open clusters is
$\sim$2 $\times 10^7$.  From equation (\ref{eq:N0}) we have $N_{0,
\textrm{bin}} = 10^6$ and $N_{0, \textrm{trip}} = 2 \times 10^5$.  Note that
with our assumed velocity dispersion we have $a_{\crit} \sim 10^4$ AU, which
is comparable to the semi-major axes of the widest binaries observed.  Thus
in an open cluster environment we are in the low-velocity regime even for
the widest systems.  Since scrambles are a dominant channel to produce high
inclination triples only at very high velocities, the contribution of
scrambles can be neglected in open clusters.

With these numbers, equation (\ref{eq:Gamma_newtrip}) gives
$\Gamma_{\textrm{new triple}} \sim 2 \times 10^{-5}$ yr$^{-1}$ for triple
scattering and $\Gamma_{\textrm{new triple}} \sim 5 \times 10^{-4}$
yr$^{-1}$ for binary scattering in all open clusters in the Galaxy.  Thus,
even in open cluster environments the new triple formation rate is below the
SN Ia rate.  These estimates suggest that it is difficult for scattering to
produce high-inclination triples at a rate consistent with the SN Ia rate.
Moreover, the estimated rate is dominated by scattering events near
$a_{\crit}$, which is $\sim$10$^4$ AU.  These separations are only smaller
than the typical separations between stars in the open cluster by a factor
of a few, so our implicit assumption that each scattering event proceeds in
dynamical isolation is likely unwarranted at the largest semi-major axes
(see Section~\ref{subsubsec:typeia_summary}). 

\subsubsection{Estimate for globular clusters}

To estimate $\Gamma_{\textrm{new triple}}$ for globular clusters we make the
following assumptions: (1) the typical mass of a globular cluster is $2
\times 10^5$ \Msun{} \citep{harris96}; (2) the typical half-mass radius is 3
pc \citep{harris96}; (3) there are 150 globular clusters in the Galaxy
\citep{harris96}; (4) 0.7 per cent of the systems in a globular cluster are
WD-WD binaries \citep{shara+hurley02}; (5) the triple fraction relative to
the binary fraction is the same as the field, i.e., 0.2
\citep{raghavan+10};\footnote{In reality the triple fraction in globular
clusters will be determined by the dynamical evolution of the cluster, so
there is no reason that the triple fraction in globular clusters should be
similar to that of the field.  However, the triple fraction in globular
clusters is poorly constrained, though it is unlikely to be very much higher
than that of the field, so we assume here a triple fraction to binary
fraction of 0.2 as a rough upper limit.} and (6) the typical velocity
dispersion of a globular cluster is 6 \kms{} \citep[p.~31]{harris96,
binney+tremaine08}.

Under these assumptions the average number density in the half-mass radius
is $\sim$2500 pc$^{-3}$, and the typical separation between stars is
$\sim$10$^4$ AU.  To maintain our assumption of dynamical isolation we
therefore take $a_{\max} = 10^3$ AU.  Given a median star mass of 0.36
\Msun{} \citep{maschberger13}, the total number of stars within the
half-mass radii of all globular clusters is $\sim$4$ \times 10^7$, the
number of WD-WD binaries is $\sim$3$ \times 10^5$, and the total number of
WD-WD binaries with tertiaries is $\sim$6$ \times 10^4$.  This then implies
that $N_{0, \textrm{bin.}} \sim 3 \times 10^6$ and $N_{0, \textrm{trip.}}
\sim 7 \times 10^5$.  We furthermore have $a_{\crit} \sim 40$ AU.

With these numbers, equation~(\ref{eq:Gamma_newtrip}) gives
$\Gamma_{\textrm{new triple}} \sim 10^{-7}$ yr$^{-1}$ for binary-binary
scattering and $\Gamma_{\textrm{new triple}} \sim 4 \times 10^{-6}$
yr$^{-1}$ for triple scattering for all globular clusters in the Galaxy.  To
compare these rates with the Galactic SN Ia rate, we note that the total
mass in globular clusters is $\sim$3$ \times 10^7$ \Msun{} whereas the total
stellar mass in the Galactic disk is $\sim$5$ \times 10^{10}$ \Msun{}
\citep{mcmillan11}.  Thus the fraction of stellar mass in globular clusters
is $\sim$5$ \times 10^{-4}$, so the expected SN Ia rate in globular clusters
assuming no enhancement is $\sim$3$ \times 10^{-6}$ yr$^{-1}$.  This is
comparable to the scattering rates that we find.  Thus, if the SN Ia rate is
not enhanced in globular clusters, scattering could be an important channel
to form WD-WD binaries with high inclination tertiaries.  If, however, the
SN Ia rate is enhanced by a factor of $\sim$25 as claimed by
\citet{graham+15} scattering would be unlikely to be the primary channel by
which to produce SN Ia progenitors in triples. 

\subsubsection{Summary of rate estimates}
\label{subsubsec:typeia_summary}

Although there are substantial uncertainties and simplifications in the
analysis of this subsection, these results imply that if the triple scenario
is correct, either scattering of very wide triples in open clusters leads to
the formation of high-inclination triple systems or a mechanism other than
scattering is responsible for the formation of high-inclination triple
systems.  It also may be that the full dynamics of open clusters beyond the
isolated scattering events we have considered here (such as the dissociation
of the cluster) are important for the formation of WD-WD binaries in high
inclination triple systems.  While we have treated scattering events in open
clusters in isolation, this is likely not a good approximation, particularly
at the large semi-major axes considered here \citep{geller+leigh15}.
It is also possible that triple-binary scattering produces quadruples, which
undergo strong KL oscillations over a wider range of parameter space
\citep{pejcha+13}.  Scattering in globular clusters is more promising, as
the rate of new triple formation can match the Galactic SN Ia rate (after
correcting for the fraction of mass in globular clusters).  Finally, we note
that our calculations of the normalized cross sections were done assuming
relatively compact triples (semi-major axis ratios of 10) and binaries of
equal size.  The distribution of semi-major axis ratios of triples in the
Galaxy will certainly be broader than we assumed, so the true normalized
cross sections will be different from the cross sections we calculated.
Indeed, they will likely be lower due to the fact that scattering is
generally more efficient in compact triples---see Section~\ref{subsec:pop}.
These issues should be investigated more fully in a future work.

We conclude that scattering in the field cannot produce WD-WD binaries in
high-inclination triples at a rate consistent with the SN Ia rate, but we
cannot rule out the possibility that scattering in open clusters and
globular clusters leads to high inclination triples at rates consistent with
the SN Ia rate.

\subsection{Application to free-floating planets}
\label{subsec:planet}

The dynamical effect of interloping stars or binaries on planetary systems
has been studied extensively (see \citealt{adams10} for a review).
\citet{malmberg+07} measured the rate of close encounters between stars in
an open cluster environment and found that dynamical processes will alter
the population of planetary systems similar to the Solar System.
\citet{malmberg+davies09} found that these close encounters can produce
eccentric orbits with an eccentricity distribution similar to the observed
distribution of eccentricities of very eccentric exoplanets.
\citet{malmberg+11} calculated the long-term effect of distant flybys on
planetary systems and showed that $\sim$10 per cent of systems are pushed
from a configuration which is stable on timescales of $\sim$10$^8$ yr to a
configuration which is unstable on these timescales.  Most recently,
\citet{li+adams15} calculated the cross sections for various outcomes and
changes to the orbital parameters from scattering events between a planetary
system and a binary star.  In this subsection we apply our results to the
free-floating planet population.

Microlensing surveys have claimed the existence of a large population of
free-floating planets \citep{zapatero+00, sumi+11}.  It is unknown whether
these sub-brown dwarf objects form in isolation or if they were ejected from
the planetary systems in which they were born.  Numerical simulations
indicate that it may be difficult for planet-planet scattering to fully
account for the observed numbers of free-floating planets
\citep{veras+raymond12}.  While there is some observational evidence for the
isolated formation of sub-brown dwarf objects in the Rosette Nebula
\citep{gahm+13}, it is nevertheless unclear if isolated formation is the
dominant channel for the production of free-floating planets.  

We here examine the efficiency of planet ionization due to scattering
events.  We perform three classes of scattering events: (1) a multi-planet
system scattering off of a single or binary star system, (2) a circumbinary
planetary system (i.e., a P-type orbit) scattering off of a single or binary
star system, and (3) a planetary system in a binary (i.e., an S-type orbit)
scattering off of a single or binary star system.  We furthermore examine
planet scattering in both open clusters and in the field.  In these
experiments the masses of the stars are set to 1 \Msun{} and the planets are
set to 0.01 \Msun{}.  In the triple systems the inner object is given a
semi-major axis of 1 AU and the outer object is given a semi-major axis of
20 AU and the orbits are set to be coplanar.  (In the case of a multi-planet
system we instead set the semi-major axis of the inner planet to 4 AU to
encourage planet-planet scattering.)  In the case of scattering off of a
binary, the incoming binary is given a semi-major axis of 100 AU.  For the
field the incoming velocity is set to 40 km s$^{-1}$ and for open clusters
it is set to 3 \kms{}.  Our assumed incoming velocity for open clusters is
larger in this section than in Sections~\ref{subsec:typeia}
and~\ref{subsec:collisions} due to the fact that the scattering experiments
we perform in this subsection are prohibitively expensive at an incoming
velocity of 0.3 \kms{}.  Consequently we calculate the cross sections at
this larger incoming velocity and later scale the results to a lower
incoming velocity.

With these initial conditions we then calculate the cross section for planet
ionization and present them in Table \ref{tbl:planet}.  In all systems in
the field the normalized cross section is within an order of magnitude of
unity.  In open clusters the cross section is enhanced by one to two orders
of magnitude because the incoming velocity is smaller by a factor of
$\sim$10.  Multi-planet--single scattering has the largest normalized cross
section due to induced planet-planet scattering.  Although this effect is
also present in multi-planet--binary scattering, its normalized cross
section is much lower simply due to the much larger area of the interloping
binary.  The physical cross section for planet ionization is slightly larger
in multi-planet--binary scattering than multi-planet--single scattering. 

\begin{table}

\caption{Cross sections for planet ionization.  We include both the
normalized cross section, $\hat{\sigma}$, and the physical cross section,
$\sigma$.  Multi-planet systems consist of two planets, circumbinary systems
consist of a planet orbiting a binary, and S-type planets consist of a
planet orbiting a star in a binary system.}

\begin{tabular}{lll}

\multicolumn{3}{l}{\textsc{Field environment}} \\

\hline

System & $\hat{\sigma}$ & $\sigma$ (AU$^2$) \\

\hline

Multi-planet--single & 6.52 $\pm$ 1.14 & 2608 $\pm$ 457 \\
Multi-planet--binary & 0.32 $\pm$ 0.05 & 3325 $\pm$ 508 \\
Circumbinary--single & 3.21 $\pm$ 0.08 & 1285 $\pm$ 31 \\
Circumbinary--binary & 4.00 $\pm$ 0.09 & 41648 $\pm$ 950 \\
S-type planet--single & 0.68 $\pm$ 0.02 & 272 $\pm$ 7 \\
S-type planet--binary & 1.26 $\pm$ 0.03 & 13147 $\pm$ 309 \\

\hline

 & & \\

\multicolumn{3}{l}{\textsc{Open cluster environment}} \\

\hline

System & $\hat{\sigma}$ & $\sigma$ (AU$^2$) \\

\hline

Multi-planet--single & 68.73 $\pm$ 14.57 & 27493 $\pm$ 5828 \\
Multi-planet--binary & 15.57 $\pm$ 4.44 & 161879 $\pm$ 46219 \\
Circumbinary--single & 53.59 $\pm$ 6.69 & 21436 $\pm$ 2675 \\
Circumbinary--binary & 15.20 $\pm$ 3.51 & 158082 $\pm$ 36466 \\
S-type planet--single & 64.43 $\pm$ 2.03 & 25771 $\pm$ 811 \\
S-type planet--binary & 7.57 $\pm$ 0.77 & 78703 $\pm$ 8046 \\

\hline

\end{tabular}
\label{tbl:planet}
\end{table}

The timescale for planet ionization can be written
\begin{multline}
t_{\textrm{scatter}} = \frac{1}{\pi n \hat{\sigma} a_{\out}^2 v}
\\
= \frac{1.1 \times 10^{12}}{\hat{\sigma}} \left( \frac{n}{10 \,
\textrm{pc}^{-3}} \right)^{-1} \left( \frac{a_{\out}}{20 \, \textrm{AU}}
\right)^{-2} \left( \frac{v}{3 \, \textrm{km} \, \textrm{s}^{-1}}
\right)^{-1} \, \textrm{yr}.
\end{multline}
Even for the outcomes with the largest cross sections ($\hat{\sigma} \sim
70$) the ionization timescale for a given system in a cluster is $\sim$20
Gyr.  We noted above that $\hat{\sigma}$ was calculated for an incoming
velocity of 3 \kms{}.  For a more realistic incoming velocity of 0.3 \kms{},
$\hat{\sigma}$ would increase by two orders of magnitude due to
gravitational focusing.  However, the scattering timescale is also
proportional to $v^{-1}$, so the overall effect of a smaller incoming
velocity would be to decrease the ionization timescale by one order of
magnitude to $\sim$2 Gyr.  This implies that planet ionization from
scattering with interloping stellar systems is rare in stellar clusters;
only $\sim$10 per cent of systems with planets on wide orbits would have
lost them in a cluster with an age of 200 Myr.  In the field, planet
ionization occurs even less frequently due to the lower densities and cross
sections.  Assuming a stellar density of 0.1 pc$^{-3}$, a typical velocity
of 40 \kms{}, and a normalized cross section of $\sim$6, the scattering
timescale is $\sim$4$\times 10^{12}$ yr.  Thus, fewer than 1 per cent of
systems with planets on wide orbits would have lost them in the lifetime of
the Galaxy.  Scattering from interloping systems can therefore produce some
free-floating planets, but not at the level claimed by \citet{sumi+11}.

\subsection{Stellar collisions during scattering events}
\label{subsec:collisions}

Although the distances between stars are typically much larger than their
radii, close encounters between stars can result in tidal interactions,
mergers, and collisions.  Candidates for observed outbursts from collisions
or mergers include V838 Mon and V1309 Sco \citep{bond+03, tylenda+soker06,
tylenda+11, kochanek+14}.  It is still unclear whether they should
preferentially occur between stars born in the same system or between stars
in different systems \citep{leigh+11, perets+kratter12}.  If collisions are
typically between stars in the same system, then these collisions would be
driven either by dynamical instabilities in the primordial system (in which
case collisions will occur not long after the protostar reaches the main
sequence, if it ever does) or due to KL oscillations.  If, however, a
substantial fraction of collisions are between stars in different systems,
then this necessarily implies that scattering would be an important channel
for stellar collisions.

\subsubsection{Numerical calculations}
\label{subsubsec:collision_numerical}

Collisions are always a possibility in resonant scattering events because
such events are chaotic.  We here rerun our model system ($a_{\textrm{in}} =
1$ AU, $a_{\out} = 10$ AU---see Section \ref{subsec:model}) but we now give
the stars radii of 1 \Rsun{}.  We do not include the effects of tides.  We
halt the calculation after a single collision, so multiple collisions are
not considered. 

To calculate these cross sections we must account for the fact that some
fraction of the observed collisions in triple-single and triple-binary
scattering are due to KL oscillations and are not induced by scattering.  To
correct for this effect we run an identical set of triples in isolation for
the same length of time and calculate the fraction of triples that collide.
This fraction ($\sim$few per cent) is then subtracted from the observed
collision fraction from the scattering experiments before the cross section
is calculated.  We find that the normalized cross section for collision at
the critical velocity is $\hat{\sigma}_0 = 0.109 \pm 0.004$ for
binary-binary scattering, $\hat{\sigma}_0 = 0.134 \pm 0.004$ for
triple-single scattering, and $\hat{\sigma}_0 = 0.235 \pm 0.007$ for
triple-binary scattering.  

Because these calculations are all done in the equal mass case, they are
biased towards scattering-induced collisions and away from KL-induced
collisions because EKM oscillations will be stronger, leading to a larger
collision fraction for triples in isolation.  It is also plausible that at
more extreme mass ratios ionizations will be a more common outcome (see
Section~\ref{subsubsec:massratio}), leading to fewer scattering-induced
collisions.  This implies that the collision cross sections derived above
are likely upper limits.  We do not calculate the velocity dependence of the
collision cross sections, however, so we cannot rigorously assert that these
cross sections are upper limits.

\subsubsection{Rate calculation}

We show in Appendix~\ref{sec:appendix} that the velocity dependence of the
collision cross section for three-body interactions scales as $\hat{v}^{-2}$
for $\hat{v} \ll 1$ and $\hat{v}^{-6}$ for $\hat{v} \gg 1$ so long as
$\hat{v}$ is much less than the escape speed of the stars.  Collisions are
therefore an exchange-like process.  Because of this, the rate of stellar
collisions from scattering may be estimated in the same way as the new
triple formation rate was estimated in Section~\ref{subsec:typeia}.  We here
follow the same analysis to estimate the rate of collisions due to
scattering with one modification.  Because we hold the stellar radii fixed,
the ratio $(R/a)$ varies as we vary $a$.  \citet{fregeau+04} found that the
collision cross section scales linearly with the ratio $(R/a)$ for
binary-binary scattering, so we incorporate this factor into the rate
calculation:
\begin{equation}
\sigma = \pi a^2 \hat{\sigma}_0 \frac{v^2_{\crit}}{v^2} \left( \frac{a_0}{a}
\right) = \frac{\pi a_0 \hat{\sigma}_0 \eta^2}{v^2}, \quad \hat{v} < 1,
\end{equation}
and
\begin{equation}
\sigma = \pi a^2 \hat{\sigma}_0 \frac{v_{\crit}^6}{v^6} \left( \frac{a_0}{a}
\right) = \frac{\pi a_0 \hat{\sigma}_0 \eta^6}{a^2 v^6}, \quad \hat{v} > 1,
\end{equation}
where $a_0$ is the outer semi-major axis at which $\hat{\sigma}_0$ was
calculated of 10 AU and $\eta$ is defined as in Section~\ref{subsec:typeia}.
Note that we assume here that the collision cross sections for triple-single
and triple-binary scattering carry the same $(R/a)$ dependence as
binary-binary scattering.  Proceeding to integrate
equation~(\ref{eq:differential_rate}) as we did before, we find
\begin{equation}
\label{eq:Gamma_coll}
\Gamma_{\textrm{collision}} = \frac{\pi N_0 n \hat{\sigma}_0 \eta^2 a_0}{v}
\left[ \ln \left( \frac{a_{\crit}}{a_{\min}} \right) + \frac{1}{2} \left(
\frac{\eta^4}{v^4} \right) \frac{1}{a_{\crit}^2} \right],
\end{equation}
where, as in equation~(\ref{eq:Gamma_newtrip}), we have neglected
$a_{\max}$. 

We now evaluate $\Gamma_{\textrm{collision}}$ for the field, open cluster
environments, and globular cluster environments using the same assumptions
as we did in Section~\ref{subsec:typeia}.  We additionally assume that the
mass of the Galactic disk is $5 \times 10^{10}$ \Msun{}
\citep{dehnen+binney98}.  Given a mean stellar mass of 0.36 \Msun{}
\citep{maschberger13}, this implies that the total number of systems in the
Galactic disk is $1.4 \times 10^{11}$.  Making use of our assumption that 10
per cent of systems are triples \citep{raghavan+10}, this implies that the
total number of triples is $1.4 \times 10^{10}$, and the normalization
constant of equation~(\ref{eq:N0}) is $N_0 = 10^9$ for the field.  We assume
the same normalization constant for open clusters as we did in
Section~\ref{subsec:typeia} of $N_0 = 2 \times 10^5$.  Furthermore, assuming
a typical stellar mass of 0.36 \Msun{} we find $\eta^2 \sim 850$ AU km$^2$
s$^{-2}$.  With these numbers we estimate $\Gamma_{\textrm{collision}} \sim
2 \times 10^{-6}$ yr$^{-1}$ in the field, $\sim$9$ \times 10^{-6}$ yr$^{-1}$
in open cluster environments.  For globular clusters we assume that the
binary fraction is 3 per cent \citep{milone+12} and the ratio of the triple
fraction to the binary fraction is 0.2 as it is in the field.  This implies
a normalization constant for globular clusters of $N_0 = 3 \times 10^6$ and
a collision rate of $\sim$8$ \times 10^{-4}$ yr$^{-1}$ in globular cluster
environments.  The ratio between the estimated rate in the open clusters and
the field is much smaller here than in the case of our SN Ia rate estimate
because $a_{\crit}$ is much larger in open clusters, leading to smaller
values of $(R/a)$, and hence smaller collision cross sections.  

These rates are much smaller than the observed rate of stellar mergers in
the Galaxy of $\sim$0.5 yr$^{-1}$ \citep{kochanek+14}.  This implies that
scattering is a relatively minor contributor to the stellar merger rate.  Of
course, since $\sim$10 per cent of all triples are formed at inclinations
$|\cos i| \leq 0.1$, many of these systems will undergo merger events, and
will therefore likely be a much more important contributor to the stellar
merger rate.  We also note that as in Section~\ref{subsec:typeia}, our
assumption of dynamical isolation is likely invalid in an open cluster
environment.  We save a treatment of the full dynamics of open clusters for
a future work. 

\subsection{How long do high inclination triples survive?}
\label{subsec:kl_lifetime}

Although KL oscillations can drive interesting behavior in a diverse set of
systems in isolation, real systems are not isolated.  Perturbations to the
gravitational potential of a hierarchical triple generally suppress the KL
resonance.  While several of these effects have already been studied in
detail (e.g., relativistic precession), scattering has not.  

Scattering will suppress KL oscillations on the timescale of an interaction
in which the triple is scattered from a high-inclination state to a
low-inclination state or some other configuration in which KL oscillations
are not present (e.g., two binaries).  Moreover, if this timescale is much
shorter than the timescale of KL oscillations themselves, KL oscillations
will not occur at all.

To calculate the lifetime of high inclination triples we performed a suite
of scattering experiments over a range of outer semi-major axes with a
semi-major axis ratio of 10.  We performed experiments in two conditions:
the field, for which we take the incoming velocity to be 40 \kms{}; and
globular clusters, for which we take the incoming velocity to be 6 \kms{}.
The scattering timescale is calculated from the cross section for the triple
to either disrupt or change its inclination from the range $80^{\circ} < i <
110^{\circ}$ (see Section~\ref{subsec:typeia}) to less than the Kozai angle
($i < 39^{\circ}$ or $i > 141^{\circ}$).  We take the number density of
stars to be 0.25 pc$^{-3}$ in the field and $10^4$ pc$^{-3}$ in globular
clusters. 

\begin{figure}
\centering
\includegraphics[width=8cm]{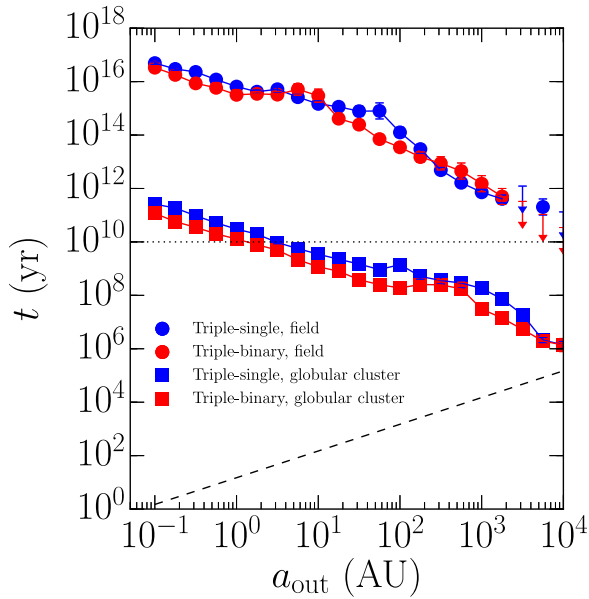}

\caption{The lifetime of high-inclination triple systems in globular
clusters and in the field.  Triple systems in the field persist for a Hubble
time out to widths of $10^5$ AU and triple systems in globular clusters
persist for a Hubble time (black dotted line) out to widths of a few AU.  We
also include the KL timescale (black dashed line).  For all but the widest
triples in globular clusters the scattering time is much longer than
$t_{\textrm{KL}}$.  All but the very widest triples in globular clusters
therefore undergo many KL oscillations before disruption by scattering.  The
scattering timescale is proportional to the inverse of the outer semi-major
axis (blue dotted line).}

\label{fig:kl_lifetime}
\end{figure}

The scattering timescales for triple-single and triple-binary scattering,
along with $t_{\textrm{KL}}$ is shown in Fig.~\ref{fig:kl_lifetime}.  The
scattering timescale, $t_{\textrm{scat}}$, is proportional to $a^{-1}$
because, for fixed incoming velocity, $\sigma / a^2 \propto
v_{\textrm{crit}}^2$ \citep{hut+bahcall83} and $v_{\textrm{crit}}^2 \propto
a$.

In the field the density of stars is so low that high inclination triples
persist for a Hubble time except for the very widest triples
($a_{\textrm{out}} \gtrsim 10^5$ AU).  Globular clusters are dense enough,
however, that high-inclination triples wider than $\sim$few AU will be
disrupted in a Hubble time.  But even in globular clusters, triples as wide
as 1000 AU will undergo many KL oscillations before disruption by
scattering.  This analysis ignores the cumulative effect of many
perturbations to the angular momentum of the outer binary, however.  At
velocities much larger than the critical velocity, perturbations to the
angular momentum of a binary become much larger than perturbations to its
energy.  Very distant passages cause the angular momentum of wide binaries
to undergo a random walk and can generally lead to extremely large
eccentricities before disruption occurs \citep{kaib+raymond14}.  In the case
of a triple with a wide outer binary, these perturbations can drive the
triple to instability in less time than it would take the triple to disrupt
due to scattering alone.  Moreover, if the cumulative effect of many distant
passages can be modeled as a global tidal field, the outer binary of a wide
triple will undergo eccentricity oscillations similar to KL oscillations
\citep{katz+dong11}.  Since the cumulative effect of multiple distant
perturbations can have a substantial impact on the orbital parameters of a
triple system on long timescales, particularly in dense environments like
globular clusters, these long-term effects should be more fully
investigated.

\section{Conclusions}
\label{sec:conclusion}

We have explored the properties of binary-binary, triple-single, and
triple-binary scattering events in a wide variety of contexts using over 400
million numerical scattering experiments.  We have calculated the cross
sections for the outcomes of these scattering events in several model
systems and grouped these outcomes into a small number of outcome classes
(Tables~\ref{tbl:tripsing}, \ref{tbl:tripbin}, \ref{tbl:binbin}, and
\ref{tbl:cat}).  These outcome classes can, in turn, be broadly grouped into
``exchange-like'' and ``ionization-like'' outcomes based on their velocity
dependence in analogy to binary-single scattering.  Ionization-like outcomes
exhibit a $\hat{v}^{-2}$ dependence at large $\hat{v}$ and exchange-like
outcomes exhibit a steeper velocity dependence at large $\hat{v}$ (often
$\hat{v}^{-6}$, but sometimes $\hat{v}^{-4}$---see Fig.~\ref{fig:vinf}).
The statistical uncertainties on these cross sections are on the order of a
few per cent or less.  The systematic uncertainty is comparable, but only
serves to increase the cross sections of ionization-like outcomes.  

We find that the cross section for new triple formation is
``exchange-like,'' and at the critical velocity the normalized cross section
for new triple formation is $\hat{\sigma}_0 = 0.1$ for both triple-single
and triple-binary scattering, and $\hat{\sigma}_0 = 0.02$ for binary-binary
scattering.  At high velocities the new triple cross section has a
$\hat{v}^{-6}$ dependence for binary-binary scattering.  For triple-single
scattering, the new triple cross section also has a $\hat{v}^{-6}$
dependence for velocities slightly in excess of the critical velocity, but
at high velocities, scrambles, which have a $\hat{v}^{-2}$ dependence, are
the dominant channel to form new triples.  Similarly, the cross section for
quadruple formation from triple-binary scattering is an exchange-like
process with a $\hat{v}^{-6}$ dependence at high velocity.  These results
imply that quadruple formation and triple formation in environments
dominated by binary-binary scattering are efficient only if the velocity is
below the critical velocity (i.e., in cluster environments).  In high
velocity environments, triple-single scattering is the dominant channel to
produce new triples.

We provide analytic fits for the velocity dependence of the ionization cross
sections in binary-single, binary-binary, and triple-single scattering
(equations~\ref{eq:exanalytic} and \ref{eq:ionanalytic}).  We also provide
analytic fits for the velocity dependence of the new triple cross sections
in binary-binary and triple-single scattering
(equations~\ref{eq:vnewtrip_bb} and \ref{eq:vnewtrip_ts}).

We measure the dependences of the cross sections on various orbital
parameters.  We find that the dependence of the cross sections on the
semi-major axis ratio can be attributed entirely to changes in the critical
velocities of the various components, at least as long as the semi-major
axis ratio is greater than $\sim$10 (Fig.~\ref{fig:alpha}).

We find that there is no eccentricity dependence on the cross sections,
except insofar as triples with more eccentric tertiaries come closer to the
boundary of stability (Fig.~\ref{fig:ecc}).  Weaker perturbations to the
eccentricity of the tertiary can therefore cause the triple to destabilize,
leading to a modest increase in the ionization cross section for large outer
eccentricities.  This increase is well modeled using our fit to the cross
section for changes to the outer eccentricity in flyby outcomes.  We also
studied the mass dependence of the cross sections and find that in general
cross sections increase for ionization-like outcomes in binary-binary and
triple-binary scattering as the mass of a single star tends toward infinity
(Fig.~\ref{fig:massratio}).  In triple-single scattering the cross sections
for all outcomes decrease in the high mass limit because the interloping
system cannot be disrupted. 

Almost all scattering events are simple flybys which leave the hierarchical
structure intact.  We calculate the cross section for changes to the orbital
parameters after flybys in triple-single and triple-binary scattering events
(Fig.~\ref{fig:cumu}).  We find that changes to the eccentricity of the
tertiary are well fit by a Gompertz function (equation~\ref{eq:gompertz}).
The normalized cross section for a large change in the inclination of the
triple ($\Delta \cos i \sim 0.5$) is $\hat{\sigma} \sim 0.1$ after
correcting for changes to the inclination due to Kozai-Lidov oscillations
(panel c of Fig.~\ref{fig:cumu}).  

We study the properties of triples formed from scattering events and find
that dynamically formed triples are extremely compact
(Figs.~\ref{fig:dyntrip} and \ref{fig:poptrip}).  Indeed, most dynamically
formed triples are very close to the stability boundary although longer-term
simulations indicate that they are stable for at least hundreds of orbits.
Because these triples are so compact, the timescale for Kozai-Lidov
oscillations is very short, sometimes just a few times the outer orbital
period (Figs.~\ref{fig:tkdist} and \ref{fig:poptkl}).  These triples should
therefore exhibit strong non-secular dynamics.  We furthermore find that the
inclination distribution of dynamically formed triples is approximately
uniform in $\cos i$ and the distribution of outer eccentricities is
approximately thermal.  These results hold both for a model system and for a
population study where we model scattering events in the field. 

Because many-body gravitational dynamics are ubiquitous in astrophysics, our
results have implications in a variety of different contexts.  We find that
scattering in the field cannot produce high inclination triple systems at a
rate consistent with the SN Ia rate, particularly in ellipticals.  This
result poses problems for the triple scenario for SNe Ia because high
inclination triples need to be produced after the stars that eventually
comprise the inner binary of the triple evolve into WDs.  If this does not
occur KL oscillations will drive the inner binary to tidal circularization
and will suppress future KL oscillations.  We show in
Section~\ref{subsec:typeia} that scattering in open clusters is more
efficient and can produce high inclination triples at a rate a factor of
$\sim$20 below the SN Ia rate.  However, it is unlikely that our assumption
of dynamical isolation in open clusters is valid, so the formation of high
inclination triples may be more efficient than we have estimated when the
full dynamics of the cluster are considered.  A triple scenario where
intermediate mass stars in binaries or triple systems evolve to WDs and then
scattering leads to high inclination tertiaries may therefore be an
important contributor to the prompt component of the SN Ia delay time
distribution.  If the time for KL oscillations to drive the WD-WD binary to
merger is much longer than the lifetime of the cluster, scattering in open
clusters could also contribute to the delayed component of the delay time
distribution.  We save a complete exploration of scattering in cluster
environments and its implications for SNe Ia for a future work.

Scattering can also lead to collisions between stars.  We show that the
velocity dependence for collisions in binary-single scattering due to
three-body interactions is exchange-like with a high-velocity dependence of
$\hat{v}^{-6}$.  We estimate the rate of stellar collisions in the Galaxy
due to triple scattering to be $\sim$2$ \times 10^{-6}$ yr$^{-1}$.  In open
cluster environments the estimated collision rate is comparable, $\sim$9$
\times 10^{-6}$ yr$^{-1}$.  In globular cluster environments the estimated
collision rate is larger, $\sim$8$ \times 10^{-4}$ yr$^{-1}$, due
principally to the higher number densities and higher velocity dispersion
relative to open clusters.  However, our assumptions about dynamical
isolation are likely not valid in an cluster environments
\citep{geller+leigh15}.  The collision rate in clusters may therefore be
much higher than our na\"ive analysis suggests.

We additionally apply our results to several different types of planetary
systems and find that scattering from external encounters is a negligible
contributor to the free-floating planet population compared to planet-planet
scattering.  However, in dense open clusters nearly 10 per cent of planets
on wide ($\sim$100 AU) orbits will be ionized due to scattering with
interloping systems. 

We finally examine the stability of KL oscillations to scattering events
(Fig.~\ref{fig:kl_lifetime}).  We find that the cross section for the triple
to move from a high-inclination to a low-inclination state is small enough
that the timescale to scatter out of a regime in which KL oscillations take
place is much longer than the Hubble time for systems in the field and all
systems with outer semi-major axes  $\lesssim$few AU in globular clusters.
Furthermore, in both the field and in globular clusters the KL timescale is
much shorter than the scattering timescale.  Thus triples with moderate
semi-major axis ratios in any environment may be considered sufficiently
isolated that at least several KL oscillations will proceed undisturbed.

\section*{Acknowledgments}

We thank John Fregeau for releasing \textsc{Fewbody} under the GNU Public
License.  JMA thanks Benjamin Shappee and Daniel Fabrycky for useful
discussion.  JMA thanks Nathan Leigh and Adrian Hamers for a careful read of
the manuscript and helpful comments.  The authors thank the referee for a
timely and helpful report.  This paper made use of \textsc{MatPlotLib}
\citep{hunter07}, \textsc{IPython} \citep{perez+granger07}, and GNU parallel
\citep{tange11}.  This research made use of Astropy, a community-developed
core Python package for Astronomy \citep{astropy13}.  This work was
supported in part by an allocation of computing time from the Ohio
Supercomputer Center.  This research was supported by the National Science
Foundation under NSF AST Award No.~1313252.

\bibliographystyle{mn2e}
\bibliography{refs}

\begin{appendix}

\section{The collision cross section}
\label{sec:appendix}

The dependence of the collision cross section in few-body scattering
problems on the incoming velocity and stellar radius has been investigated
by a number of studies \citep[e.g.,][]{fregeau+04, leigh+geller15}, but an
analytic fit to these dependences does not exist in the literature.  We
present such a fit here in the case of equal mass stars.

Collisions can occur in two different ways: through a three-body interaction
or through a two-body interaction.  In a three-body interaction the
interloping star exchanges momentum with one star of the binary before
eventually colliding with the other.  In a two-body interaction the
interloping star collides into one of the stars of the binary without
interacting with the other at all. 

For three-body interactions, the collision cross section scales with
velocity in the same way that the cross section for exchange does.  That is,
for $\hat{v} < v_{\crit}$ the velocity scaling is $\hat{v}^{-2}$ and for
$\hat{v} > v_{\crit}$ the velocity scaling is $\hat{v}^{-6}$.  The
low-velocity scaling is due to gravitational focusing, whereas the
high-velocity scaling is likely because the mechanism by which collisions
occur in three-body scattering events is similar to the mechanism by which
exchanges occur.  Exchanges occur when two stars undergo a close approach
and exchange momenta \citep{hut83}.  This ejects one star formerly in the
binary and replaces it with the interloping star on a similar orbit.  A
collision may occur if the momentum exchange places the interloping star on
an orbit eccentric enough to lead to a collision.  Since the mechanism for
collision is similar, we would then expect the collision cross section to
display the same $\hat{v}^{-6}$ velocity dependence as the exchange cross
section.  Note, however, that this is simply a plausibility argument.  We do
not attempt here to show rigorously that this is the mechanism responsible
for the observed velocity dependence.  Nevertheless, the numerical
calculations indicate that the cross sections indeed show a $\hat{v}^{-6}$
velocity dependence in this regime just like exchange. 

We may combine these velocity dependences as we did in
Section~\ref{subsec:binsingapprox} to find the overall cross section:
\begin{equation}
\frac{1}{\hat{\sigma}_{\textrm{3-body}}} = \frac{\hat{v}^2}{\hat{\sigma}_a}
+ \frac{\hat{v}^6}{\hat{\sigma}_b},
\end{equation}
where $\hat{\sigma}_a$ and $\hat{\sigma}_b$ are constants of order unity
that must be fit to numerical data. 

Let us consider now two-body interactions.  Like three-body interactions,
these exhibit two velocity regimes.  However, in this case the two velocity
regimes are separated by the escape velocity of the stars, $v_{\esc}$.  For
$\hat{v} \gg v_{\esc}$, gravitational focusing from the individual stars is
negligible and the cross section for collision will approach a constant
value equal to the geometric cross section of the stars.  For velocities
less than $v_{\esc}$, gravitational focusing from individual stars will lead
to a $\hat{v}^{-2}$ dependence.  For two-body interactions, then, the
overall cross section will be
\begin{equation}
\hat{\sigma}_{\textrm{2-body}} = \hat{\sigma}_{\textrm{geometric}} +
\frac{\hat{\sigma}_c}{\hat{v}^2},
\end{equation}
where the geometric cross section is just given by
\begin{equation}
\hat{\sigma}_{\textrm{geometric}} = 2 \pi \left( \frac{R}{a} \right)^2,
\end{equation}
and $\hat{\sigma}_c$ must be fit to numerical data, though it will be of
order $\hat{\sigma}_{\textrm{geometric}} \hat{v}_{\esc}^2$.

\begin{figure}
\centering
\includegraphics[width=8cm]{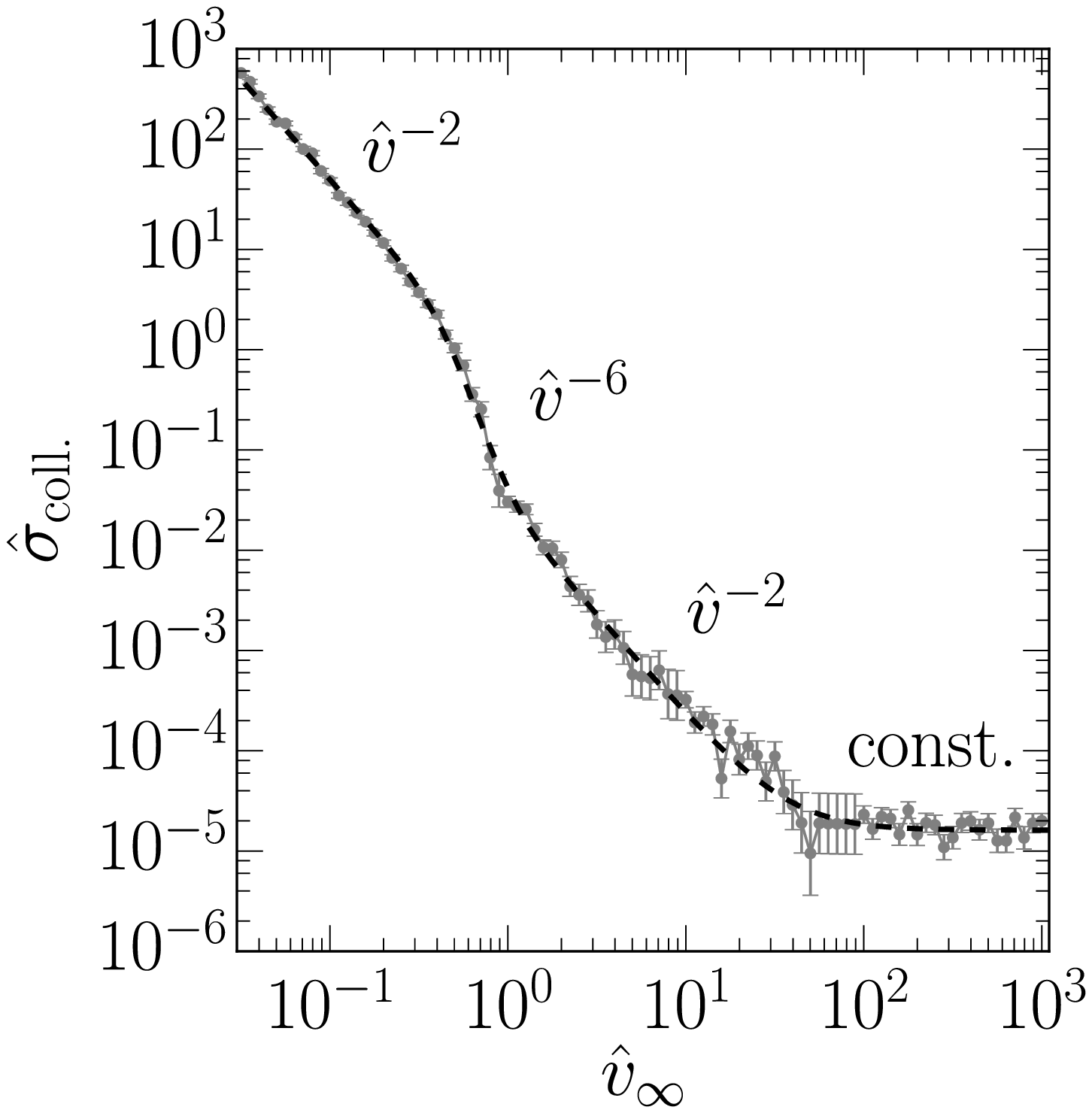}

\caption{The cross section for a collision to occur in binary-single
scattering as a function of the normalized incoming velocity.  We show both
numerical data (gray points) and a fit to equation (\ref{eq:coll_vdep})
(black dashed line).  This system consists of three 1 $M_{\odot}$ stars each
with a radius of 0.3 $R_{\odot}$.  The binary orbit is circular with a
semi-major axis of 1 AU.  The best-fitting parameters are $\hat{\sigma}_a
\approx 0.475$, $\hat{\sigma}_b \approx 2.00 \times 10^{-2}$,
$\hat{\sigma}_c \approx 2.25 \times 10^{-2}$, and
$\hat{\sigma}_{\textrm{geometric}} \approx 1.61 \times 10^{-5}$.}

\label{fig:coll_vdep}
\end{figure}

The collision cross sections for two- and three-body interactions may then
be combined additively to yield the overall collision cross section:
\begin{equation}
\label{eq:coll_vdep}
\hat{\sigma}_{\textrm{coll.}} = \hat{\sigma}_{\textrm{2-body}} +
\hat{\sigma}_{\textrm{3-body}} = \left( \frac{\hat{v}^2}{\hat{\sigma}_a} +
\frac{\hat{v}^6}{\hat{\sigma}_b} \right)^{-1} +
\frac{\hat{\sigma}_c}{\hat{v}^2} + \hat{\sigma}_{\textrm{geometric}}.
\end{equation}
A fit of equation (\ref{eq:coll_vdep}) to numerical data is shown in
Fig.~\ref{fig:coll_vdep}.  There is an excellent match between the numerical
data and the analytic fit.

\end{appendix}

\end{document}